\documentclass{sig-alternate}
\usepackage{times,amsmath,epsfig}

\long\def\comment#1{}

\usepackage{color}
\usepackage{graphicx}
\usepackage{xspace}
\usepackage{subfigure}
\usepackage{amsmath}
\usepackage{float}
\usepackage{amssymb}
\usepackage{latexsym}
\usepackage{graphicx}
\usepackage{url}
\usepackage{epsfig}
\usepackage{mathrsfs}
\usepackage{times}
\usepackage{multirow}
\usepackage{color}
\usepackage{epstopdf}
\usepackage{algorithm}
\usepackage{algorithmic}

\newcounter{example}[section]
\renewcommand{\theexample}{\nthesection.\arabic{example}}
\newenvironment{example}{
     \refstepcounter{example}
     {\vspace{1ex} \noindent\bf  Example  \theexample:}}{
     \vspace{1ex}} 

\newcounter{definition}[section]
\renewcommand{\thedefinition}{\nthesection.\arabic{definition}}

\newcounter{theorem}[section]
\renewcommand{\thetheorem}{\nthesection.\arabic{theorem}}
\newenvironment{theorem}{\begin{em}
        \refstepcounter{theorem}
        {\vspace{1ex} \noindent\bf  Theorem  \thetheorem:}}{
        \end{em}\vspace{1ex}} 

\newcounter{lemma}[section]
\renewcommand{\thelemma}{\nthesection.\arabic{lemma}}
\newenvironment{lemma}{\begin{em}
        \refstepcounter{lemma}
        {\vspace{1ex}\noindent\bf Lemma \thelemma:}}{
        \end{em}\vspace{1ex}} 

\newcounter{remark}[section]
\renewcommand{\theremark}{\nthesection.\arabic{remark}}

\newcommand{\proofsketch}{\noindent{\bf Proof Sketch: }}

\newcommand{\nthesection}{\arabic{section}}

\newcommand{\eop}{\hspace*{\fill}\mbox{$\Box$}}

\newcommand{\stitle}[1]{\vspace{1ex} \noindent{\bf #1}}

\newcommand{\kw}[1]{{\ensuremath {\mathsf{#1}}}\xspace}

\newcommand{\kwnospace}[1]{{\ensuremath {\mathsf{#1}}}}

\newcommand{\len}{\kw{len}}
\newcommand{\simpleAlg}{{\sl BF-U}\xspace}
\newcommand{\scc}{\kw{SCC}}
\newcommand{\sccs}{\kwnospace{SCC}s\xspace}

\newcommand{\DFSSPFA}{{\sl FindNC}\xspace}
\newcommand{\DFS}{{\sl DFS}\xspace}
\newcommand{\BFS}{{\sl BFS}\xspace}
\newcommand{\DFSEVEN}{{\sl DS-U}\xspace}
\newcommand{\refineAlg}{{\sl GR-U}\xspace}
\newcommand{\greedy}{{\sl Greedy}\xspace}
\newcommand{\greedyD}{{\sl Greedy-D}\xspace}
\newcommand{\greedyR}{{\sl Greedy-R}\xspace}
\newcommand{\refine}{{\sl Refine}\xspace}
\newcommand{\lengthLD}{{\sl PN-path-D}\xspace}
\newcommand{\lengthL}{{\sl PN-path}\xspace}
\newcommand{\lengthLR}{{\sl PN-path-R}\xspace}
\newcommand{\lsubgraph}{{\sl $l$-Subgraph}\xspace}
\newcommand{\agony} {{\sl agony}\xspace}

\newcommand{\vlabel}{\kw{label}}
\newcommand{\vlevel}{\kw{level}}
\newcommand{\rvlevel}{\kw{rlevel}}
\newcommand{\pnpath}{\kwnospace{pn}-\kw{path}}
\newcommand{\pnpaths}{\kwnospace{pn}-\kwnospace{path}s\xspace}

\newcommand{\vrank}{\kw{rank}}

\newcommand{\ppath}{\kwnospace{p}-\kw{path}}
\newcommand{\ppaths}{\kwnospace{p}-\kwnospace{path}s\xspace}
\newcommand{\pedge}{\kwnospace{p}-\kw{edge}}
\newcommand{\pedges}{\kwnospace{p}-\kwnospace{edge}s\xspace}
\newcommand{\pcycle}{\kwnospace{positive}-\kw{cycle}}

\newcommand{\npath}{\kwnospace{n}-\kw{path}}
\newcommand{\npaths}{\kwnospace{n}-\kwnospace{path}s\xspace}
\newcommand{\nedge}{\kwnospace{n}-\kw{edge}}
\newcommand{\nedges}{\kwnospace{n}-\kwnospace{edge}s\xspace}
\newcommand{\ncycle}{\kwnospace{negative}-\kw{cycle}}
\newcommand{\ncycles}{\kwnospace{negative}-\kwnospace{cycle}s\xspace}

\newcommand{\parent}{\kw{parent}}
\newcommand{\kcycle}{\kwnospace{k}-\kw{cycle}}
\newcommand{\kcycles}{\kwnospace{k}-\kwnospace{cycle}s\xspace}
\newcommand{\eulerian}{{\cal U}}
\newcommand{\appeulerian}{\widetilde{\cal U}}

\newcommand{\refineAlgD}{{\sl GR-U-D}\xspace}
\newcommand{\refineAlgR}{{\sl GR-U-R}\xspace}

\newcommand{\shigher} {{\sl{strictly}-\sl{higher}}\xspace}
\newcommand{\whigher} {{\sl{weakly}-\sl{higher}}\xspace}

\begin{document}

\title{Exploring Hierarchies in Online Social Networks}

\author{%
{Can Lu, Jeffrey Xu Yu, Rong-Hua Li$^*$, Hao Wei}%
\vspace{1.6mm}\\
\fontsize{10}{10}\selectfont\itshape
The Chinese University of Hong Kong, Hong Kong, China \\
\vspace{1.6mm}
\fontsize{10}{10}\selectfont\itshape
$^{*}$Guangdong Province Key Laboratory of Popular High Performance
Computers, Shenzhen University, China \\
\fontsize{9}{9}\selectfont\ttfamily\upshape
\{lucan,yu,rhli,hwei\}@se.cuhk.edu.hk
}


\def\thepage{\arabic{page}}
\pagestyle{plain}

\maketitle

\thispagestyle{plain}

\begin{abstract}
Social hierarchy (i.e., pyramid structure of societies) is a
fundamental concept in sociology and social network analysis.  The
importance of social hierarchy in a social network is that the
topological structure of the social hierarchy is essential in both
shaping the nature of social interactions between individuals and
unfolding the structure of the social networks. The social hierarchy
found in a social network can be utilized to improve the accuracy of
link prediction, provide better query results, rank web pages, and
study information flow and spread in complex networks.  In this paper,
we model a social network as a directed graph $G$, and consider the
social hierarchy as DAG (directed acyclic graph) of $G$, denoted as
$G_D$. By DAG, all the vertices in $G$ can be partitioned into
different levels, the vertices at the same level represent a disjoint
group in the social hierarchy, and all the edges in DAG follow one
direction. The main issue we study in this paper is how to find DAG
$G_D$ in $G$. The approach we take is to find $G_D$ by removing all
possible cycles from $G$ such that $G = \eulerian(G) \cup G_D$ where
$\eulerian(G)$ is a maximum Eulerian subgraph which contains all
possible cycles.  We give the reasons for doing so, investigate the
properties of $G_D$ found, and discuss the applications.  In addition,
%
%
we develop a novel two-phase algorithm, called Greedy-\&-Refine, which
greedily computes an Eulerian subgraph and then refines this greedy
solution to find the maximum Eulerian subgraph. We give a bound
between the greedy solution and the optimal.
The quality of our greedy approach is high.
We conduct comprehensive
experimental studies over 14 real-world datasets. The results show
that our algorithms are at least two orders of magnitude faster than
the baseline algorithm.
\end{abstract}

\section{Introduction}

Social hierarchy refers to the pyramid structure of societies, with
minority on the top and majority at the bottom, which is a prevalent
and universal feature in organizations. Social hierarchy is also
recognized as a fundamental characteristic of social interactions,
being well studied in both sociology and
psychology~\cite{gould2002origins}. In recent years, social hierarchy
has attracted considerable attention and generates profound and
lasting influence in various fields, especially social networks. This
is because the hierarchical structure of a population is essential in
shaping the nature of social interactions between individuals and
unfolding the structure of underlying social networks.  Gould
in~\cite{gould2002origins} develops a formal theoretical model to
model the emergence of social hierarchy, which can accurately predict
the network structure. By the social status theory
in~\cite{gould2002origins}, individuals with low status typically
follow individuals with high status.  Clauset et
al. in~\cite{clauset2008hierarchical} develop a technique to infer
hierarchical structure of a social network based on the degree of
relatedness between individuals. They show that the hierarchical
structure can explain and reproduce some commonly observed topological
properties of networks and can also be utilized to predict missing
links in networks.
Assuming that underlying hierarchy is the primary
factor guiding social interactions, Maiya and Berger-Wolf
in~\cite{maiya2009inferring} infer social hierarchy from undirected
weighted social networks based on maximum likelihood.  All these
studies imply that social hierarchy is a primary organizing
principle of social networks, capable of shedding light on many
phenomena. In addition, social hierarchy is also used in many aspects
of social network analysis and data mining. For instance, social
hierarchy can be utilized to improve the accuracy of link
prediction~\cite{liben2007link}, provide better query
results~\cite{kleinberg1999authoritative}, rank web
pages~\cite{granovetter1973strength}, and study information flow and
spread in complex
networks~\cite{acemoglu2010spread,almendral2003information}.


In this paper, we focus on social networks that can be modeled by directed graphs, because in many social networks (e.g., Google+, Weibo,
Twitter) information flow and influence propagate follow certain
directions from vertices to vertices.  Given a social network as a
directed graph $G$, its social hierarchy can be represented as a
directed acyclic graph (DAG).  By DAG, all the vertices in $G$ are
partitioned into different levels (disjoint groups), and all the edges
in the cycle-free DAG follow one direction, as observed in social
networks that prestige users at high levels are followed by users at
low levels and the prestige users typically do not follow their
followers. Here, a level in DAG represents the status of a vertex in
the hierarchy the DAG represents.

The issue we study in this paper is how to find hierarchy as a DAG in
a general directed graph $G$ which represents a social network.  Given
a graph $G$, there are many possible ways to obtain a DAG. First,
converting graph $G$ into a DAG, by contracting all vertices in a
strongly connected component in $G$ as a vertex in DAG, does not serve
the purpose, because all vertices in a strongly connected component do
not necessarily belong to the same level in a hierarchy. Second, a
random DAG does not serve the purpose, because it heavily relies on
the way to select the vertices as the start to traverse and the way to
traverse. Therefore, two random DAGs can be significantly different
topologically. Third, finding the maximum DAG of $G$ is not only
NP-hard but also NP-approximate~\cite{guruswami2008beating}.  The way
we do is to find the DAG by removing all possible cycles from $G$
following \cite{gupte2011finding}.  In~\cite{gupte2011finding} Gupte
et al. propose a way to decompose a directed graph $G$ into a maximum
Eulerian subgraph $\eulerian(G)$ and DAG $G_D$, such that $G =
\eulerian(G) \cup G_D$. Here, all possible cycles in $G$ are in
$\eulerian(G)$, and all edges in $G_D$ do not appear in
$\eulerian(G)$.
%
%
We take the same approach to find DAG $G_D$ for a graph $G$ by finding
the maximum Eulerian subgraph $\eulerian(G)$ of $G$ such that $G =
\eulerian(G) \cup G_D$, as given in~\cite{gupte2011finding}.

\stitle{Main contributions}: We summarize the main contributions of
our work as follows.
%
%
%
First, unlike~\cite{gupte2011finding} which studies a measure between
0 and 1 to indicate how close a given directed graph is to a perfect
hierarchy, we focus on the hierarchy (DAG). In addition to the
properties investigated in \cite{gupte2011finding}, we show that $G_D$
found is representative,
exhibits the pyramid rank distribution. In addition, $G_D$ found
can be used to study social mobility and recover hidden directions of social relationships.  Here, social mobility is a
fundamental concept in sociology, economics and politics, and refers
to the movement of individuals from one status to another.
Second, we significantly improve the efficiency of computing the
maximum Eulerian subgraph $\eulerian(G)$.  Note that the time
complexity of the \simpleAlg algorithm~\cite{gupte2011finding} is $O(nm^2)$,
where $n$ and $m$ are the numbers of vertices and edges,
respectively. Such an algorithm is impractical, because it can only
work on small graphs.  We propose a new algorithm with time complexity
$O(m^2)$, and propose a novel two-phase algorithm, called
Greedy-\&-Refine, which greedily computes an Eulerian subgraph in $O(n
+ m)$ and then refines this greedy solution to find the maximum
Eulerian subgraph in $O(cm^2)$ where $c$ is a very small constant less
than 1.
%
%
The quality of our greedy approach is high.
Finally, we conduct
extensive performance studies using 14 real-world datasets to evaluate
our algorithms, and confirm our findings.

\stitle{Further related work}:
%
%
%
Ball and Newman~\cite{ball2013friendship} analyze directed networks
between students with both reciprocated and unreciprocated friendships
and develop a maximum-likelihood method to infer ranks between
students such that most unreciprocated friendships are from
lower-ranked individuals to higher-ranked ones, corresponding to
status theory~\cite{gould2002origins}. Leskovec et
al. in~\cite{leskovec2010signed,leskovec2010predicting} investigate
signed networks and develop an alternate theory of status in replace
of the balance theory frequently used in undirected and unsigned
networks to both explain edge signs observed and predict edge signs
unknown. Influence has been widely studied~\cite{weiChen13}, finding
social hierarchy provides a new perspective to explore the influence
given the existence of a social hierarchy.

Eulerian graphs have been well studied in the theory community
\cite{fleischner1990eulerian,fleischner2001some,
  catlin1992supereulerian,chen1995reduction,li2004maximum}. For
example, in \cite{fleischner1990eulerian}, Fleischner gives a
comprehensive survey on this topic. In \cite{fleischner2001some}, the
same author surveys several applications of Eulerian graphs in graph
theory. Another closely related concept is super-Eulerian graph, which
contains a spanning Eulerian subgraph
\cite{catlin1992supereulerian,chen1995reduction,li2004maximum}, here a spanning Eulerian subgraph means an Eulerian subgraph that includes all vertices. The
problem of determining whether or not a graph is super-Eulerian is
NP-complete \cite{chen1995reduction}. Most of these work mainly focus
on the properties of Eulerian subgraphs. There are no much related
work on computing the maximum Eulerian subgraphs for large graphs. To
the best of our knowledge, the only one in the literature is done by
Gupte, et al.\ in \cite{gupte2011finding}. However, the time
complexity of their algorithm is $O(nm^2$), which is clearly
impractical for large graphs.


\stitle{Organization}: In Section~\ref{sec:pre}, we focus on the
properties of the social hierarchy found after giving some useful
concepts on maximum Eulerian subgraph, and discuss the
applications. In Section~\ref{sec:bf}, we discuss an existing
algorithm \simpleAlg \cite{gupte2011finding}. In
Section~\ref{sec:simplealgo}, we propose a new algorithm \DFSEVEN of time complexity $O(m^2)$, and
treat it as the baseline algorithm.  We present a new two-phase
algorithm \refineAlg for finding the maximum Eulerian subgraph, as
well as its analysis in Section~\ref{sec:gr}. Extensive experimental
studies are reported in Section~\ref{sec:performance}. Finally, we
conclude this work in Section~\ref{sec:conclusion}.

\section{The Hierarchy}
\label{sec:pre}

Consider an unweighted directed graph $G = (V,E)$, where $V(G)$ and
$E(G)$ denote the sets of vertices and directed edges of $G$,
respectively. We use $n=|V(G)|$ and $m=|E(G)|$ to denote the number of
vertices and edges of graph $G$, respectively. In $G$, a path $p = (v_1,v_2,\cdots, v_k)$ represents a sequence of
edges such that $(v_i,v_{i+1}) \in E(G)$, for each $v_i~{}(1 \leq i
<k)$. The length of path $p$, denoted as $\len(p)$, is the number of
edges in $p$.  A simple path is a path $(v_1,v_2,\cdots, v_k)$ with
$k$ distinct vertices. A cycle is a path where a same vertex appears
more than once, and a simple cycle is a path $(v_1,v_2,\cdots,
v_{k-1}, v_k)$ where the first $k-1$ vertices are distinct while $v_k
= v_1$. For simplicity, below, we use $V$ and $E$ to denote $V(G)$ and
$E(G)$ of $G$, respectively, when they are obvious.
For a vertex $v_i \in V(G)$, the in-neighbors of $v_i$, denoted as
$N_{I}(v_i)$, are the vertices that link to $v_i$, i.e., $N_{I}(v_i) =
\{v_j~{}|~{}(v_j,v_i)$ $\in E(G)\}$, and the out-neighbors of $v_i$,
denoted as $N_{O}(v_i)$, are the vertices that $v_i$ links to, i.e.,
$N_{O}(v_i) = \{v_j~{}|~{}(v_i,v_j)$ $\in E(G)\}$.  The in-degree
$d_I(v_i)$ and out-degree $d_O(v_i)$ of vertex $v_i$ are the numbers
of edges that direct to and from $v_i$, respectively, i.e.,
$d_I(v_i) = |N_I(v_i)|$ and $d_O(v_i) = |N_O(v_i)|$.

A strongly connected component (\scc) is a maximal subgraph of a
directed graph in which every pair of vertices $v_i$ and $v_j$ are
reachable from each other.

A directed graph $G$ is an Eulerian graph (or simply Eulerian) if for
every vertex $v_i \in V(G)$, $d_I(v_i) = d_O(v_i)$. An Eulerian graph
can be either connected or disconnected. An Eulerian subgraph of a
graph $G$ is a subgraph of $G$, which is Eulerian, denoted as
$G_U$. The maximum Eulerian subgraph of a graph $G$ is an Eulerian
subgraph with the maximum number of edges, denoted as
$\eulerian(G)$. Given a directed graph $G$, we focus on the problem of
finding its maximum Eulerian subgraph, $\eulerian(G)$, which does not
need to be connected.  Note that the problem of finding the maximum
Eulerian subgraph ($\eulerian(G)$) in a directed graph can be solved
in polynomial time, whereas the problem of finding the maximum
connected Eulerian subgraph is NP-hard
\cite{cai2011parameterized}. The following example illustrates the
concept of maximum Eulerian subgraph.

\begin{figure}[t]
\centering
\includegraphics[scale=0.3]{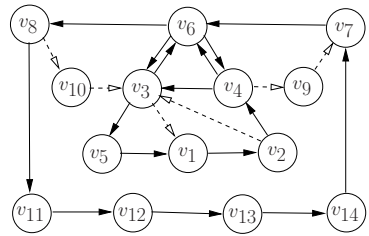}
\vspace*{-0.2cm}
\caption{Illustration of the maximum Eulerian subgraph}
\vspace*{-0.6cm}
\label{fig:fig1}
\end{figure}

\begin{example}
Fig.~\ref{fig:fig1} shows a graph $G = (V, E)$ with 14 vertices and 22 edges. Its maximum Eulerian subgraph $\eulerian(G)$ is a subgraph of $G$, where its edges are in solid lines: $E(\eulerian(G)) = \{(v_1,v_2),$ $(v_2,v_4),$ $(v_4,v_3),$ $(v_3,v_5),$ $(v_5,v_1),$
$(v_4,v_6),$ $(v_6,v_4),$ $(v_3,$ $v_6),$ $(v_6,$ $v_3),$ $(v_6,$
$v_8),(v_8,v_{11}), (v_{11},v_{12}),$ $(v_{12},$ $v_{13}),$ $(v_{13},$
$v_{14}),(v_{14},v_7),$ $(v_7,v_6)\}$, and $V(\eulerian(G))$ is the set of
vertices that appear in $E(\eulerian(G))$.
\end{example}

The main issue here is to find a hierarchy of a directed graph $G$ as
DAG $G_D$ by finding the maximum Eulerian subgraph $\eulerian(G)$ for
a directed graph $G$. With $\eulerian(G)$ found, $G_D$ can be
efficiently found due to $G = \eulerian(G) \cup G_D$, and
$E(\eulerian(G)) \cap E(G_D) = \emptyset$. We discuss the properties
of the hierarchy $G_D$ and the applications.

\stitle{The representativeness}: The maximum Eulerian subgraph
$\eulerian(G)$ for a general graph $G$ is not unique. A natural
question is how representative $G_D$ is as the hierarchy. Note that
$G_D$ is only unique w.r.t $\eulerian(G)$ found. Below, we show $G_D$
identified by an arbitrary $\eulerian(G)$ is representative based on a
notion of \shigher defined between two vertices in $G_D$, over a
ranking $r(\cdot)$ where $r(u)<r(v)$ for each edge $(u,v) \in
G_D$. Here, for two vertices $u$ and $v$, a larger rank implies a
vertex is in a higher status in a follower relationship, and $u$ is
\shigher than $v$ if $r(u)>r(v)$ and $u$ is reachable from $v$,
i.e. there is a directed path from $v$ to $u$ in $G_D$.


\begin{theorem} Let $G_{D_1}$ and $G_{D_2}$ be two
DAGs for $G$ such that $G = \eulerian_1(G) \cup G_{D_1} =
\eulerian_2(G) \cup G_{D_2}$. There are no vertices $u$ and $v$ such
that $u$ is \shigher than $v$ in $G_{D_1}$ whereas $v$ is \shigher
than $u$ in $G_{D_2}$.
\label{them:uni}
\end{theorem}

\proofsketch Assume the opposite. We can construct an auxiliary graph $G'=G \cup
\{(u,v)\}$. Then finding the maximum Eulerian subgraph for $G'$ can be
done in two steps.  In the first step, find the maximum Eulerian
subgraph $\eulerian(G)$, and in the second step, find the maximum
Eulerian subgraph for $G$ plus the additional edge $(u, v)$.  Since $G
= \eulerian_1(G) \cup G_{D_1} = \eulerian_2(G) \cup G_{D_2}$, there
are supposed to be at least two corresponding relaxing orders, when the
first phase terminates, namely, identifying $\eulerian_1(G)$ and
$\eulerian_2(G)$. For one relaxing order,
we can show that the added edge $(u,v)$ can be relaxed, which results
in finding $\eulerian(G')$ such that $|E(\eulerian(G'))| >
|E(\eulerian(G))|$. For the other relaxing order,
we can also show that the added edge $(u,v)$ cannot be relaxed and
$\eulerian(G') = \eulerian(G)$.  It leads to a contradiction, because it
can find two different maximum Eulerian subgraphs for $G'$ with
different sizes.

Alternatively, let the ranking in $G_{D_1}$ and $G_{D_2}$ be
$r_1(\cdot)$ and $r_2(\cdot)$. Assume there are two vertices $u$ and
$v$ such that $u$ is \shigher than $v$ by $r_1$ whereas $v$ is
\shigher than $u$ by $r_2$. We prove this cannot achieve based on the
finding in \cite{gupte2011finding}. In \cite{gupte2011finding}, it
gives a total score on $G$ which measures how $G$ is different from
DAG $G_D$ based on a ranking $r(\cdot)$. The total score, denoted as
$A(G,r)$, is obtained by summing up the weights assigned to edges,
$max\{r(u)-r(v)+1,0\}$ for edge $(u,v)$. The finding in
\cite{gupte2011finding} is that the minimum total score equals to the
number of edges in the maximum Eulerian subgraph, $min_r\{A(G, r)\}
=|E(\eulerian(G))|$. Choose $r_1$ and $r_2$ satisfying that
$A(G,r_1)=|E(\eulerian_1(G))|$ and $A(G,r_2)=|E(\eulerian_2(G))|$.
Since $u$ is \shigher than $v$ in $G_{D_1}$, there is a directed path
from $v$ to $u$ in $G_{D_1}$. We can construct an auxiliary graph $G'=
G \cup \{(u,v)\}$, then $|E(\eulerian(G'))|>|E(\eulerian_1(G))|$. On the
other hand, over the same $G'$, since $v$ is \shigher than $u$ in
$G_{D_2}$, we can show $|E(\eulerian(G'))| \leq A(G',r_2)=A(G,r_2)$,
which leads to a contradiction.  \eop


\stitle{A case study}: With the hierarchy (DAG $G_D$) found, suppose
we assign every vertex $u$ a minimum non-negative rank $r(u)$ such
that $r(u)<r(v)$ for any edge $(u,v) \in G_D$, where $r(\cdot)$ is a
strictly-higher rank. To show whether such ranking reflects the ground
truth, as a case study, we conduct testing using Twitter, where the
celebrities are known, for instance, refer to Twitter Top 100
(\url{http://twittercounter.com/pages/100}).
%
%
%
We sample a subgraph among 41.7 million users (vertices) and 1.47
billion relationships (edges) from Twitter social graph ${\bf G}$
crawled in 2009~\cite{kwak2010twitter}.  In brief, we randomly sample
5 vertices in the celebrity set given in Twitter, and then sample
1,000,000 vertices starting from the 5 vertices as seeds using random
walk sampling~\cite{DBLP:dblp_conf/kdd/LeskovecF06}.  We construct an
induced subgraph $G'$ of the 1,000,000 vertices sampled from ${\bf
  G}$, and we uniformly sample about 10,000,000 edges from $G'$ to
obtain the sample graph $G$, which contains 759,105 vertices and
11,331,061 edges. In $G$, we label a vertex $u$ as a celebrity, if $u$
is a celebrity and has at least 100,000 followers in ${\bf G}$. There
are 430 celebrities in $G$ including Britney Spears, Oprah Winfrey,
Barack Obama, etc. We compute the hierarchy ($G_D$) of $G$ using our
approach and rank vertices in $G_D$. The hierarchy reflects the truth:
88\% celebrities are in the top 1\% vertices and 95\% celebrities in
the top 2\% vertices.  In consideration of efficiency, we can
approximate the exact hierarchy with a greedy solution obtained by
\greedy in Section~\ref{sec:gr}. In the approximate hierarchy, 85\%
celebrities are in the top 1\% vertices and 93\% celebrities in the
top 2\% vertices.

\comment{
By the maximum Eulerian subgraph $\eulerian(G)$ and DAG $G_D$ found,
all vertices in $G$ can be partitioned into different levels where all
vertices at the same level represent a disjoint group in the social
hierarchy.  For a specific ranking $r$, vertex $u$'s rank is denoted
as $r(u)$. Gupte et al. in \cite{gupte2011finding} introduce a concept
\agony, where each edge $(u,v)$ produces an agony of
$max(r(u)-r(v)+1,0)$, and the total agony in the graph $G$ for a
specific ranking $r$ is
\begin{displaymath}
A(G,r)=\sum_{(u,v)\in E} max(r(u)-r(v)+1,0)
\end{displaymath}
They prove the problem of finding $r$ that minimize the total agony in the graph is equivalent to finding the maximum Eulerian subgraph $\eulerian(G)$
\begin{displaymath}
\begin{aligned}
A(G)&=min_{r \in Rankings}( \sum_{(u,v)\in E} max(r(u)-r(v)+1,0) ) \\
&=|E(\eulerian(G))|
\end{aligned}
\end{displaymath}
On the other hand, in this paper, we consider DAG $G_D$ as the core
hierarchy structure of $G$ and produce a ranking $r$ that for any edge
$(u,v) \in E(G_D)$, $r(u)<r(v)$. Following, we will first prove the
stability of $G_D$ and then demonstrate that our approach is more
generalized than that of \cite{gupte2011finding} and apply our ranking
approach to some applications in social network analysis in section
~\ref{sec:performance}.
}

\comment{
\begin{figure}[h]
\begin{center}
    \includegraphics[width=0.8\columnwidth,height=3cm]{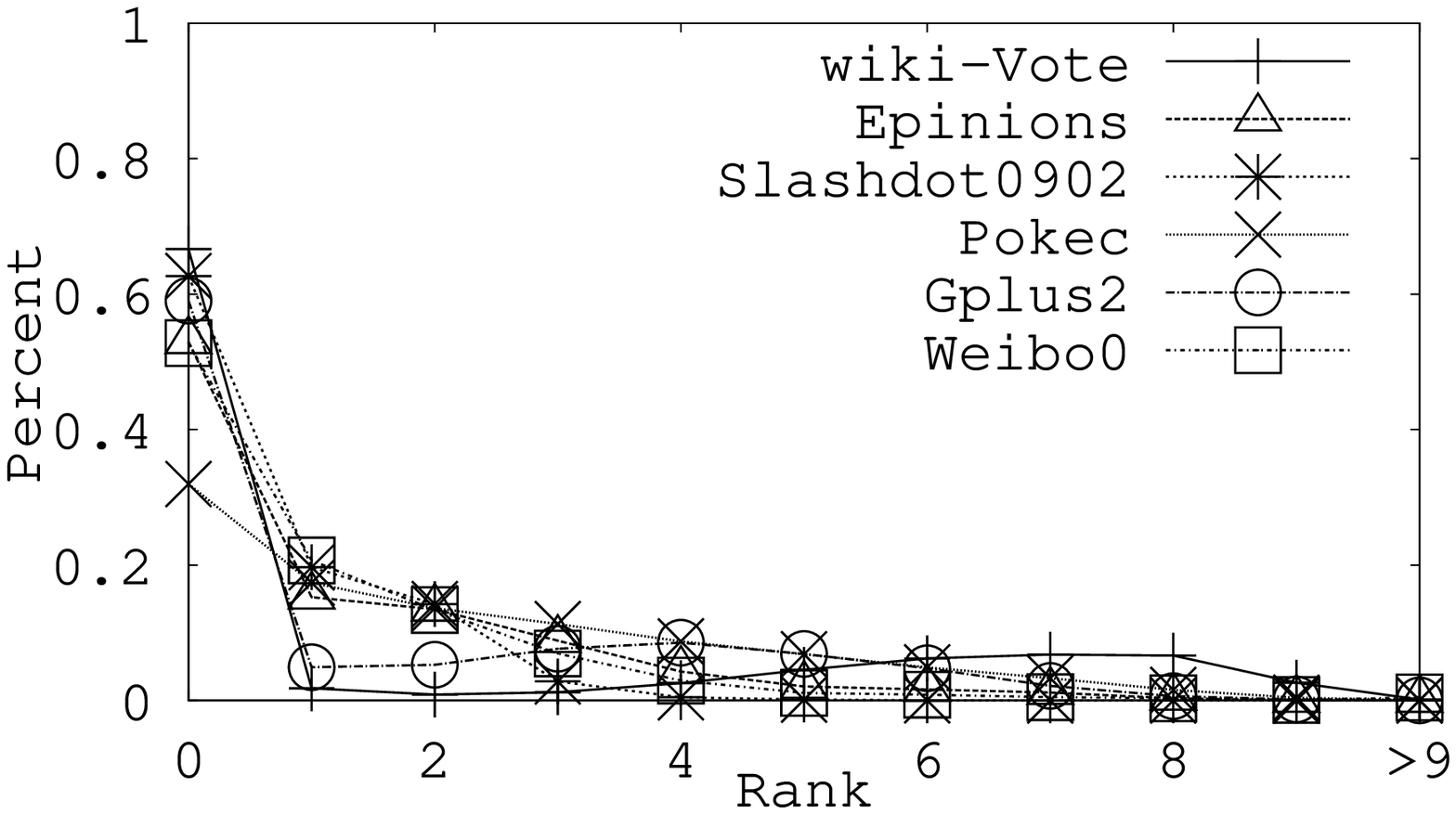}
\end{center}
\vspace*{-0.6cm}
\caption{Pyramid rank distribution}
\vspace*{-0.4cm}
\label{fig:rankdis}
\end{figure}
}

\begin{figure}[t]
\begin{center}
\begin{tabular}[t]{c}
  \hspace*{-0.5cm}
    \subfigure[The hierarchy]{
\includegraphics[width=0.33\columnwidth,height=2cm]{chart/RankDis}
         \label{fig:hierarchyRD}
    }
    \subfigure[Random DAG]{
\includegraphics[width=0.33\columnwidth,height=2cm]{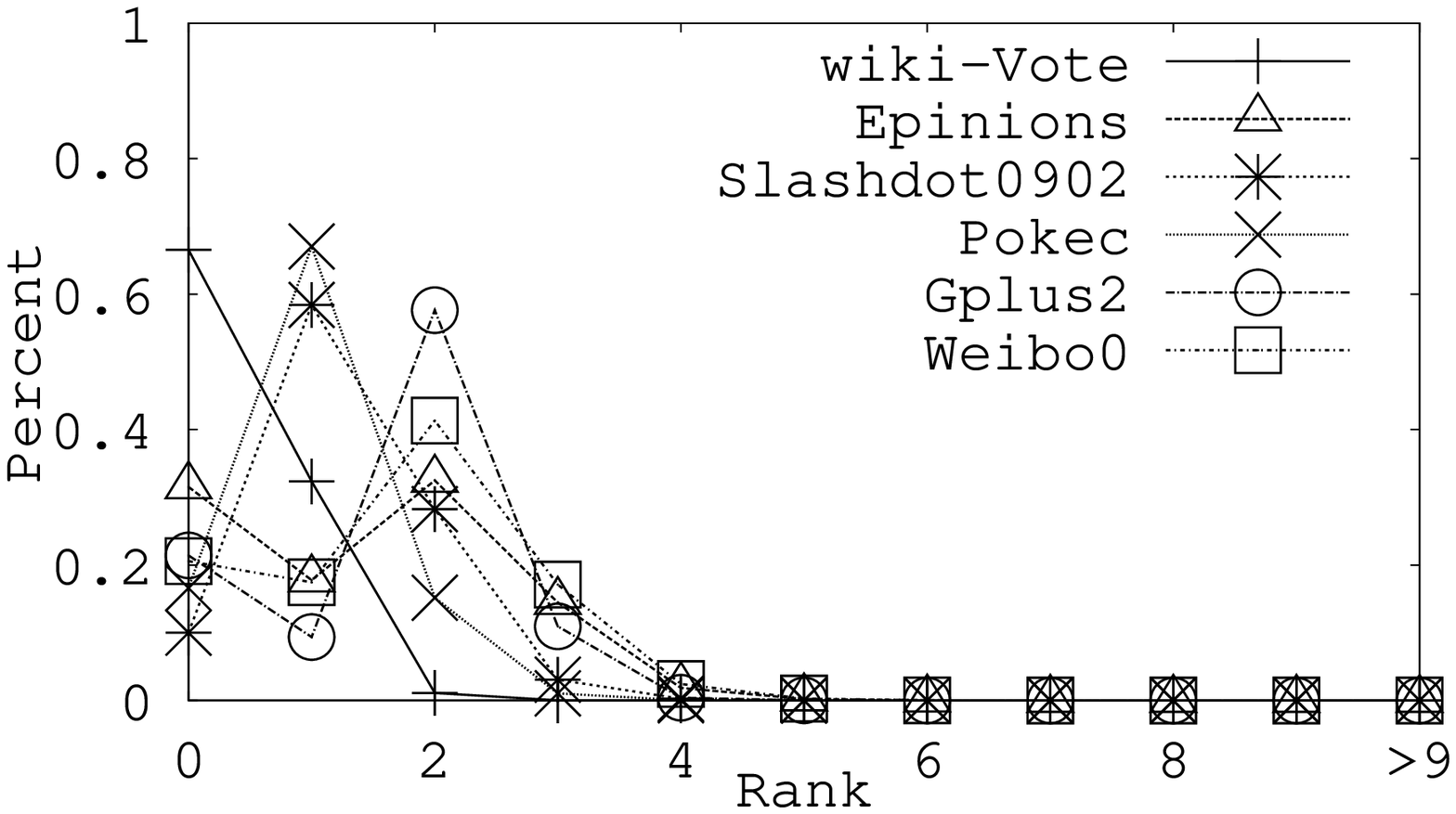}
       \label{fig:RDRD}
    }
    \subfigure[\scc]{
\includegraphics[width=0.33\columnwidth,height=2cm]{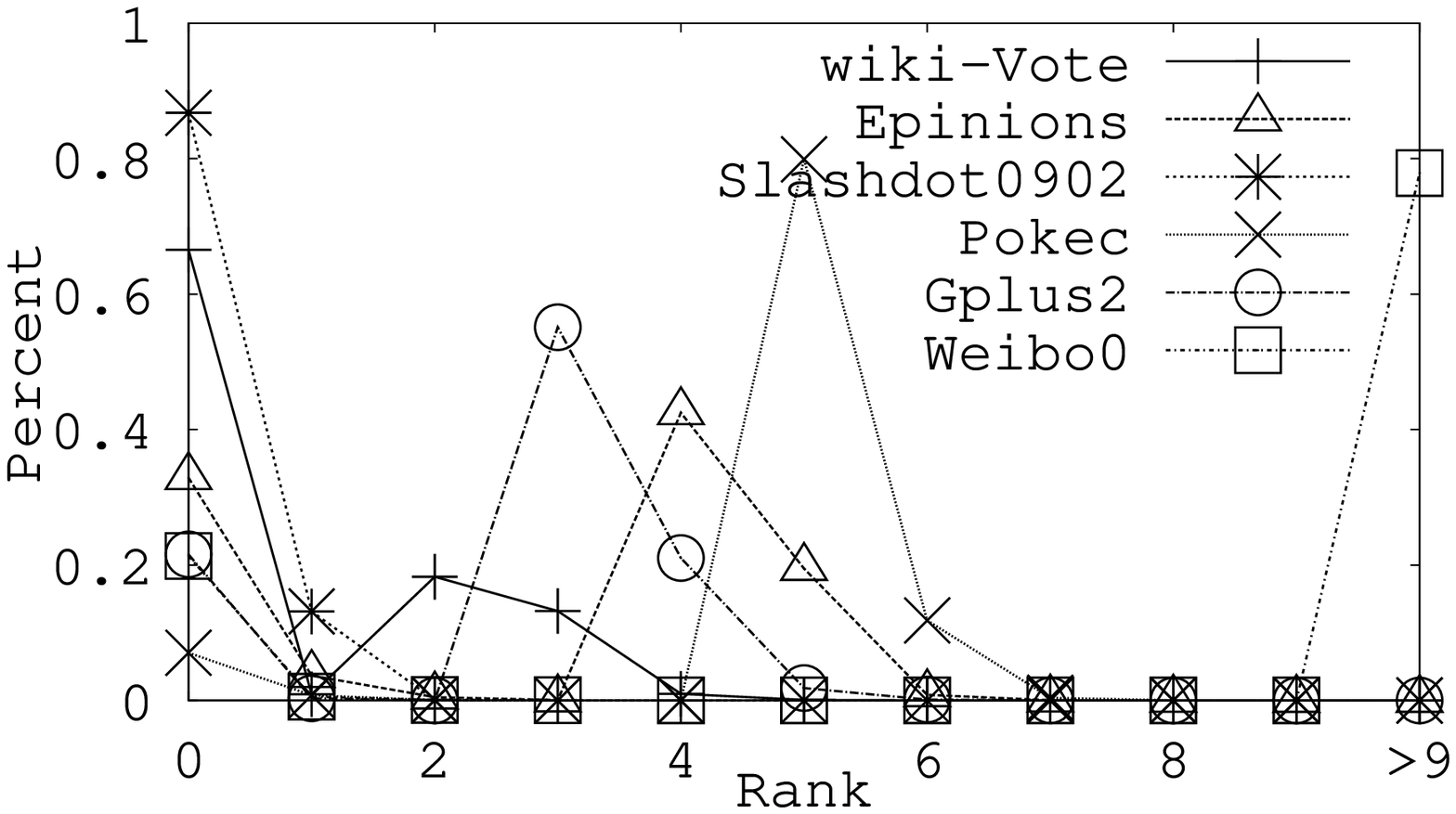}
       \label{fig:SCCRD}
    }
\end{tabular}
\end{center}
\vspace*{-0.6cm}
\caption{Rank Distribution}
\vspace*{-0.4cm}
\label{fig:rankdis}
\end{figure}


\comment{
\stitle{The pyramid structure of rank distribution} is one of the most
fundamental characteristics of social hierarchy.  We test the social
networks: wiki-Vote, Epinions, Slashdot0902, Pokec, Google+,
Weibo. The details about the datasets are in Table~\ref{tbl:evolving}
and Table~\ref{tbl:realdataset}.  The rank distribution derived from
$G_D$ indicates the existence of pyramid structure.
Fig.~\ref{fig:rankdis} shows some social networks.  Here, the x-axis
is the rank where a high rank means a high status, and the y-axis is
the percentage in a rank over all vertices. By analyzing the vertices,
$u$, in $G$ over the difference between in-degree and out-degree,
i.e. $d_I(u)-d_O(u)$, it reflects the fact that those vertices $u$
with negative $d_I(u)-d_O(u)$ are always at the bottom of $G_D$,
whereas those vertices in the higher rank are typically with large
positive $d_I(u)-d_O(u)$ values.}

\stitle{The pyramid structure of rank distribution} is one of the most
fundamental characteristics of social hierarchy.  We test the social
networks: wiki-Vote, Epinions, Slashdot0902, Pokec, Google+,
Weibo. The details about the datasets are in Table~\ref{tbl:evolving}
and Table~\ref{tbl:realdataset}. The rank distribution derived from
hierarchy $G_D$, shown in Fig.~\ref{fig:hierarchyRD}, indicates the
existence of pyramid structure, while the rank distributions derived
from a random DAG (Fig.~\ref{fig:RDRD}) and by contracting \sccs
(Fig.~\ref{fig:SCCRD}) are rather random. Here, the x-axis is the rank
where a high rank means a high status, and the y-axis is the
percentage in a rank over all vertices. By analyzing the vertices,
$u$, in $G$ over the difference between in-degree and out-degree,
i.e. $d_I(u)-d_O(u)$, it reflects the fact that those vertices $u$
with negative $d_I(u)-d_O(u)$ are always at the bottom of $G_D$,
whereas those vertices in the higher rank are typically with large
positive $d_I(u)-d_O(u)$ values.


\stitle{The social mobility}: With the DAG $G_D$ found, we can further
study social mobility over the social hierarchy $G_D$
represents. Here, social mobility is a fundamental concept in
sociology, economics and politics, and refers to the movement of
individuals from one status to another. It is important to
identify individuals who jump from a low status (a level in $G_D$) to
a high status (a level in $G_D$).  We conduct experimental studies
using the social network Google+ (\url{http://plus.google.com})
crawled from Jul. 2011 to Oct. 2011 \cite{zhenqiang2012evolution,
  zhenqiang2011jointly}, and Sina Weibo (\url{http://weibo.com})
crawled from 28 Sep. 2012 to 29 Oct. 2012 \cite{zhang2013social}.  For Google+ and Weibo, we
randomly extract 100,000 vertices respectively, and then extract all
edges among these vertices in 4 time intervals during the period the
datasets are crawled, as shown in Table~\ref{tbl:evolving}.

\comment{
We extract two
snapshots at two different time intervals, Jul. and Aug., 2011.  Let
$G_1$ and $G_2$ represent two subgraphs extracted from these two
snapshots.  $G_1$ has 1 million vertices and 1 million edges, and
$G_2$ has the same 1 million vertices in $G_1$ but 4.4 million
edges. Over $G_1$ and $G_2$, two DAGs $G_{D_1}$ and $G_{D_2}$ can be
found using our approach. In brief, there are in total 12 levels in
both DAGs, $G_{D_1}$ and $G_{D_2}$.  In $G_{D_1}$, there are 837,771
users at the bottom level, among them 8 users jumped to the top level
in $G_{D_2}$.  The 8 users are the most interesting to be further
investigated.

With the DAG $G_D$ and maximum Eulerian subgraph $\eulerian(G)$ found,
we can further study diverse problems in social network analysis, for
instance, social mobility and social relationship directionality
\cite{zhang2014proposed}. Social mobility is a fundamental concept in
sociology, economics and politics, and refers to the movement of
individuals from one status to another.
}

\begin{table}[t]
{\scriptsize 
\begin{center}
    \begin{tabular}{|l|r|r|r|r|} \hline
    {\bf Graph}  & $|V|$ & $|E|$ & $|V(\eulerian(G))|$ & $|E(\eulerian(G))|$\\ \hline\hline
    Gplus0 & 100,000 &   115,090 &   2,833 &     6,271\\\hline
    Gplus1 & 100,000 &   512,281 &  14,797 &    70,537\\\hline
    Gplus2 & 100,000 &  2,867,781 &  51,605 &   770,854\\\hline
    Gplus3 & 100,000 &  8,289,203 &  87,941 & 3,644,147\\\hline\hline
    Weibo0 & 100,000 & 2,431,525 &  96,765 &   850,136\\\hline
    Weibo1 & 100,000 & 2,446,002 &  96,833 &   855,131\\\hline
    Weibo2 & 100,000 & 2,463,050 &  96,902 &   861,729\\\hline
    Weibo3 & 100,000 & 2,479,140 &  96,969 &   868,044\\\hline

    \end{tabular}
\end{center}
}
\vspace*{-0.4cm}
\caption{To study social mobility}
\vspace*{-0.2cm}
\label{tbl:evolving}
\end{table}

\begin{figure}[t]
\begin{center}
\begin{tabular}[t]{c}
    \subfigure[Google+ (exact)]{
\includegraphics[width=0.45\columnwidth,height=2.5cm]{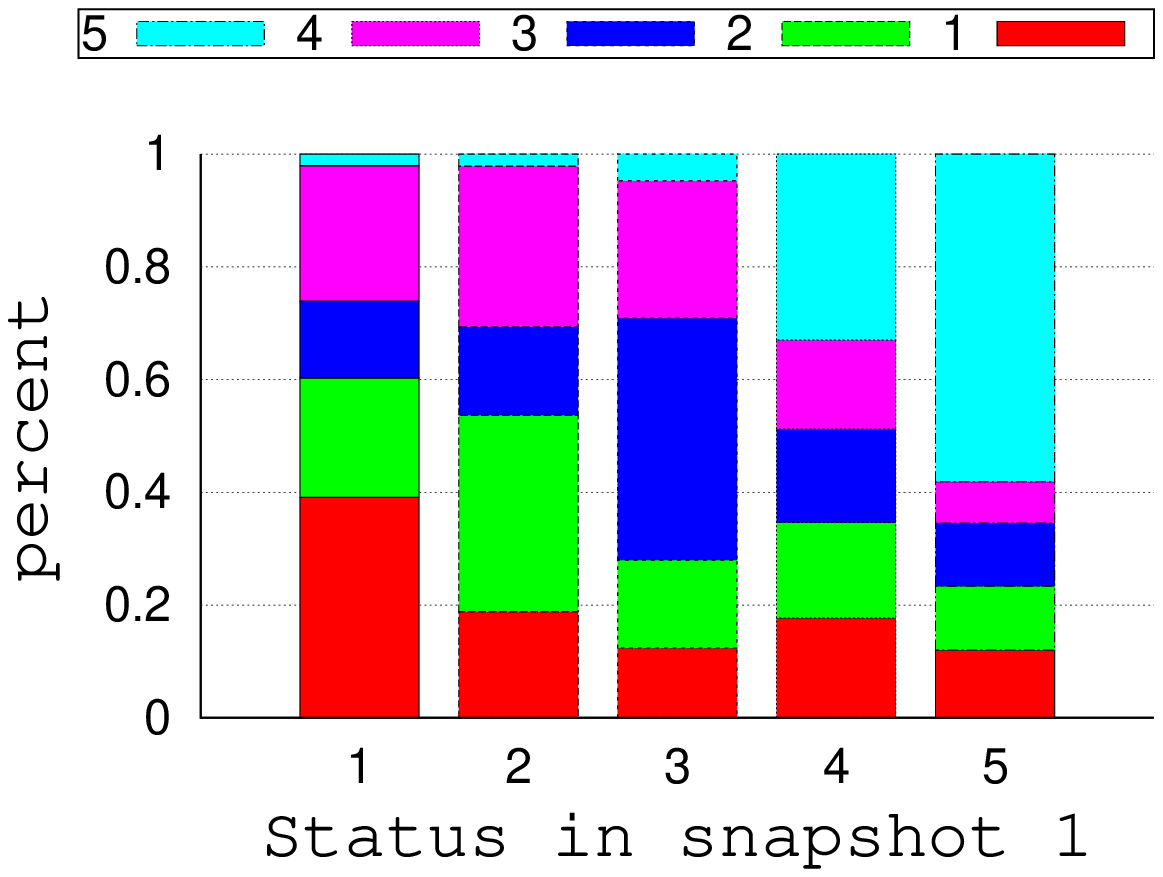}
         \label{fig:GplusMob}
    }
    \subfigure[Weibo (exact)]{
\includegraphics[width=0.45\columnwidth,height=2.5cm]{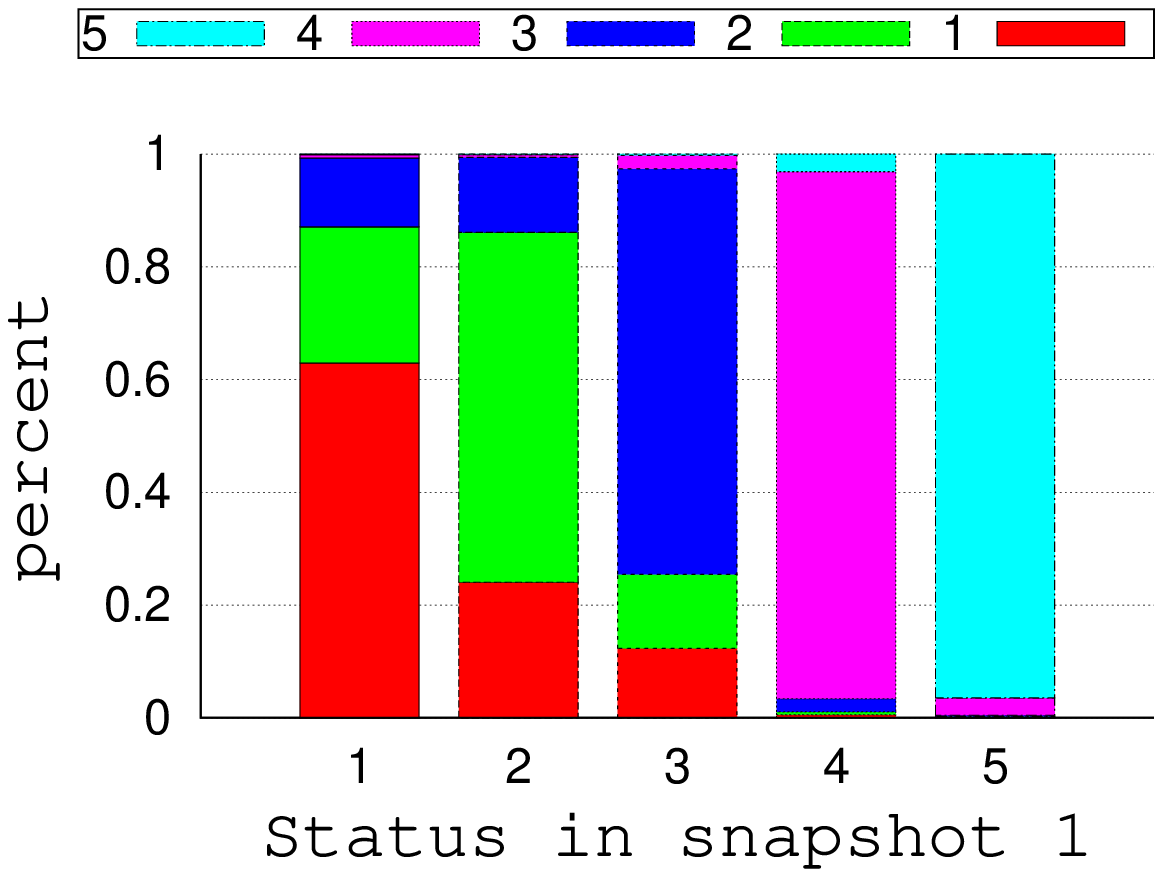}
       \label{fig:weiboMob}
    }
\\
    \subfigure[Google+ (approx)]{
\includegraphics[width=0.45\columnwidth,height=2.5cm]{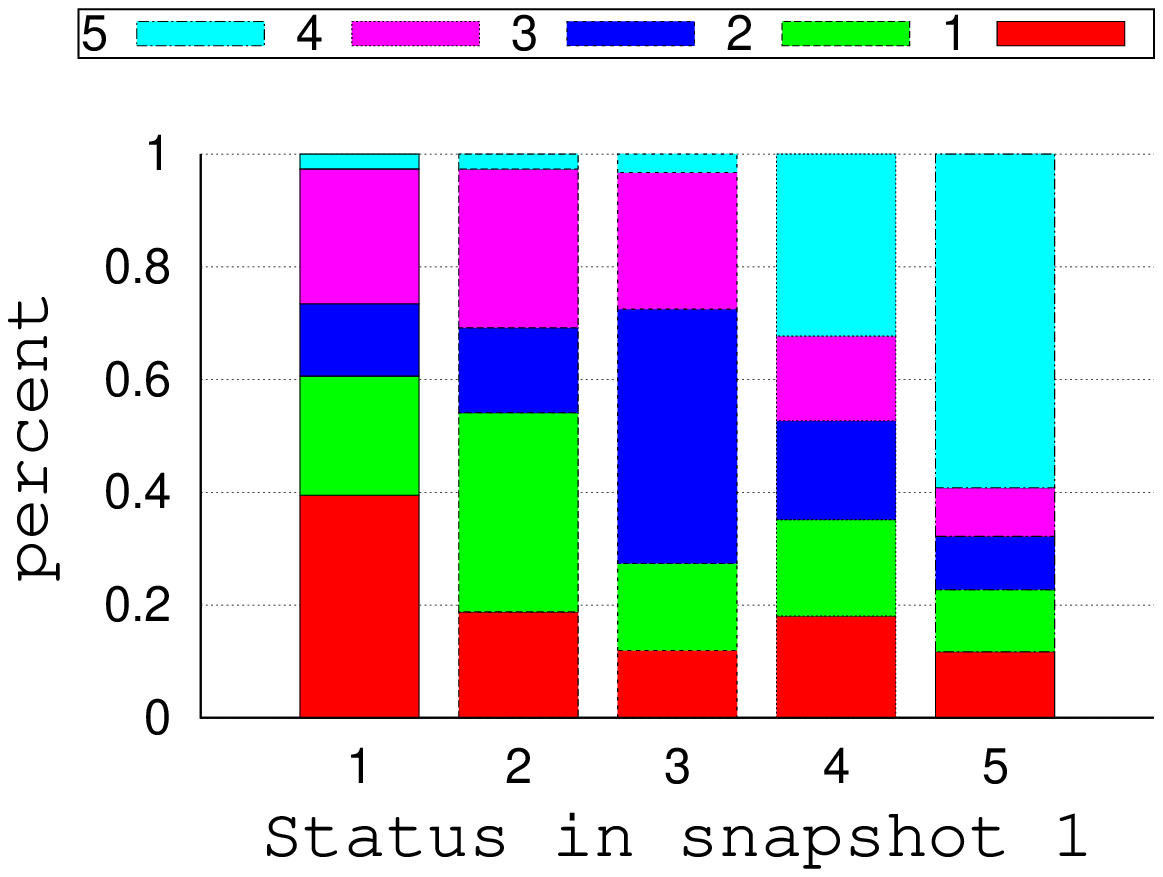}
         \label{fig:GplusMobgr}
    }
    \subfigure[Weibo (approx)]{
\includegraphics[width=0.45\columnwidth,height=2.5cm]{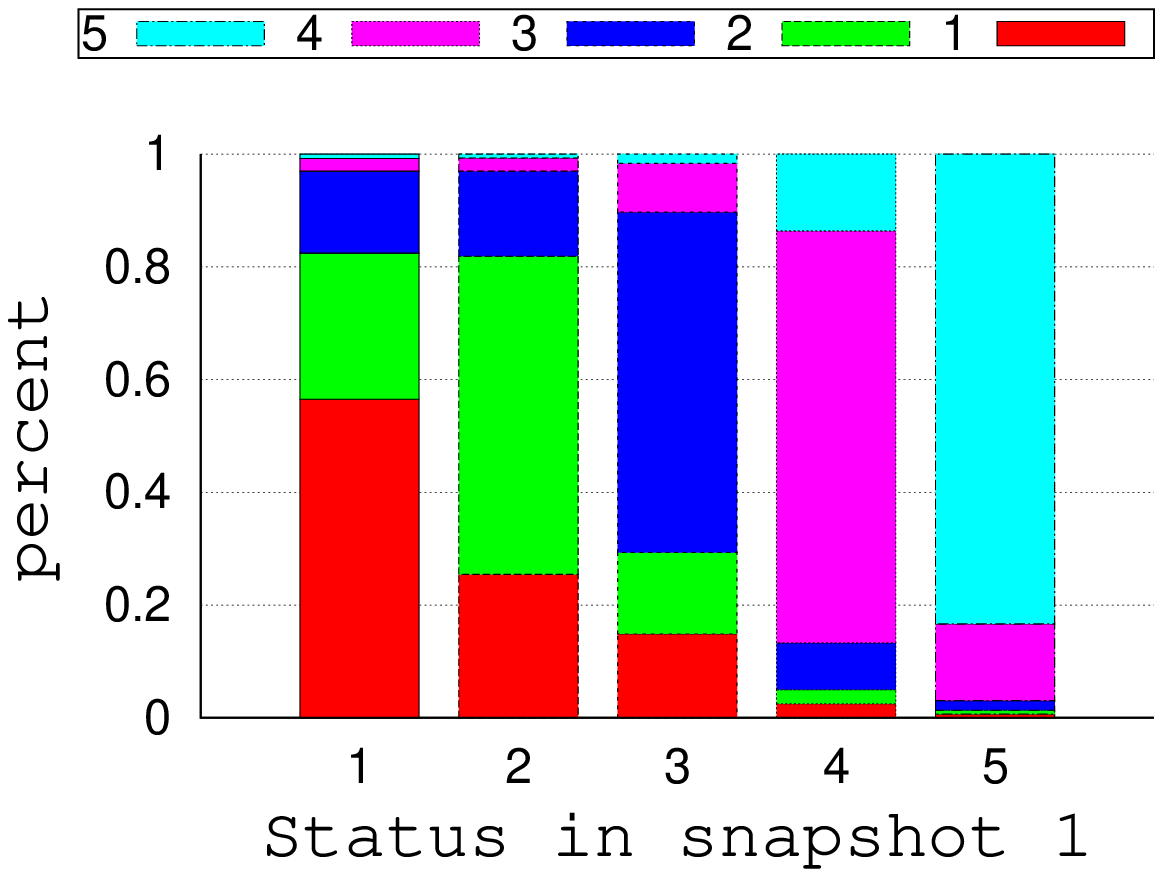}
       \label{fig:weiboMobgr}
    }

\end{tabular}
\end{center}
\vspace*{-0.6cm}
\caption{Social mobility result from hierarchy}
\vspace*{-0.4cm}
\label{fig:mobility}
\end{figure}

We show social mobility in Fig.~\ref{fig:mobility}.  We compare two
snapshots, $G_1$ and $G_2$, and investigate the social mobility from
$G_1$ to $G_2$.  For Google+, $G_1$ and $G_2$ are Gplus0 and Gplus1,
and for Weibo, $G_1$ and $G_2$ are Weibo0 and Weibo1.  For $G_1$, we
divide all vertices into 5 equal groups. The top 20\%  go into
group 5, and the second 20\% go to group 4, for example.  In
Fig.~\ref{fig:mobility}, the x-axis shows the 5 groups for $G_1$.
%
%
Consider the number of vertices in a group as
100\%. In Fig.~\ref{fig:mobility}, we show the percentage of vertices
in one group moves to another group in $G_2$. Fig.~\ref{fig:GplusMob}
and Fig.~\ref{fig:weiboMob} show the results for Google+ and Weibo.
Some observations can be made. Google+ is a new social network when crawled since
it starts from Jun. 29, 2011, and Weibo is a rather mature social
network since it starts from Aug. 14, 2009. From
Fig.~\ref{fig:GplusMob}, many vertices move from one status to
another, whereas from Fig.~\ref{fig:weiboMob}, only a very small number
of vertices move from one status to another.
Similar results can be observed from approximate hierarchies, by our
greedy solution \greedy given in Section~\ref{sec:gr}, as shown in
Fig.~\ref{fig:mobility}(c) and Fig.~\ref{fig:mobility}(d).
Those moved to/from the highest level deserve to be investigated.

\comment{
We show social mobility in Fig.~\ref{fig:mobility}.  We compare two
snapshots, $G_1$ and $G_2$, and investigate the social mobility from
$G_1$ to $G_2$.  For Google+, $G_1$ and $G_2$ are Gplus0 and Gplus1,
and for Weibo, $G_1$ and $G_2$ are Weibo0 and Weibo1.  For $G_1$, we
divide all vertices into 5 equal groups. The top 20\% go into group 5,
and the second 20\% go to group 4, for example.  In each figure, the
x-axis shows the 5 groups for $G_1$. Consider the number of vertices
in a group as 100\%, we show the percentage of vertices in one group
moves to another group in $G_2$. Fig.~\ref{fig:GplusMobHier} and
Fig.~\ref{fig:weiboMobHier} shows the results for Google+ and Weibo
based on our concept of hierarchy. Some observations can be
made. Google+ is a new social network when crawled since it starts
from Jun. 29, 2011, and Weibo is a rather mature social network since
it starts from Aug. 14, 2009. From Fig.~\ref{fig:GplusMobHier}, many
vertices move from one status to another, whereas from
Fig.~\ref{fig:weiboMobHier}, only a very small number of vertices move
from one status to another. Fig.~\ref{fig:GplusMobHier} and
Fig.~\ref{fig:weiboMobHier} shows the results for Google+ and Weibo
based on random DAG and Fig.~\ref{fig:GplusMobHier} and
Fig.~\ref{fig:weiboMobHier} shows the results for Google+ and Weibo by
contracting \sccs. Above observations can also be made, however, this
is mainly due to the fact that most new edges are among the same
\scc. Furthermore, those moved to/from the highest level deserve to be
investigated.
}

\stitle{Recovering the hidden directions} is to identify the direction
of an edge if the direction of the edge is
unknown~\cite{zhang2014proposed}. The directionality of edges in
social networks being recovered is important in many social analysis
tasks. We show that our approach has advantage over the
semi-supervised approach (SM-ReDirect) in \cite{zhang2014proposed}.
Here, the task is using the given 20\% directed edges as training data
to recover the directions for the remaining edges. In our approach,
we construct a graph $G$ from the 20\% training data, and identify
$G_D$ by $G = \eulerian(G) \cup G_D$. With the ranking $r(\cdot)$ over
the vertices, we predict the direction of an edge $(u, v)$ is from $u$
to $v$ if $r(v) > r(u)$. Take Slashdot and Epinion datasets used
\cite{zhang2014proposed}, our approach outperforms the matrix-factorization based SM-ReDirect both in
terms of accuracy and efficiency.  For Slashdot, our prediction accuracy is 0.7759 whereas SM-ReDirect is 0.6529.  For Epinion, ours is 0.8285 whereas SM-Redirect is 0.7118. Using approximate hierarchy, our accuracy is 0.7682 for Slashdot and 0.8277 for Epinion, respectively.

\begin{algorithm}[t]
\caption{\simpleAlg($G$)}
{\small
{\bf Input}: A graph $G = (V, E)$ \\
{\bf Output}: Two subgraphs of $G$, $\eulerian(G)$ and $G_D$ ($G = \eulerian(G) \cup G_D$)
\label{alg:simple}
\begin{algorithmic}[1]
\STATE $w(v_i, v_j) \leftarrow -1$ for each edge $(v_i, v_j) \in E$;
\WHILE { there is a negative cycle $p_c$ in $G$}
   \FOR {every edge $(v_i,v_j)$ in the negative cycle $p_c$}
        \STATE $w(v_i, v_j) \leftarrow -w(v_i, v_j)$;
        \STATE Reverse the direction of the edge $(v_i, v_j)$ to be
               $(v_j, v_i)$;
\ENDFOR
\ENDWHILE
\STATE $G_D$ is a subgraph that contains all edges with weight -1;
\STATE $\eulerian(G)$ is a subgraph containing the reversed edges with weight +1;
\end{algorithmic}
}
\end{algorithm}

\section{The Existing Algorithm}
\label{sec:bf}

To find the maximum Eulerian subgraph, Gupte, et
al.~in~\cite{gupte2011finding} propose an iterative algorithm based on
the Bellman-Ford algorithm, which we call \simpleAlg
(Algorithm~\ref{alg:simple}). Let $w(v_i, v_j)$ be a weight assigned
to an edge $(v_i, v_j)$ in $G$. Initially, \simpleAlg assigns an
edge-weight with a value of -1 to every edge in graph $G$ (Line~1).
Let a negative cycle be a cycle with a negative sum of edge weights.
In every iteration~(Lines~2-7), \simpleAlg finds a negative cycle
$p_c$ repeatedly until there are no negative cycles. For every edge
$(v_i, v_j)$ in the negative cycle $p_c$ found, it changes the weight of
$(v_i, v_j)$ to be $-w(v_i, v_j)$ and reverses the direction of the
edge (Lines~4-5). As a result, it finds a maximum Eulerian subgraph
$\eulerian(G)$ and a directed acyclic graph (DAG) of $G$, denoted as
$G_D$, such that $G = \eulerian(G) \cup G_D$.
Since the number of edges with weight +1 increases by at least one
during each iteration, there are at most $O(m)$ iterations
(Line~2-7). In every iteration it has to invoke the Bellman-Ford
algorithm to find a negative cycle (Line~2), or to determine whether
there is a negative cycle. The time complexity of Bellman-Ford
algorithm is $O(nm)$. Therefore, in the worst case, the total time
complexity of \simpleAlg is $O(nm^2)$, which is too expensive for
real-world graphs.

\begin{algorithm}[t]
\caption{\DFSEVEN($G$)}
{\small
{\bf Input}: A graph $G = (V, E)$ \\
{\bf Output}: Two subgraphs of $G$, $\eulerian(G)$ and $G_D$ ($G = \eulerian(G) \cup
G_D$)
\label{alg:FMES}
\begin{algorithmic}[1]
\STATE {\bf for} each edge $(v_i, v_j)$ in $E(G)$ {\bf do}
      $w(v_i, v_j) \leftarrow -1$;
\STATE {\bf for} each vertex $u$ in $V(G)$ {\bf do}
      $dst(u) \leftarrow 0$, $relax(u) \leftarrow true$, $ pos(u)
\leftarrow 0$;
\WHILE  {there is  a vertex $u \in V(G)$ such that $relax(u) = true$}
       \STATE $S_V \leftarrow \emptyset$, $S_E \leftarrow \emptyset$,
              $NV \leftarrow \emptyset$;
       \IF {\DFSSPFA($G$, $u$)}
            \WHILE{$S_V$.top() $\neq NV$}
                \STATE $S_V$.pop();
                       $(v_i, v_j) \leftarrow S_E$.pop();
                \STATE $w(v_i, v_j) \leftarrow -w(v_i, v_j)$;
                \STATE Reverse the direction of the edge $(v_i, v_j)$
                       to be $(v_j, v_i)$;
            \ENDWHILE

            \STATE $S_V$.pop();
                   $(v_i, v_j) \leftarrow S_E$.pop();
            \STATE $w(v_i, v_j) \leftarrow -w(v_i, v_j)$;
            \STATE Reverse the direction of the edge $(v_i, v_j)$
                       to be $(v_j, v_i)$;
        \ENDIF
\ENDWHILE
\STATE $G_D$ is a subgraph that contains all edges with weight -1;
\STATE $\eulerian(G)$ is a subgraph containing the reversed edges with weight +1;
\end{algorithmic}
}
\end{algorithm}

\begin{algorithm}[t]
\caption{ \DFSSPFA($G$, $u$)}
\label{alg:dfs-spfa}
{\small
\begin{algorithmic}[1]
\STATE $S_V$.push($u$);
\FOR { each edge $(u,v)$ starting at $pos(u)$ in $E(G)$ }
    \STATE $pos(u) \leftarrow pos(u)+1$;
    \IF { $dst(u)+w(u,v)<dst(v)$}
           \STATE $dst(v) \leftarrow dst(u)+w(u,v)$;
           \STATE $relax(v) \leftarrow true$, $pos(v) \leftarrow 0$;
           \IF { $v$ is not in $S_V$}
                 \STATE $S_E$.push($(u,v)$);
                 \STATE {\bf if} {\DFSSPFA($G$, $v$)} {\bf then}
                        {\bf return} $true$; {\bf endif}
           \ELSE
                \STATE $S_E$.push($(u,v)$);
                       $NV \leftarrow v$;
                       {\bf return} $true$;
           \ENDIF
     \ENDIF
\ENDFOR
\STATE $relax(u) \leftarrow false$;
\STATE $S_V$.pop();
       $S_E$.pop() if $S_E$ is not empty;
       {\bf return} $false$;
\end{algorithmic}
}
\end{algorithm}

\section{A New Algorithm}
\label{sec:simplealgo}

To address the scalability problem of \simpleAlg, we propose a new
algorithm, called \DFSEVEN. Different from \simpleAlg which starts by
finding a negative cycle using the Bellman-Ford algorithm in every
iteration, \DFSEVEN finds a negative cycle only when necessary with
condition. In brief, in every iteration, when necessary, \DFSEVEN
invokes an algorithm \DFSSPFA (short for find a negative cycle) to
find a negative cycle while relaxing vertices following \DFS order.
%
%
Applying amortized analysis~\cite{tarjan1985amortized}, we prove the
time complexity of \DFSEVEN, is $O(m^2)$ to
find the maximum Eulerian subgraph $\eulerian(G)$.

The \DFSEVEN algorithm is outlined in Algorithm~\ref{alg:FMES}, which
invokes \DFSSPFA (Algorithm~\ref{alg:dfs-spfa}) to find a negative
cycle. Here, \DFSSPFA is designed based on the same idea of relaxing
edges as used in the Bellman-Ford algorithm.
In addition to edge weight $w(v_i,v_j)$, we use three variables for
every vertex $u$, $relax(u)$, $pos(u)$, and $dst(u)$.  Here,
$relax(u)$ is a Boolean variable indicating whether there are
out-going edges from $u$ that may need to relax to find a negative
cycle. It will try to relax an edge from $u$ further when $relax(u) =
true$. When relaxing from $u$, $pos(u)$ records the next vertex $v$ in
$N_{O}(u)$ (maintained as an adjacent list) for the edge $(u, v)$ to
be relaxed next. It means that all edges from $u$ to any vertex before
$pos(u)$ has already been relaxed.  $dst(u)$ is an estimation on the
vertex $u$ which decreases when relaxing. When $dst(u)$ decreases,
$relax(u)$ is reset to be $true$ and $pos(u)$ is reset to be 0, since
all its out-going edges can be possibly relaxed again.
Initially, in \DFSEVEN, every edge weight $w(v_i, v_j)$ is initialized
to -1, and the three variables, $relax(u)$, $pos(u)$, and $dst(u)$, on
every vertex $u$ are initialized to $true$, $0$, and $0$,
respectively.
All $w(v_i, v_j)$, $relax(u)$, $pos(u)$, and $dst(u)$ are used in
\DFSSPFA to find a negative cycle following the main idea of
Bellman-Ford algorithm in DFS order.
A negative cycle, found by \DFSSPFA while relaxing edges, is
maintained using a vertex stack $S_V$ and an edge stack $S_E$ together
with a variable $NV$, where $NV$ maintains the first vertex of a
negative cycle.
In \DFSEVEN, by popping vertex/edges from $S_v$/$S_E$
until encountering the vertex in $NV$, a negative cycle can be
recovered.
As shown in Algorithm~\ref{alg:FMES}, in the while statement
(Lines~3-15), for every vertex $u$ in $V(G)$, only when there is a
possible relax ($relax(u) = true$) and there is a negative cycle found by
the algorithm \DFSSPFA, it will reverse the edge direction and update
the edge weight, $w(v_i, v_j)$, for each edge $(v_i, v_j)$ in the
negative cycle (Lines~6-13).

\begin{figure}[h]
\begin{center}
\includegraphics[scale=0.3]{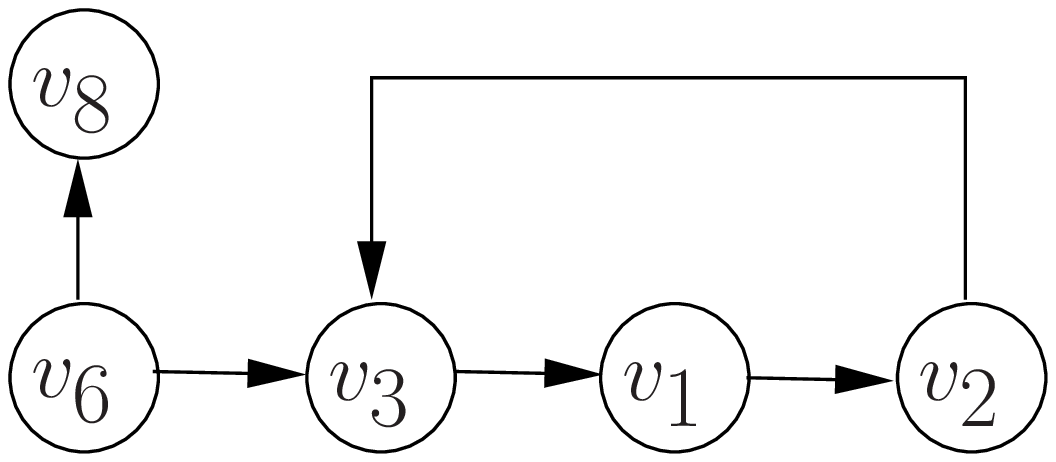}
\end{center}
\vspace*{-0.6cm}
\caption{A subgraph $G$ of Fig.~\ref{fig:fig1}}
\vspace*{-0.2cm}
\label{fig:fig0}
\end{figure}

\begin{example}
We explain \DFSEVEN (Algorithm~\ref{alg:FMES}) using an example graph
$G$ in Fig.~\ref{fig:fig0}.  For ever vertex $u$, there is an adjacent
list to maintain its out-neighbors. Initially, for every vertex $u$,
$dst(u)=0, relax(u)=true, pos(u)=0$; and for every edge $(v_i,v_j)$,
$w(v_i,v_j)=-1$. Suppose we process $v_6, v_3, v_1, v_2, v_8$ in such
an order. In the first iteration, $relax(v_6)=true$ and
\DFSSPFA($G,v_6$) returns true, which implies a negative cycle is
found.  Here, for all vertices in $G$, we have $dst(v_6)=0,
relax(v_6)=true, pos(v_6)=1$; $dst(v_3)=-4, relax(v_3)=true,
pos(v_3)=0$; $dst(v_1)=-2, relax(v_1)=true, pos(v_1)=0$; $dst(v_2)=-3,
relax(v_2)=true, pos(v_2)=0$; $dst(v_8)=0, relax(v_8)=true,
pos(v_8)=0$. In addition, $NV=v_3$, $S_V=\{v_6,v_3,v_1,v_2\}$ ,
$S_E=\{(v_6,v_3),$ $(v_3,v_1),$ $(v_1,v_2), (v_2,v_3)\}$. Following
Lines~6--13 in Algorithm~\ref{alg:FMES}, we find and reverse
negative cycle $(v_3,v_1,v_2,v_3)$ and make
$w(v_3,$ $v_2)=w(v_2,v_1)=w(v_1,v_3)=1$. In the second iteration, the
out-neighbors of $v_6$ are relaxed from $pos(v_6)=1$ in $v_6$'s
adjacent list, i.e. from edge $(v_6,v_8)$.  \DFSSPFA($G,v_6$) returns
false. We have, $dst(v_6)=0, relax(v_6)=false, pos(v_6)=2$, and
$dst(v_8)=-1, relax(v_8)=false, pos(v_8)=1$. In the following
iterations (\DFSSPFA($G,v_3$), \DFSSPFA($G,v_1$), and
\DFSSPFA($G,v_2$)), all return false. Finally, for vertex $v_8$, since
$relax(v_8)=false$, \DFSSPFA($G,v_8$) is unnecessary, and
\DFSEVEN($G$) terminates. It finds the maximum Eulerian subgraph
$\eulerian(G)=\{(v_3,v_1), (v_1,v_2),$ $(v_2,v_3)\}$.
\end{example}

\begin{lemma}
In Algorithm~\ref{alg:FMES}, if there is a negative cycle $C$,
$relax(u)$ $=true$ holds for at least one vertex $ u \in V(C)$.
\label{lem:-4}
\end{lemma}

\proofsketch Assume the opposite, i.e., there exists a negative cycle
$C$ such that for every vertex $ u \in V(C)$, $relax(u)=false$.
Let $C=(v_0, v_1, \dots, v_{k-1}, v_k=v_0)$, then $ \sum_{i=0}^{k-1}
w(v_i,v_{i+1}) < 0$.
Since $relax(v_i)=false$ holds for $i=0,1,\dots,k-1$, then
$dst(v_{i+1}) \leq dst(v_i) + w(v_i, v_{i+1})$.
It leads to a contradiction, if summing both sides from $i=0$ to
$i=k-1$, then $dst(v_0)=dst(v_k) \leq dst(v_0)+\sum_{i=0}^{k-1}
w(v_i,v_{i+1}) <dst(v_0)$.  \eop

\begin{theorem}
Algorithm~\ref{alg:FMES} correctly finds the maximum Eulerian subgraph
$\eulerian(G)$ when it terminates.
\label{them:correctness}
\end{theorem}

\proofsketch It can be proved by Lemma~\ref{lem:-4}.
\eop

\begin{lemma}
Given an Eulerian graph $G$, when \DFSEVEN($G$) terminates, for each
vertex $u$, $dst(u) \in [-2m, 0]$, where $m=|E(G)|$.
\label{lem:-1}
\end{lemma}

\proofsketch We do mathematical induction on the maximum number of
cycles the Eulerian graph $G$ contains.
\begin{enumerate}
\item If $G$ contains only one cycle, i.e. $G$ is a simple cycle
  itself, it is easy to see that for each vertex $u$, $dst(u) \geq  -m
  \in [-2m,0]$.
\item Assume Lemma~\ref{lem:-1} holds when $G$ contains no more than
  $k$ cycles, we prove it also holds when $G$ contains at most $k+1$
  cycles.  We first decompose $G$ into a simple cycle $C$ which is the
  last negative cycle found during \DFSEVEN($G$) and the remaining is
  an Eulerian graph $G'$ containing at most $k$ cycles. We explain the
  validation of this decomposition as follows. If the last negative
  cycle found contains some positive edges, then the resulting maximum
  Eulerian subgraph $\eulerian(G)$ will contain some negative edges,
  it is against the fact that $G$ itself is given as Eulerian. Next,
  we decompose \DFSEVEN($G$) into two phases, it finds $G'$ as an
  Eulerian subgraph in the first phase while cycle $C$ is identified in the
  second phase. According to the assumption, when the first phase
  completes, for each vertex $u \in G', dst(u) \in [-2|E(G')|-|E(C)|,
    0]$, where $-2|E(G')|$ is by \DFSEVEN($G'$) and $-|E(C)|$ is the
  result by relaxing $C$. There are two cases for the second phase.
    \begin{enumerate}
    \item If $V(C) \bigcap V(G') = \emptyset$, the two phases are
      independent. Therefore, when the second phase terminates, for
      each vertex $u \in V(C)$, $dst(u)  \in [-2|E(C)|, 0]$, and for
      each vertex $u \in V(G')$, $dst(u) \in [-2|E(G')|, 0]$,
      Lemma~\ref{lem:-1} holds.
    \item If $V(C) \bigcap V(G') \neq \emptyset$, suppose $w \in V(C)
      \bigcap V(G')$, then $dst(w) \in [-2|E(G')|-|E(C)|, 0]$ when the
      first phase completes. During the second phase, $dst(w)$
      decreases by $|E(C)|$, then $dst(w) \geq -2|E(G')|-2|E(C)|$ $\in
      [-2m,0]$. For any vertex $v \in V(G') \setminus V(C)$, $dst(v)$
      can only change along a path $p=(w_0, w_1,$ $ \dots, w_{k-1},$
      $w_k=v)$, where $w_0 \in V(C)$ and $( w_i, w_{i-1})$ $ \in
      E(G')$, for $i=1,\ldots,k$.
%
%
Then $d(v)=d(w_0)+\sum_{i=0}^{k-1} w(w_i,w_{i+1})
      >d(w_0)\geq -2m \in [-2m, 0]$. Therefore, Lemma~\ref{lem:-1}
      holds.
    \end{enumerate}
\vspace*{-0.4cm}
\eop
\end{enumerate}

\begin{figure}[h]
\centering
\hspace*{0.8cm}
\includegraphics[scale=0.3]{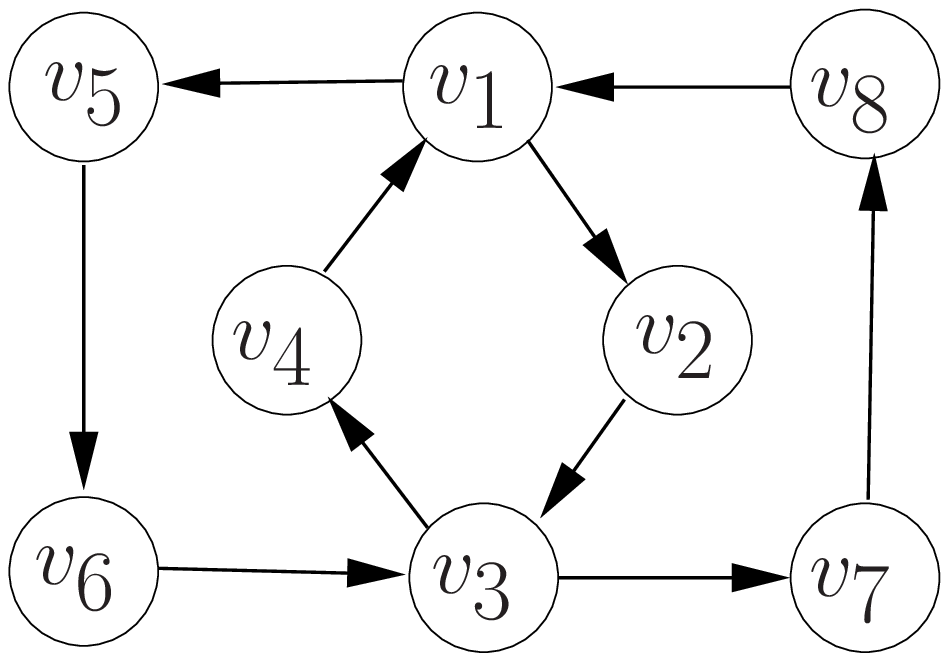}
\vspace*{-0.2cm}
\caption{Example graph to explain 2(b) in Lemma~\ref{lem:-1}}
\vspace*{-0.2cm}
\label{fig:fig-1}
\end{figure}

\begin{example}
We explain the proof of 2(b) in Lemma~\ref{lem:-1} using
Fig.~\ref{fig:fig-1}. Fig.~\ref{fig:fig-1} shows an Eulerian graph $G$
containing simple cycles. Suppose the first negative cycle found is
$C_1 = (v_1, v_2, v_3, v_4,$ $v_1)$, with the resulting $dst(v_1)=-4,
dst(v_2)=-1, dst(v_3)=-2, dst(v_4)=-3$. In a similar way, suppose the
second negative cycle found is $C_2 = (v_1, v_5, v_6, v_3, v_2, v_1)$
by relaxing $dst(v_1)=-5, dst(v_5)=-5, dst(v_6)=-6, dst(v_3)=-7,
dst(v_2)=-6$, and the third negative cycle is $C_3 = (v_3, v_7, v_8,
v_1, v_4, v_3)$ with $dst(v_1)=-10, dst(v_4)=-9, dst(v_3)=-8,
dst(v_7)=-8, dst(v_8)=-9$. By reversing these three negative cycles,
we have a cycle $C=(v_1, v_2, v_3, v_4)$, and a graph $G'$ which is a
simple cycle with $(v_1, v_5, v_6, v_3, v_7, v_8, v_1)$. As can be
seen, the current $\min\{dst(u)\}=dst(v_1)=-10$ is in the range of $
[-2|E(G')|-|E(C)|,0]$. When \DFSEVEN($G$) terminates, $\min\{dst(u)|u
\in V(C)\}$ $=dst(v_1)=-14$ is in the rage of $[-2|E(G)|, 0]$, and
$\min\{dst(u)|$ $u \in V(G') \setminus V(C)\}=dst(v_8)=-13$ is in the
range of $[-2|E(G)|,$ $0]$. It shows that Lemma~\ref{lem:-1} holds for
this example.
\end{example}

\begin{lemma}
Given a general graph $G$, when \DFSEVEN($G$) terminates, for each
vertex $u$, $dst(u) \in [-4m,0]$, where $m=|E(G)|$.
\label{lem:-2}
\end{lemma}

\proofsketch For a general graph $G$, we can add edges $(u,v)$ from
vertices $u$ with $d_I(u) > d_O(u)$ to vertices $v$ with $d_I(v) <
d_O(v)$ and $(u,v) \notin E(G)$. Obviously, the resulting augment
graph $G^A$ has at most $2m$ edges.

Based on Lemma~\ref{lem:-1}, when \DFSEVEN($G^A$) terminates, each
vertex $u$ satisfies $d(u) \in [-4m,0]$. On the other hand,
\DFSEVEN($G^A$) can be decomposed into two phases, \DFSEVEN($G$) and
further relaxations exploiting $E(G^A) \setminus E(G)$, implying that
for each vertex $u$, $dst(u) \in [-4m,0]$ holds when \DFSEVEN($G$)
terminates.  \eop

\begin{lemma}
For each value of $dst(u)$ of every vertex $u$, the out-neighbors of
$u$, i.e. $N_O(u)$, are relaxed at most once.
\label{lem:-3}
\end{lemma}

\proofsketch As shown in Algorithm.~\ref{alg:dfs-spfa}, $dst(u)$ is
monotone decreasing, and $pos(u)$ is monotone increasing for a
particular $dst(u)$ value. So Lemma~\ref{lem:-3} holds.  \eop

\begin{theorem}
Time complexity of \DFSEVEN($G$) is $O(m^2)$.
\label{them:simpleTC}
\end{theorem}

\proofsketch Given
Lemma~\ref{lem:-1}, Lemma~\ref{lem:-2}, and Lemma~\ref{lem:-3}, since
every edge $(u,v)$ is checked at most $|dst(u)|+|dst(v)| \leq 8m$
times for relaxations. By applying amortized
analysis~\cite{tarjan1985amortized}, the time complexity of
\DFSEVEN($G$) is $O(m^2)$.  \eop

\comment{
Consider Algorithm~\ref{alg:FMES}. Each iteration of the while loop
repeats until \DFSSPFA returns true when it finds a negative
cycle. In other words, it consists of two parts: all \DFSSPFA return
false and the first \DFSSPFA that returns true.  The total time
complexity of the first part is bounded as $O(m)$ since no edge can
be visited more than twice. For the second part, it only traverses a
small part of the graph, to be approximated as $O(m)$. Therefore,
the time complexity of \DFSEVEN is approximated as $O(K\cdot m)$, where
$K$ is the number of iterations, bounded by $|E(\eulerian(G))| \leq
m$. In the following discussion, we will analyze the time complexity
of algorithms based on the number of iterations.
}

Consider Algorithm~\ref{alg:FMES}. During each iteration of the while loop, only a small part of the graph can be traversed and most edges are visited at most twice. Therefore, each iteration can be approximately bounded as $O(m)$, and the time complexity of \DFSEVEN is approximated as $O(K\cdot m)$, where
$K$ is the number of iterations, bounded by $|E(\eulerian(G))| \leq m$. In the following discussion, we will analyze the time complexity of algorithms based on the number of iterations.

\section{The Optimal: Greedy-\&-Refine}
\label{sec:gr}

\DFSEVEN reduces the time complexity of \simpleAlg to $O(m^2)$, but it
is still very slow for large graphs. To further reduce the running
time of \DFSEVEN, we propose a new two-phase algorithm which is shown
to be two orders of magnitude faster than \DFSEVEN. Below, we first
introduce an important observation which can be used to prune many
unpromising edges. Then, we will present our new algorithms as well as
theoretical analysis.

Let ${\cal S}$ be a set of strongly connected components (\sccs) of
$G$, such that ${\cal S} = \{G_1, G_2, \cdots\}$, where $G_i$ is an
\scc of $G$, $G_i \subseteq G$, and $G_i \cap
G_j = \emptyset$ for $i \neq j$. We show that for any edge, if it is not included in any \scc $G_i$ of $G$, then it cannot be contained in the maximum Eulerian subgraph $\eulerian(G)$. Therefore, the problem of
finding the maximum Eulerian subgraph of $G$ becomes a problem of finding the maximum
Eulerian subgraph of each $G_i \in {\cal S}$, since the union of the maximum
Eulerian subgraph of $G_i \in {\cal S}$, $1 \leq i \leq |{\cal S}|$, is
the maximum Eulerian subgraph of $G$.

\comment{
\begin{figure}[t]
\centering
\includegraphics[scale=0.3]{fig/figure}
\vspace*{-0.3cm}
\caption{A toy example graph}
\vspace*{-0.6cm}
\label{fig:even}
\end{figure}

}

\begin{lemma}
An Eulerian graph $G$ can be divided into several edge disjoint simple
cycles.
\label{lem:1}
\end{lemma}

\proofsketch It can be proved if there is a process that we can
repeatedly remove edges from a cycle found in an Eulerian graph $G$, and $G$ has no edges after the last cycle being removed. Note that $d_I(u) =
d_O(u)$ for every $u$ in $G$. Let a subgraph of $G$, denoted as $G_c$,
be such a cycle found in $G$. $G_c$ is an Eulerian subgraph, and $G
\ominus G_c$ is also an Eulerian subgraph. The lemma is established. \eop

\begin{theorem}
Let $G$ be a directed graph, and ${\cal S} = \{G_1, G_2,$
$\cdots\}$ be a set of \sccs of $G$.  The
maximum Eulerian subgraph of $G$, $\eulerian(G) = \bigcup_{G_i \in {\cal S}}
  \eulerian(G_i)$.
\label{them:1}
\end{theorem}

\proofsketch For each edge $e=(u,v) \in \eulerian(G)$, there is at least one cycle containing this edge, given by Lemma~\ref{lem:1}. Therefore, $u$ and $v$ belong to the same \scc, i.e., for any edge $e' \in G - {\cal S}$, it cannot be included in $\eulerian(G)$. The theorem is established.
\eop

Below, we discuss how to find the maximum Eulerian subgraph for each
strongly connected component (\scc) $G_i$ of $G$. In the following
discussion, we assume that a graph $G$ is an \scc itself.

We can use \DFSEVEN to find the maximum Eulerian subgraph for an \scc
$G$. However, \DFSEVEN is still too expensive to deal with large
graphs. The key issue is that the number of iterations in \DFSEVEN (Algorithm~\ref{alg:FMES}, Lines~3-15), can be very
large when the graph and its maximum Eulerian subgraph are both very
large. Since in most iterations, the number of edges with weight +1
increases only by 1, and thus it takes almost $|\eulerian(G)|$ iterations to get the
optimal number of edges in the maximum Eulerian subgraph
$\eulerian(G)$.

\begin{figure}[t]
\centering
\includegraphics[width=0.7\columnwidth, scale=0.3]{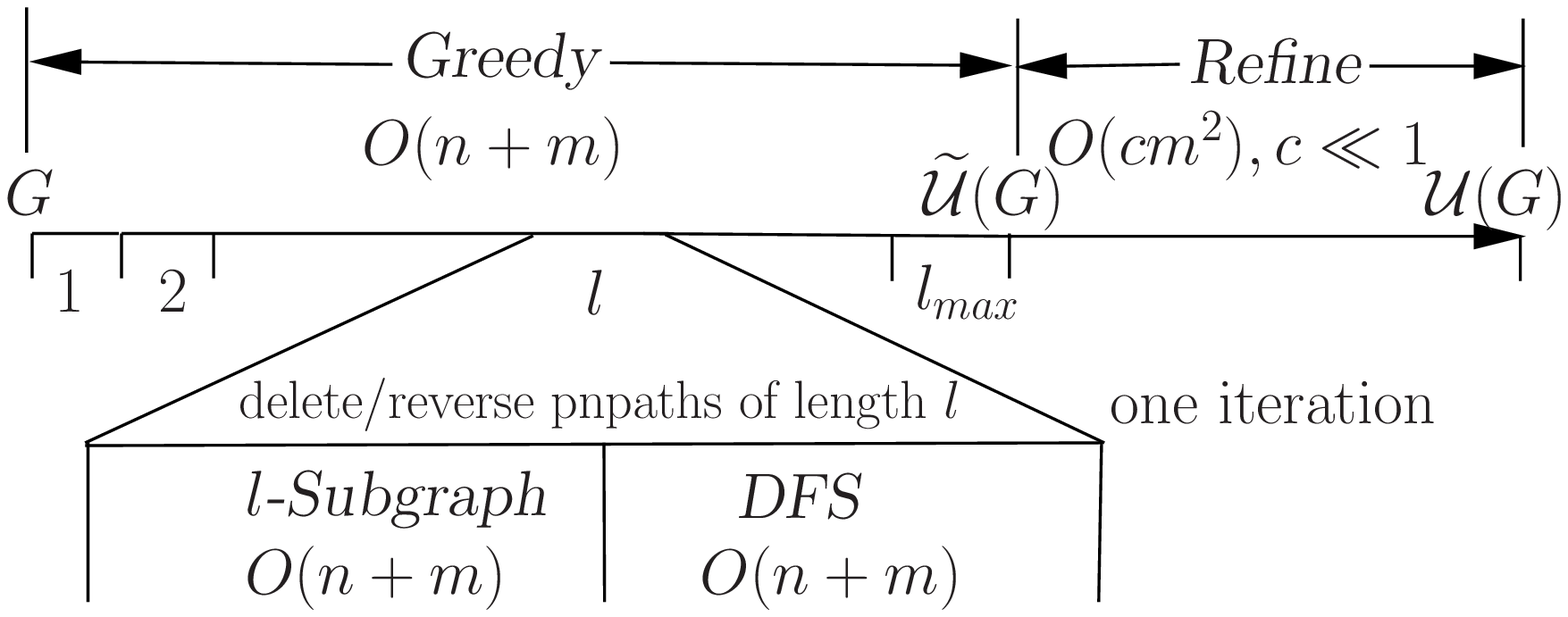}
\vspace*{-0.2cm}
\caption{An Overview of Greedy-\&-Refine}
\vspace*{-0.2cm}
\label{fig:overview}
\end{figure}

In order to reduce the number of iterations, we propose a two-phase
Greedy-\&-Refine algorithm, abbreviated by \refineAlg.  Here, a
\greedy algorithm computes an Eulerian subgraph of $G$, denoted as
$\appeulerian(G)$, and a \refine algorithm refines the greedy solution
$\appeulerian(G)$ to get the maximum Eulerian subgraph $\eulerian(G)$,
which needs at most $|E(\eulerian(G))|-|E(\appeulerian(G))|$
iterations.
The \refineAlg algorithm is given in Algorithm~\ref{alg:gr}, and an
overview is shown in Fig.~\ref{fig:overview}.  In
Algorithm~\ref{alg:gr}, it first computes all \sccs (Line~1). For each
\scc $G_i$, it computes an Eulerian subgraph using \greedy, denoted as
$\appeulerian(G_i)$ (Line~3).
In \greedy, in every iteration $l$ ($ 1 \leq l \leq l_{max}$), it
identifies a subgraph by an \lsubgraph algorithm, and further
deletes/reverses all specific length-$l$ paths called
\pnpaths which we will discuss
in details by \DFS. Note $l_{max}$ is a small number.
After computing $\appeulerian(G_i)$, $G_i-\appeulerian(G_i)$ is near
acyclic, and it moves all cycles from $G_i-\appeulerian(G_i)$ to
$\appeulerian(G_i)$
(Line~4).
%
Finally, it refines $\appeulerian(G_i)$ to obtain the optimal
$\eulerian(G_i)$ by calling \refine (Line~5). The union of all
$\eulerian(G_i)$ is the maximum Eulerian subgraph for $G$. Below, we
first list some important concepts introduced in the algorithm and
analysis parts in Table.~\ref{tbl:concept}, and
then we shall detail the greedy algorithm and refine algorithm,
respectively.

\begin{table*}[t]
{\small
\begin{center}
    \begin{tabular}{|l|l|l|}\hline
{\bf Used-In} & {\bf Symbol} & {\bf Meaning} \\ \hline\hline
Greedy &  $\vlabel(u)$ & $\vlabel(u) = d_O(u) - d_I(u)$ \\ \cline{2-3} 
&  \pnpath$(u, v)$ & path$(u=v_1,v_2,\cdots, v_l=v)$, $\vlabel(u) > 0$,
   $\vlabel(v) <0$, and $\vlabel(v_i) = 0$ for $1 < i <
   l$ \\ \cline{2-3} 
&  $G_l$ &     ($l$-Subgraph) subgraph of $G$ contains all \pnpaths of
   length $l$ \\ \cline{2-3} 
&  $G^T$ & $V(G^T)=V(G)$, $E(G^T) = \{(u,v)~{}|~{}(v, u) \in E(G)\}$
   \\ \cline{2-3} 
&  $\vlevel(v)$ & the shortest
   distance from any vertex $u$ with a positive label, $\vlabel(u)> 0$, in $G$
   \\ \cline{2-3} 
& $\rvlevel(v)$ &
  the shortest distance to any vertex $u$ with a negative label,
  $\vlabel(u) < 0$, in $G$ \\ \hline \hline
Refine &  $\overline{G}$ & $V(\overline{G})=V(G)$,  $\forall (v_i, v_j) \in
   E(G)$, $(v_j, v_i) \in E(\overline{G})$, and $w(v_j, v_i)=-w(v_i,
   v_j)$    \\ \cline{2-3} 
Analysis & \ppath/\npath & a path where every edge is with a
positive/negative weight \\ \cline{2-3}
   & \kcycle & $(v_1^+,v_1^-,v_2^+,\ldots,v_k^+,v_k^-,v_1^+)$,  where
   $(v_i^+,v_i^-)$ are \npaths, and $(v_i^-,v_{i+1}^+)$ plus
   $(v_k^-,v_1^+)$ are \ppaths  \\ \cline{2-3} 
& $\triangle_k$ & the total weight of \nedges for a \kcycle
   ($\triangle_k=\sum_{i=1,\ldots,k}w(v_i^+,v_i^-)$)    \\ \cline{2-3}
& $\triangle_k'$ & the total weight of \pedges for a \kcycle
   ($\triangle_k'=\sum_{i=1, \ldots,k-1} w(v_i^-,v_{i+1}^+) +
   w(v_k^-,v_1^+)$)   \\ \cline{2-3} 
&  ${\cal G}$ & ${\cal G} =\overline{G_P }\oplus G_N$, $G_P = G
   \ominus \appeulerian(G)$ and $G_N =G \ominus \eulerian(G)$   \\ \hline
    \end{tabular}
\end{center}
}
\vspace*{-0.4cm}
\caption{Notations}
\vspace*{-0.2cm}
\label{tbl:concept}
\end{table*}

\comment{
It is important to note that in the following discussions, we use
several subgraphs of $G$, to explain our algorithm. For example, in
Algorithm~\ref{alg:gr}, when we call \refine, we pass two inputs,
$\appeulerian(G)$ and $G$. They are easy for understanding. In
practice, all can be done by manipulating $G$ itself.
}

\begin{algorithm}[t]
\caption{\refineAlg($G$)}
\label{alg:gr}
{\small
\begin{algorithmic}[1]
\STATE Compute \sccs of $G$,  ${\cal S} = \{G_1, G_2, \cdots\}$;
\FOR{each $G_i \in {\cal S}$}
   \STATE $\appeulerian(G_i) \leftarrow$ \greedy($G_i$);
    \STATE Move all cycles found in $G_i-\appeulerian(G_i)$ to $\appeulerian(G_i)$;
    \COMMENT{Make $G_i-\appeulerian(G_i)$ acyclic}
   \STATE $\eulerian(G_i) \leftarrow$ \refine($\appeulerian(G_i)$, $G_i$);
\ENDFOR
\RETURN $\bigcup_{i=1}^{|{\cal S}|} \eulerian(G_i)$;
\end{algorithmic}
}
\end{algorithm}

\subsection{The Greedy Algorithms}

Given a graph $G$, we propose two algorithms to obtain an initial
Eulerian subgraph $\appeulerian(G)$. The first algorithm is denoted as
\greedyD (Algorithm~\ref{alg:greedyD}), which deletes edges from $G$
to make $d_I(v) = d_O(v)$ for every vertex $v$ in
$\appeulerian(G)$. The second algorithm is denoted as \greedyR
(Algorithm~\ref{alg:greedyR}), which reverses edges instead of
deletion to the same purpose. We use \greedy when we refer to either
of these two algorithms.  By definition, the resulting
$\appeulerian(G)$ is an Eulerian subgraph of $G$. The more edges we
have in $\appeulerian(G)$, the closer the resulting subgraph
$\appeulerian(G)$ is to $\eulerian(G)$. We discuss some notations
below


\stitle{The vertex label}: For each vertex $u$ in $G$, we define a
vertex label on $u$, $\vlabel(u) = d_O(u) - d_I(u)$. If $\vlabel(u) =
0$, it means that $u$ can be a vertex in an Eulerian subgraph without
any modifications. If $\vlabel(u) \neq 0$, it needs to delete/reverse
some adjacent edges to make $\vlabel(u)$ being zero.

\stitle{The pn-path}: We also define a positive-start and negative-end
path between two vertices, $u$ and $v$, denoted as \pnpath$(u,
v)$. Here, \pnpath$(u,v)$ is a path $p = (v_1,v_2,\cdots, v_l)$, where
$u = v_1$ and $v =v_l$ with the following conditions: $\vlabel(u) >
0$, $\vlabel(v) < 0$, and all $\vlabel(v_i) = 0$ for $1 < i <
l$. Clearly, by this definition, if we delete all the edges in
\pnpath$(u,v)$, then $\vlabel(u)$ decreases by 1, $\vlabel(v)$
increases by 1, and all intermediate vertices in \pnpath$(u,v)$ will
have their labels as zero. To make all vertex labels being zero, the
total number of such \pnpaths to be deleted/reversed is $N =
\sum_{\vlabel(u)>0} \vlabel(u)$.

\stitle{The transportation graph $G^T$}: A transportation graph $G^T$
of $G$ is a graph such that $V(G^T) = V(G)$ and $E(G^T) = \{
(u, v)~{}|~{} (v, u) \in E(G)\}$.

\stitle{The \vlevel and \rvlevel}: $\vlevel(v)$ is the shortest
distance from any vertex $u$ with a positive label, $\vlabel(u) > 0$,
in $G$.  $\rvlevel(v)$ is the shortest distance from any vertex $u$
with a positive label, $\vlabel(u) > 0$, in $G^T$.  Note $\rvlevel(v)$
is the shortest distance to any vertex $u$ with a negative label,
$\vlabel(u) < 0$, in $G$.


\begin{algorithm}[t]
\caption{\greedyD($G$)}
\label{alg:greedyD}
{\small
\begin{algorithmic}[1]
\STATE $l \leftarrow 1$;
       $G' \leftarrow G$;
\WHILE {some vertex $u \in G'$ with $\vlabel(u) > 0$}
   \STATE $G' \leftarrow$ \lengthLD($G'$, $l$);
          $l \leftarrow l + 1$;
\ENDWHILE
\RETURN $G'$;
%
%
\end{algorithmic}
}
\end{algorithm}

\subsubsection{The Greedy-D Algorithm}
\label{sec:greedyD}

Below, we first concentrate on \greedyD
(Algorithm~\ref{alg:greedyD}).
Let $G'$ be $G$ (Line~1). In the \textbf{while} loop (Lines~2-4), it
repeatedly deletes all \pnpaths starting from length $l = 1$ by
calling an algorithm \lengthLD (Algorithm~\ref{alg:lengthLD}) until no
vertex $u$ in $G'$ with a positive value ($\vlabel(u) > 0$).

\begin{algorithm}[t]
\caption{\lengthLD($G$, $l$)}
\label{alg:lengthLD}
{\small
\begin{algorithmic}[1]
\STATE $G_l \leftarrow$ \lsubgraph($G$, $l$);
\STATE Enqueue all vertices $u \in V(G_l)$ with $\vlabel(u)>0$ into queue $Q$;
\WHILE {$Q \neq \emptyset$}
    \STATE $u \leftarrow$ $Q$.top();
\STATE Following \DFS starting from $u$ over $G_l$, traverse unvisited
edges and mark them ``visited''; let the path from
$u$ to $v$ be \pnpath$(u, v)$, when it reaches the first vertex $v$ in
$G_l$ with $\vlevel(v) = l$;
\IF {\pnpath$(u, v) \neq \emptyset$}
     \STATE delete all edges in \pnpath$(u, v)$ from $G$;
     \STATE $\vlabel(u) \leftarrow \vlabel(u)-1$;
            $\vlabel(v) \leftarrow \vlabel(v)+1$;
     \STATE {\bf if} $\vlabel(u) = 0$ {\bf then}
            $Q$.dequeue();
\ELSE
\STATE $Q$.dequeue();
\ENDIF
\ENDWHILE
\RETURN $G$;
\end{algorithmic}
}
\end{algorithm}

\begin{figure}[t]
\centering
\includegraphics[width=2in,height=1in]{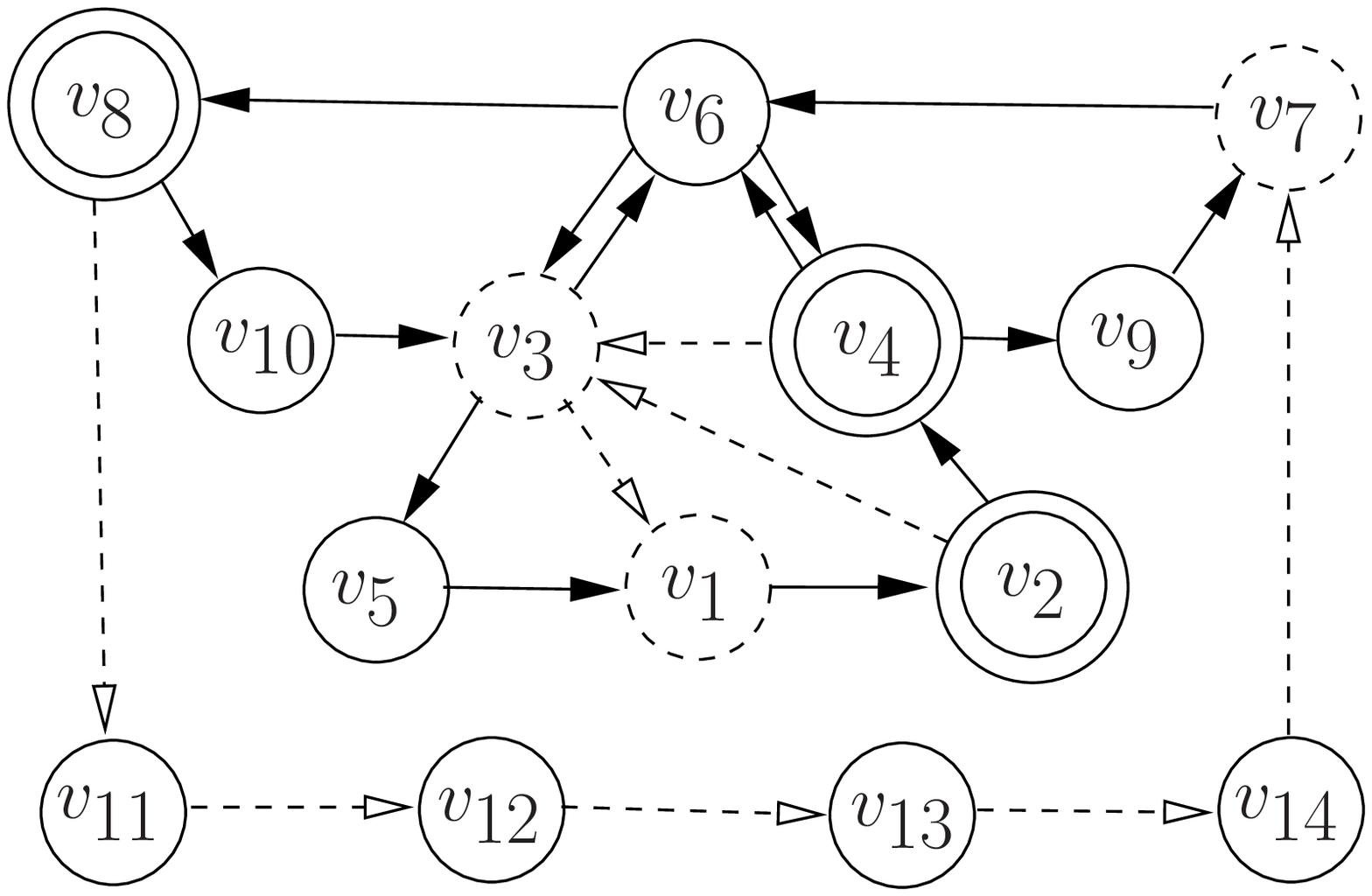}
\vspace*{-0.2cm}
\caption{An Eulerian subgraph obtained by \greedy for graph $G$ in
  Fig.~\ref{fig:fig1}}
\vspace*{-0.6cm}
\label{fig:fig2}
\end{figure}

\begin{example}
Consider graph $G$ in Fig.~\ref{fig:fig2}.  Three vertices,
$v_2$, $v_4$, and $v_8$, in double cycles, have a \vlabel +1, and
three other vertices, $v_1$, $v_3$, and $v_7$, in dashed cycles,
have a \vlabel -1. Initially, $l = 1$, \greedyD
(Algorithm~\ref{alg:greedyD}) deletes \pnpath$(v_2,v_3)$, making
$\vlabel(v_2) = \vlabel(v_3) = 0$. When $l = 2$, \pnpath$(v_4, v_1)
= (v_4,v_3,v_1)$ is deleted. Finally, when $l = 5$, \pnpath$(v_8,
v_7) = (v_8,v_{11},v_{12},v_{13},v_{14},v_7)$ will be deleted.  In
Fig.~\ref{fig:fig2}, the graph with solid edges is
$\appeulerian(G)$ or the graph $G'$ returned by
Algorithm~\ref{alg:greedyD}.  It is worth mentioning that for the same
graph $G$, \DFSEVEN needs 10 iterations. From the Eulerian subgraph
$\appeulerian(G)$ obtain by \greedy, it only needs at most 2
additional iterations to get the maximum Eulerian subgraph.
\end{example}

It is worth noting that $\appeulerian(G)$ is not optimal. Some edges
in $\appeulerian(G)$ may not be in the maximum Eulerian subgraph,
while some edges deleted should appear in the maximum Eulerian
subgraph. In next section, we will discuss how to obtain the maximum
Eulerian subgraph $\eulerian(G)$ from the greedy solution
$\appeulerian(G)$.

\stitle{Finding all pn-paths with length $l$}: The \lengthLD
algorithm is shown in Algorithm~\ref{alg:lengthLD}. In brief, for a
given graph $G$, \lengthLD first extracts a subgraph $G_l \subseteq
G$ which contains all \pnpaths of length $l$ that are possible
to be deleted from $G$ by calling an algorithm \lsubgraph
(Algorithm~\ref{alg:LSS}) in Line~1. In other words, all edges in
$E(G)$ but not in $E(G_l)$ cannot appear in any \pnpaths with a
length $\leq l$. Based on $G_l$ obtained, \lengthLD deletes \pnpaths
from $G$ (not from $G_l$) with additional conditions (in
Lines~2-13). Let $G'_l$ be a subgraph of $G_l$ that
includes all edges appearing in \pnpaths of length $l$ to be
deleted in \lengthLD. \lengthLD will return a subgraph $G \setminus
G'_l$ as a subgraph of $G$, which will be used in the next run in
\greedyD for deleting \pnpaths with length $l+1$.

We discuss the \lsubgraph algorithm (Algorithm~\ref{alg:LSS}), which
extracts $G_l$ from $G$ by \BFS (breadth-first-search) traversing $G$
twice.  In the first \BFS (Lines~4-6), it adds a virtual vertex $s$,
and adds an edge $(s, u)$ to every vertex $u$ with a positive label
($\vlabel(u) > 0$) in $G$. Then, it assigns a \vlevel to every vertex
in $G$ as follows.  Let $\vlevel(s)$ be $-1$.  By \BFS, it assigns
$\vlevel(u)$ to be $\vlevel(\parent(u)) + 1$, where $\parent(u)$ is
the parent vertex of $u$ following \BFS.  In the second \BFS
(Lines~7-10), it conceptually considers the transposition graph $G^T$
of $G$ by reversing every edge $(v, u) \in E(G)$ as $(u, v) \in
E(G^T)$ (Line~7).  Then, it assigns a different \rvlevel to every
vertex in $G$ using the transposition graph $G^T$. Like the first
\BFS, it adds a virtual vertex $t$, and adds an edge $(t, u)$ to every
vertex $u$ with a negative label ($\vlabel(u) < 0$) in $G^T$. Then, it
assigns \rvlevel to every vertex in $G^T$ as follows.  Let
$\rvlevel(t)$ be $-1$.  By \BFS, it assigns $\rvlevel(u)$ to be
$\rvlevel(\parent(u)) + 1$, where $\parent(u)$ is the parent vertex of
$u$ in $G^T$ following \BFS.  The resulting subgraph $G_l$ to be
returned from \lsubgraph is extracted as follows. Here, $V(G_l)$
contains all vertices $u$ in $G$ if $\vlevel(u) + \rvlevel(u) = l$ for
the given length $l$, and $E(G_l)$ contains all edges $(u, v)$ if both
$u$ and $v$ appear in $V(G_l)$, $(u, v)$ is an edge in the given graph
$G$, and $\vlevel(u) + 1 = \vlevel(v)$ (Lines~11-13).
The following example illustrates how \lsubgraph algorithm works.

\comment{
We give a further interpretation of \lsubgraph as follows. By the
first \BFS, all vertices are arranged by \vlevel. Here, for a vertex
$v$, its $\vlevel(v)$ shows its shortest distance from a vertex $u$
with a positive label ($\vlabel(u)$ $> 0$). The following
conclusions are easy to be made. 1) Every vertex $u$ with a positive
label ($\vlabel(u) > 0$) appears at level 0, because it has a zero
distance to itself. 2) Every vertex $v$ with zero label ($\vlabel(v)
= 0$) can appear at any level $i>0$ while only those at level
$i=1,2,\ldots,l-1$ may be included in $G_l$. 3) Every vertex $v$
with a negative label ($\vlabel(v) < 0$) can only appear at level
$\geq l$. But we are only interested in those at level $l$. In a
similar manner, by the second \BFS, all vertices are arranged by
\rvlevel. Here, for a vertex $v$, its $\rvlevel(v)$ shows its
shortest distance to a vertex $u$ with a negative label ($\vlabel(u)
< 0$) in $G$. Also, we have the following facts. 1) every vertex $u$
with a negative label ($\vlabel(u) < 0$) appears at \rvlevel $0$,
because it has a zero distance to itself. 2) Every vertex $v$ with
zero label ($\vlabel(v) = 0$) appears at \rvlevel $i>0$. 3) Every
vertex $v$ with a positive label ($\vlabel(v) > 0$) can appear at
\rvlevel $\geq l$.
}

\begin{figure}[t]
\begin{center}
\begin{tabular}[t]{c}
    \subfigure[\BFS-Tree of \BFS$(G,s)$ ]{
         \includegraphics[width=0.48\columnwidth,height=2.5cm]{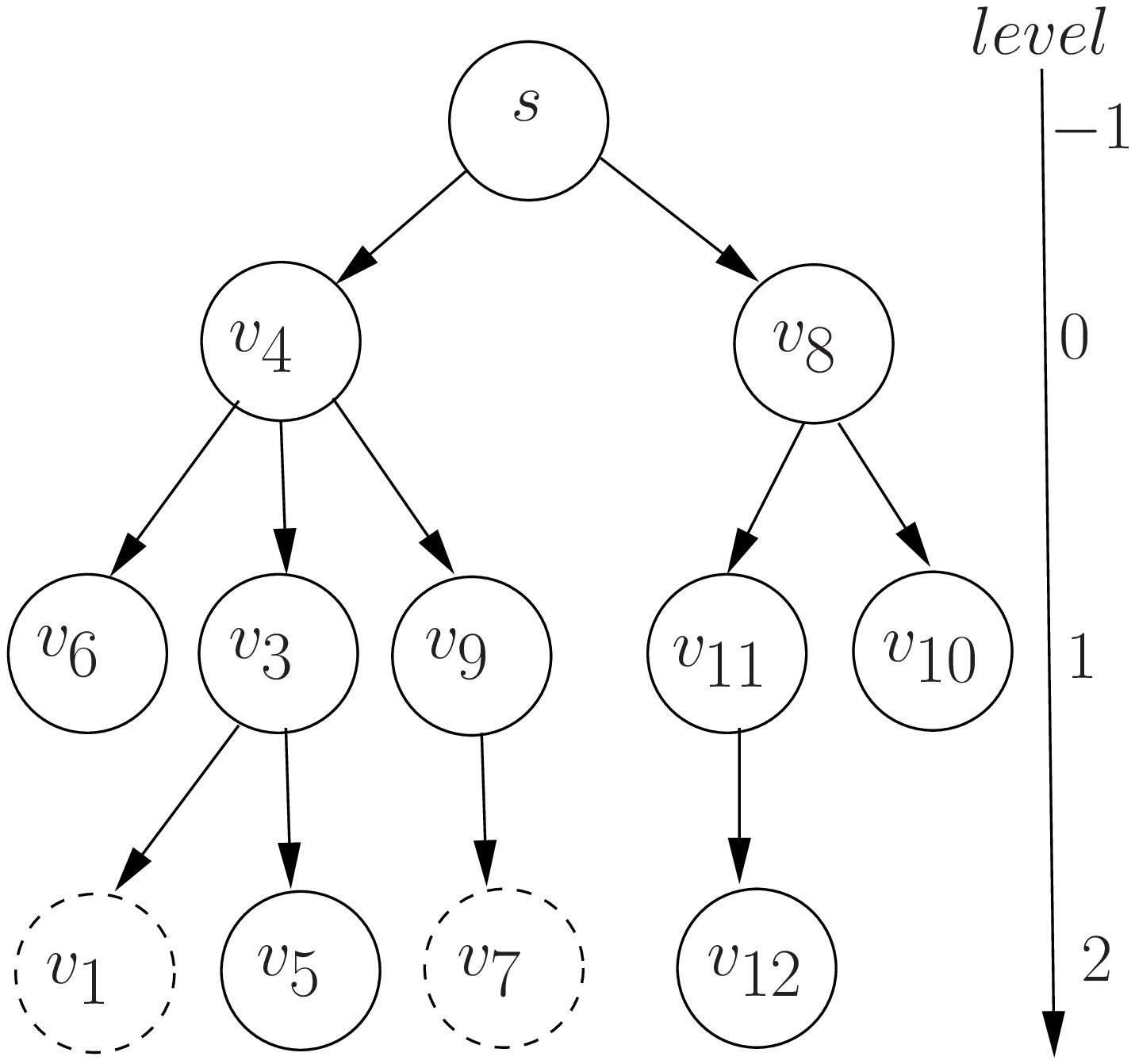}
         \label{fig:level}
    }
    \subfigure[\BFS-Tree of \BFS$(G^T,t)$]{
    \includegraphics[width=0.48\columnwidth,height=2.5cm]{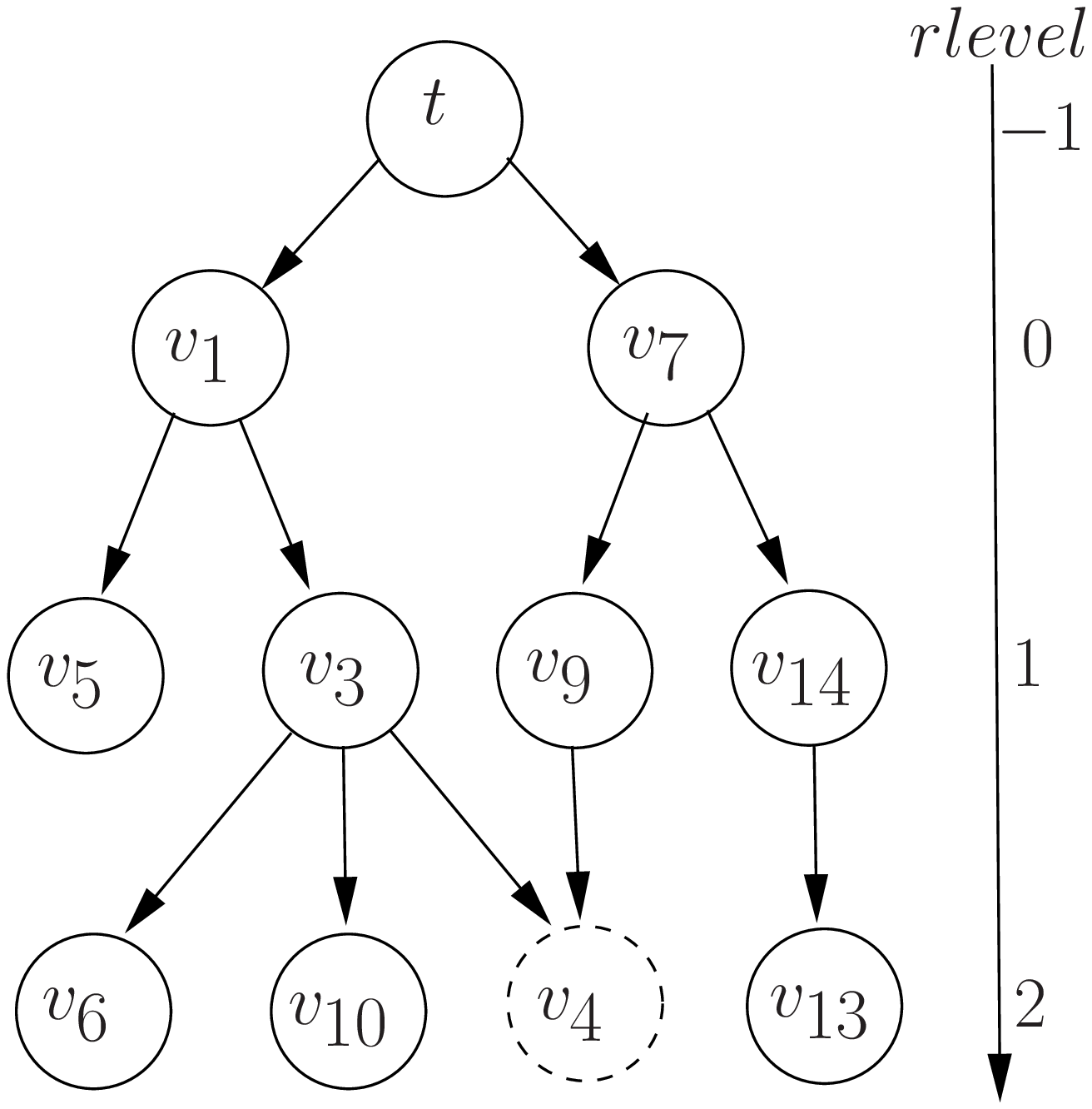}
    \label{fig:rlevel}
    }
\end{tabular}
\end{center}
\vspace*{-0.4cm}
\caption{\BFS-Trees used for constructing \lsubgraph}
\vspace*{-0.4cm}
\label{fig:lsubgraph}
\end{figure}

\begin{figure}[t]
\centering
\includegraphics[width=2in,height=1in]{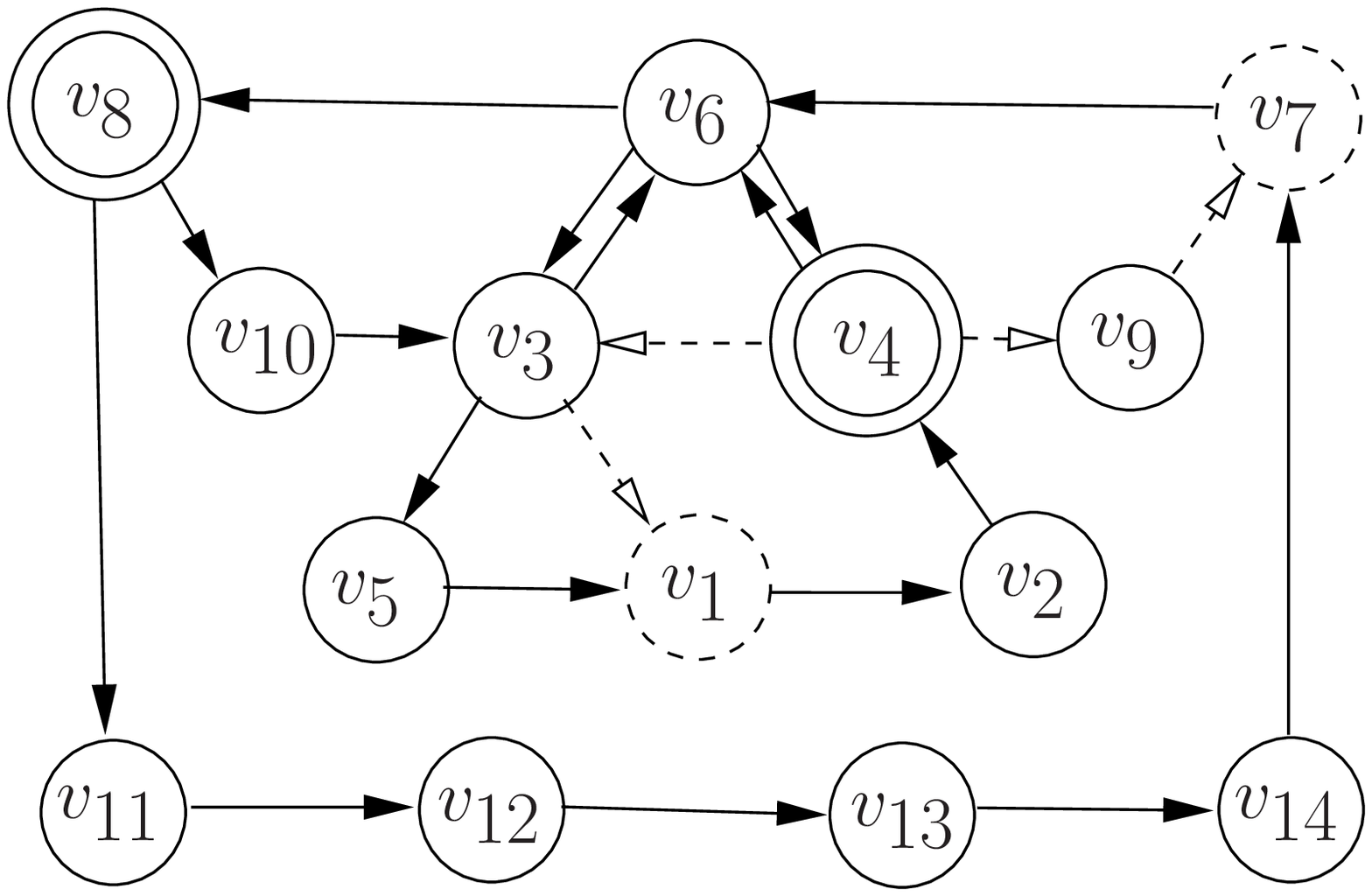}
\vspace*{-0.2cm}
\caption{An \lsubgraph for length $l = 2$}
\vspace*{-0.4cm}
\label{fig:fig4}
\end{figure}

\begin{example}
Fig.~\ref{fig:fig4} illustrates the $G_l$ returned by \lsubgraph
(Algorithm~\ref{alg:LSS}) when $l = 2$. It is constructed using two
\BFS, i.e., \BFS($G,s$) and \BFS($G^T,t$), and the associated
\BFS-trees with level $\leq 2$ and \rvlevel $\leq 2$ are shown in
Fig.~\ref{fig:level} and Fig.~\ref{fig:rlevel}, respectively.  In
Fig.~\ref{fig:level}, vertices $v_1$ and $v_7$ are the only vertices
with $\vlabel$ $< 0$.  In Fig.~\ref{fig:rlevel}, vertex $v_4$ is the
only one with $\vlabel$ $> 0$. Therefore, $G_l$ contains only four
edges, in dashed lines, which is much smaller than the original graph
$G$ to be handled.
\end{example}

\begin{algorithm}[t]
\caption{\lsubgraph($G,l$)}
\label{alg:LSS}
{\small
\begin{algorithmic}[1]
\FOR {each vertex $u$ in $V(G)$}
    \STATE $\vlevel(u) \leftarrow \infty$, $\rvlevel(u) \leftarrow \infty$;
\ENDFOR
\STATE Add a virtual vertex $s$ and an edge $(s,u)$ from $s$ to every
       vertex $u$ in $G$ if  $\vlabel(u)>0$;
\STATE $\vlevel(s) \leftarrow -1$;
\STATE $\vlevel(u) \leftarrow \vlevel(\parent(u))+1$ for all
       vertices $u$ in $G$ following \BFS staring from $s$;
\STATE Construct a graph $G^T$ where $V(G^T) = V(G)$
and $E(G^T) =
       \{(u, v)| (v, u) \in E(G)\}$;
\STATE Add a virtual vertex $t$ and an edge $(t,u)$ from $s$ to every
       vertex $u$ in $G^T$ if  $\vlabel(u) < 0$;
\STATE $\rvlevel(t) \leftarrow -1$;
\STATE $\rvlevel(u) \leftarrow \rvlevel(\parent(u))+1$ for all
       vertices $u$ in $G^T$ following \BFS staring from $t$;
\STATE Extract a subgraph $G_l$;
\STATE $V(G_l) = \{u~{}|~{}\vlevel(u) + \rvlevel(u) = l\}$;
\STATE $E(G_l) = \{(u, v)~{}|~{}u \in V(G_l), v \in V(G_l), (u, v) \in E(G),$
       $\vlevel(u) + 1 = \vlevel(v)\}$;
\RETURN $G_l$;
\end{algorithmic}
}
\end{algorithm}

\comment{
\begin{figure}[t]
\begin{center}
\begin{tabular}[t]{c}
    \subfigure[$G_l$]{
         \includegraphics[width=0.5\columnwidth,height=6cm]{fig/level1.eps}
         \label{fig:level1}
    }
    \subfigure[A \pnpath]{
         \includegraphics[width=0.5\columnwidth,height=6cm]{fig/level2.eps}
         \label{fig:level2}
    }
\end{tabular}
\end{center}
\vspace*{-0.4cm}
\caption{The \lsubgraph Algorithm}
\vspace*{-0.2cm}
\label{fig:lsubgraph}
\end{figure}

\begin{figure}[t]
\centering
\includegraphics[width=3in,height=2in]{fig/level1.eps}
\caption{$G_l$}
\label{fig:level1}
\end{figure}

\begin{figure}[t]
\centering
\includegraphics[width=3in,height=2in]{fig/level2.eps}
\caption{Path p}
\label{fig:level2}
\end{figure}
}

\comment{
\begin{figure}[t]
\centering
\includegraphics[width=2.5in,height=1.5in]{fig/gl.eps}
\vspace*{-0.2cm}
\caption{$G_l$}
\label{fig:gl}
\vspace*{-0.2cm}
\end{figure}
}

\begin{lemma}
\label{lemma:the2}
By \lsubgraph, the resulting subgraph $G_l$ includes all \pnpaths of
length $l$ in $G$.
\end{lemma}

\proofsketch
Recall that \lsubgraph returns a graph $G_l$ where $V(G_l) = \{u~{}|~{}\vlevel(u) +
\rvlevel(u) = l\}$ and $E(G_l) = \{(u, v)~{}|~{}u \in V(G_l), v \in V(G_l),
(u, v) \in E(G),$ $\vlevel(u) + 1 = \vlevel(v)\}$.
It implies the following. All vertices in $G_l$ are on
at least one shortest path from a positive label vertex $u$ ($\vlabel(u)
> 0$) to a negative label vertex $v$ ($\vlabel(v) < 0$) of length
$l$. All edges are on such shortest paths.
No any edge in a \pnpath of length $l$ will be excluded from $G_l$. In
other words, there does not exist an edge $(u', v')$ on \pnpath$(u,
v)$ of length $l$, which does not appear in $E(G_l)$.  \eop

We explain \lengthLD (Algorithm~\ref{alg:lengthLD}). Based on $G_l$
obtained from $G$ using \lsubgraph (Algorithm~\ref{alg:LSS}), in
\lengthLD, we delete all possible \pnpaths of length $l$ from $G$
(Lines~2-13). The deletion of all \pnpaths of length $l$ from the
given graph $G$ is done using \DFS over $G_l$ with a queue $Q$.  It
first pushes all vertices $u$ in $V(G_l)$ with a positive label
($\vlabel(u) > 0$) into queue $Q$, because they
are the starting vertices of all \pnpaths with length $l$. We check
the vertex $u$ on the top of queue $Q$. With the vertex $u$, we do
\DFS starting from $u$ over $G_l$, traverse unvisited edges in $G_l$,
and mark the edges visited as ``visited''. Let $p$ be the first
\pnpath$(u, v)$ with length $l$ along \DFS.
We delete all edges on $p$, and adjust the labels as to reduce
$\vlabel(u)$ by 1 and increase $\vlabel(v)$ by 1. We dequeue $u$ from
queue $Q$ until we cannot find any more \pnpaths of length $l$
starting from $u$, i.e. $p$ returned by \DFS$(u)$ is empty. It is
important to note that we only visit each edge at most once. There are
two cases. One is that the edges visited will be deleted and there is
no need to revisit. The other is that they are marked as ``visited''
but not included in any \pnpaths with length $l$. For this case,
these edges will not appear in any other \pnpaths starting from any
other vertices.

\comment{
that does not appear in $G_l$ which we have constructed in
Fig.~\ref{fig:level1}. Then, as shown in Fig.~\ref{fig:level2},
there is at least one edge $(u,v)$ in $p$ that does not appear
in $G_l$, where $\vlevel(v) + 1 \neq \vlevel(u)$.
%
%
Without loss of generality, we assume $\vlevel(u) < l$ and $\vlevel(v)
<l$.  There are two cases.
\begin{enumerate}
\item if $\vlevel(v) > \vlevel(u)$, then $\vlevel(v) > \vlevel(u)+1$,
  which contradicts with the definition of level as we can easily find
  a path of length $u.L+1$ from a positive label vertex to $v$ through
  $u$ and $(u,v)$.
\item if $\vlevel(v) \leq \vlevel(u)$, then the path $p$ can be
  represented as $p=<v_1,u,v,v_3>$, with$l(v_1)>0, l(v_3)<0$, as shown in
  Fig.~\ref{fig:level2}. In addition, we can find another path
  $p'=<v_2,v,v_3>$ and $|p'|=|<v_2,v>|+|<v,v_3>|=v.L+k-1-u.L<k$, which
  contradicts to Theorem.~\ref{the:the1}
\end{enumerate}
Therefore, such edge $e=(u,v)$ cannot exist, which implies  all
\pnpaths of length $l$ in $G$ are in $G_l$.
}

\begin{lemma}
\label{lemma:the3}
By \lengthLD, all \pnpaths of length $l$ are deleted.
\end{lemma}

\proofsketch It can be proved based on \DFS over $G_l$ obtained
from \lsubgraph.

\begin{lemma}
\label{lemma:the1}
By \lengthLD, the resulting $G$ does not
include any \pnpaths of length $\leq l$.
\end{lemma}

\proofsketch Let $G_i'$ be the resulting graph of \lengthLD after
deleting all \pnpaths of length $i$ from $G$. It is trivial when
$i=1$. Assume that it holds for $G_i'$ when $i < l$.  We prove that
$G_i'$ holds when $i = l$. First, there are no \pnpaths of length
$\leq l-1$ in graph $G_{l-1}'$ as a result of \lengthLD by
assumption.
%
%
Second, $G_l' \subseteq G_{l-1}'$ because $G_l'$ is obtained by deleting
\pnpaths of length $l$ from $G_{l-1}'$, as given in the \greedyD
algorithm (Algorithm~\ref{alg:greedyD}).
%
%
Furthermore, in \lengthLD, every vertex $u$ with $\vlabel(u) = 0$ in
$G_{l-1}'$ keeps $\vlabel(u) = 0$ in $G_l'$. If there is a
\pnpath$(u,v)$ of length $\leq l-1$ found in $G_l'$, then it must be
in $G_{l-1}'$, which contradicts the assumption.
%
%
Therefore, $G_l'$ does not include any \pnpaths of length $\leq l$.
%
\eop

\begin{theorem}
The \lengthLD algorithm correctly identifies a subgraph $G_l$ which
contains all \pnpaths of length $l$ and returns a graph includes no
\pnpaths of length $ \leq l$.
\label{the:lsubgraph}
\end{theorem}

\proofsketch It can be proved by Lemma~\ref{lemma:the2} and
Lemma~\ref{lemma:the1}.


\comment{
Except for bounding search space, Level-based Search Subgraph $S$ also
benefits searching paths for deletion in that each edge in $S$ can be
visited once and only once. Specifically, for a positive labeled
vertex $v$, we can apply DFS($S,v$) to search for a path started from
$v$ and "delete" all the edges visited until visit a negative labeled
vertex. For the correctness, without loss of generality, we choose an
arbitrary edge $(v_i,v_{i+1})$, and $v_i.L=i$. It is trivial when this
edge is on a path for deletion. Otherwise, there exist no path of
length $k-i$ from $v_i$ to a negative label vertex through edge
$(v_i,v_{i+1})$, otherwise there is a length $k$ path from $v$ to this
negative vertex. Since $i$, i.e. $v_i.L$, is the shortest length from
any positive vertex to $v_i$, there is also no length $k$ path go
through edge $(v_i,v_{i+1})$.
}

We discuss the time complexity of the \greedyD algorithm.  In our
experiments, we show that more than 99.99\% \pnpaths deleted in most real-world datasets are with a length less than or
equal to 6. We take the maximum length $l_{max}$ in the \greedyD algorithm, which is
equivalent to the iterations of calling \lengthLD, as
a constant, since it is always less than 100 in our extensive experiments.
Here, both \lengthLD and \lsubgraph cost $O(n + m)$, because
\lsubgraph invokes \BFS twice and \lengthLD performs \DFS once in addition.  Given $l_{max}$
as a constant, the time complexity of the \greedyD algorithm is $O(n +
m)$.

\begin{algorithm}[t]
\caption{\greedyR($G$)}
\label{alg:greedyR}
{\small
\begin{algorithmic}[1]
\STATE $l \leftarrow 1$;
\STATE Assign an initial value of $-1$ to the weight $w(v_i, v_j)$ for
every edge $(v_i, v_j) \in E$;
\WHILE {some vertex $u \in G$ with $\vlabel(u) > 0$}
   \STATE $G \leftarrow$ \lengthLR($G$, $l$);
   \COMMENT{\lengthLR is the same as \lengthLD
     (Algorithm~\ref{alg:lengthLD}) except that in
     Algorithm~\ref{alg:lengthLD}, Line~7 is changed to be
     ``reverse all edges in \pnpaths$(u, v)$ in $G$, both weights and directions''}
   \STATE $l \leftarrow l + 1$;
\ENDWHILE
\STATE Remove edges $(v_i, v_j)$ from $G$ if $w(v_i, v_j) = +1$;
\RETURN $G$;
%
%
\end{algorithmic}
}
\end{algorithm}


\subsubsection{The Greedy-R Algorithm}

The \greedyR algorithm is shown in Algorithm~\ref{alg:greedyR}. Like
\greedyD, \greedyR will result in an Eulerian subgraph. Unlike
\greedyD, it reverses the edges on \pnpaths of length $l$ from $l =
1$ until there does not exist a vertex $u$ in $G$ with $\vlabel(u) >
0$. Initially, \greedyR assigns every edge, $(v_i, v_j)$, in $G$
with a weight $w(v_i, v_j) = -1$. Then, in the while loop, it calls
\lengthLR. \lengthLR is the same as \lengthLD
(Algorithm~\ref{alg:lengthLD}) except that in
Algorithm~\ref{alg:lengthLD} Line~7 is changed to be ``reverse all
edges in \pnpath$(u, v)$ in $G$, both weights and directions''. As a
result, \greedyR identifies an Eulerian subgraph of $G$,
$\appeulerian(G)$. Here, $E(\appeulerian(G))$ contains all edges
with a weight $= -1$ and $V(\appeulerian(G))$ contains all the
vertices in $E(\appeulerian(G))$. Below, we give two lemmas to prove
the correctness of \greedyR.

\begin{lemma}
By \lengthLR, the resulting $G$ does not
include any \pnpaths of length $\leq l$.
\label{lemma:5.5}
\end{lemma}

\begin{figure}[t]
\centering
\includegraphics[scale=0.25]{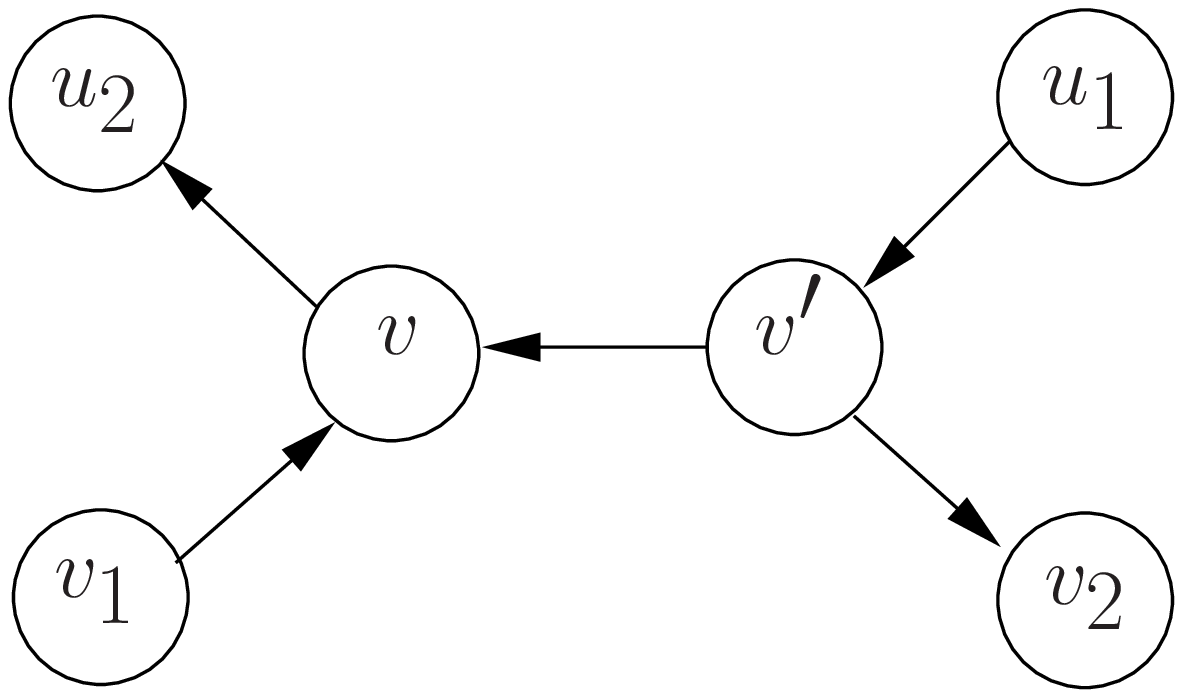}
\vspace*{-0.2cm}
\caption{An example to explain \lengthLR.}
\vspace*{-0.4cm}
\label{fig:fig11}
\end{figure}

\proofsketch Let $G_i'$ be the resulting graph of \lengthLR$(G,i)$, i.e.
after reversing all \pnpaths of length $i$ from $G$. It is trivial when
$i=1$. Assume that it holds for $G_i'$ when $i < l$.  We prove that $G_i'$ holds
when $i = l$. Otherwise, suppose that there is a \pnpath$(v_1,v_2)$ of length $<l$ in
$G_l'$, then there exists at least one edge $e=(v,v')$ in
\pnpath$(v_1,v_2)$ that has been reversed during \lengthLR$(G,l)$,
otherwise, \pnpath$(v_1,v_2)$ will be fully included in
$G_{l-1}'$. Without loss of generality, assume that edge $(v',v)$ is a part of
\pnpath$(u_1,u_2)= (u_1,v',v,u_2)$ of length $l$. Fig.~\ref{fig:fig11} shows $G_{l-1}'$ (before calling \lengthLR$(G,l)$). Then, we can easily construct \pnpath$(u_1,v_2)=(u_1,v',v_2)$ and \pnpath$(v_1,u_2)= (v_1,v,u_2)$, and at least one of them is of length $\leq l-1$, contradicting the assumption. In
addition, there can not exist any \pnpath of length $l$ in $G_l'$. As
a consequence, by \lengthLR, the resulting $G$ does not include any
\pnpaths of length $\leq l$.  \eop

Similar to Theorem.~\ref{the:lsubgraph}, \lengthLR algorithm correctly
identifies a subgraph $G_l$ which contains all \pnpaths of length $l$
and returns a graph includes no \pnpaths of length $ \leq l$.

\begin{theorem}
The \lengthLR algorithm correctly identifies a subgraph $G_l$ which
contains all \pnpaths of length $l$ and returns a graph includes no
\pnpaths of length $ \leq l$.
\label{the:lsubgraph-LR}
\end{theorem}

We omit the proof of Theorem.~\ref{the:lsubgraph-LR} since it can be
proved in a similar manner like Theorem.~\ref{the:lsubgraph} using
Lemma~\ref{lemma:5.5}.

\comment{
\begin{lemma}
The resulting graph of \greedyR is an Eulerian subgraph of $G$.
\end{lemma}

\proofsketch The proof is omitted due to space limit. \eop
}

\comment{
\proofsketch Since the resulting graph of \lengthLR contains edges
weight -1 only, for each vertex $v$, we focus on its adjacent edges of
weight -1 when consider its in-degree $d_I(v)$ and out-degree
$d_O(v)$, i.e. $d_I(v)=|N_I(v)|=|\{u|(u,v) \in E, w(u,v)=-1\}|$,
$d_O(v)=|N_O(v)|=|\{u|(v,u) \in E,$ $w(v,u)=-1\}|$. Then, the lemma is
equivalent to the statement that, in the resulting graph of \lengthLR,
$d_I(v)=d_O(v)$ for each vertex $v$.

If every time we reverse a \pnpath$(u,v)$, any vertex $v' \in $
\pnpath$(u,v)$ remains $d_I(v')=d_O(v')$, then the lemma
holds. Consider an arbitrary vertex $v'$ in \pnpath$(u,v)$, and
suppose that $v'$ has an in-coming edge $e_1=(v_1,v')$ and an
out-going edge $e_2=(v',v_2)$. There are three cases regarding their
weights.
\begin{itemize}
\item $w(e_1)=w(e_2)=1$: After reversing \pnpath$(u,v)$,
  $d_I(v')$ $=d_I(v')+1, d_O(v')=d_O(v')+1$, $d_I(v')=d_O(v')$ hold.
\item $w(e_1)=w(e_2)=-1$: After reversing \pnpath$(u,v)$,
  $d_I(v')$ $=d_I(v')-1, d_O(v')=d_O(v')-1$, $d_I(v')=d_O(v')$ hold.
\item Without loss of generality, assume that $w(e_1)=1, w(e_2)=-1$, both
  $d_I(v')$ and $d_O(v')$ remain unchanged, thus $d_I(v')=d_O(v')$
  holds.
\end{itemize}
Put it all together, the resulting graph of \greedyR is an Eulerian subgraph of $G$.
\eop
}

\begin{figure}[h]
\vspace*{-0.4cm}
\begin{center}
\begin{tabular}[t]{c}
    \subfigure[$G'$ returned by \greedyD]{
         \includegraphics[width=0.48\columnwidth,height=2.5cm]{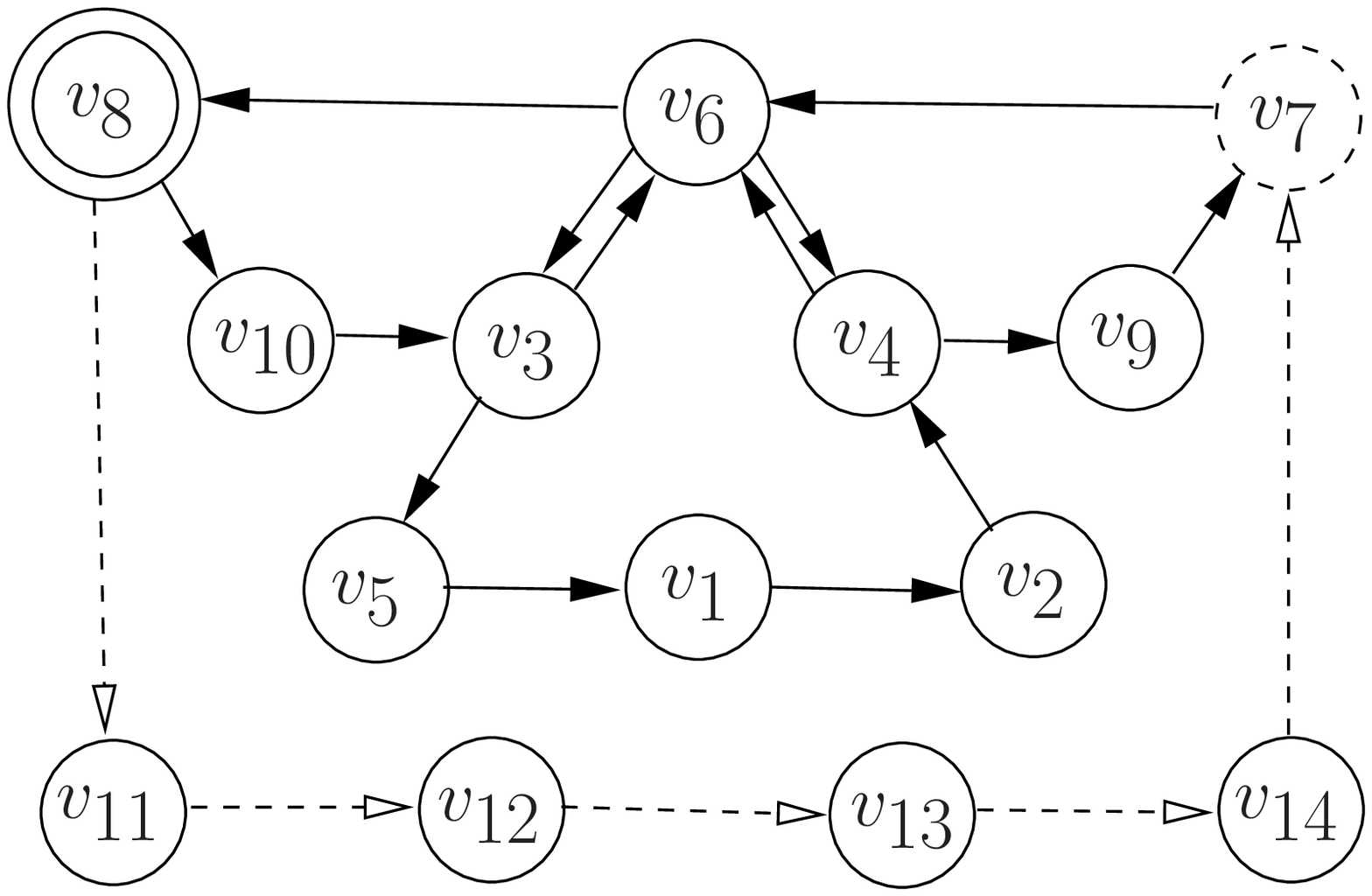}
         \label{fig:greedyD}
    }
    \subfigure[$G$ returned by \greedyR]{
    \includegraphics[width=0.48\columnwidth,height=2.5cm]{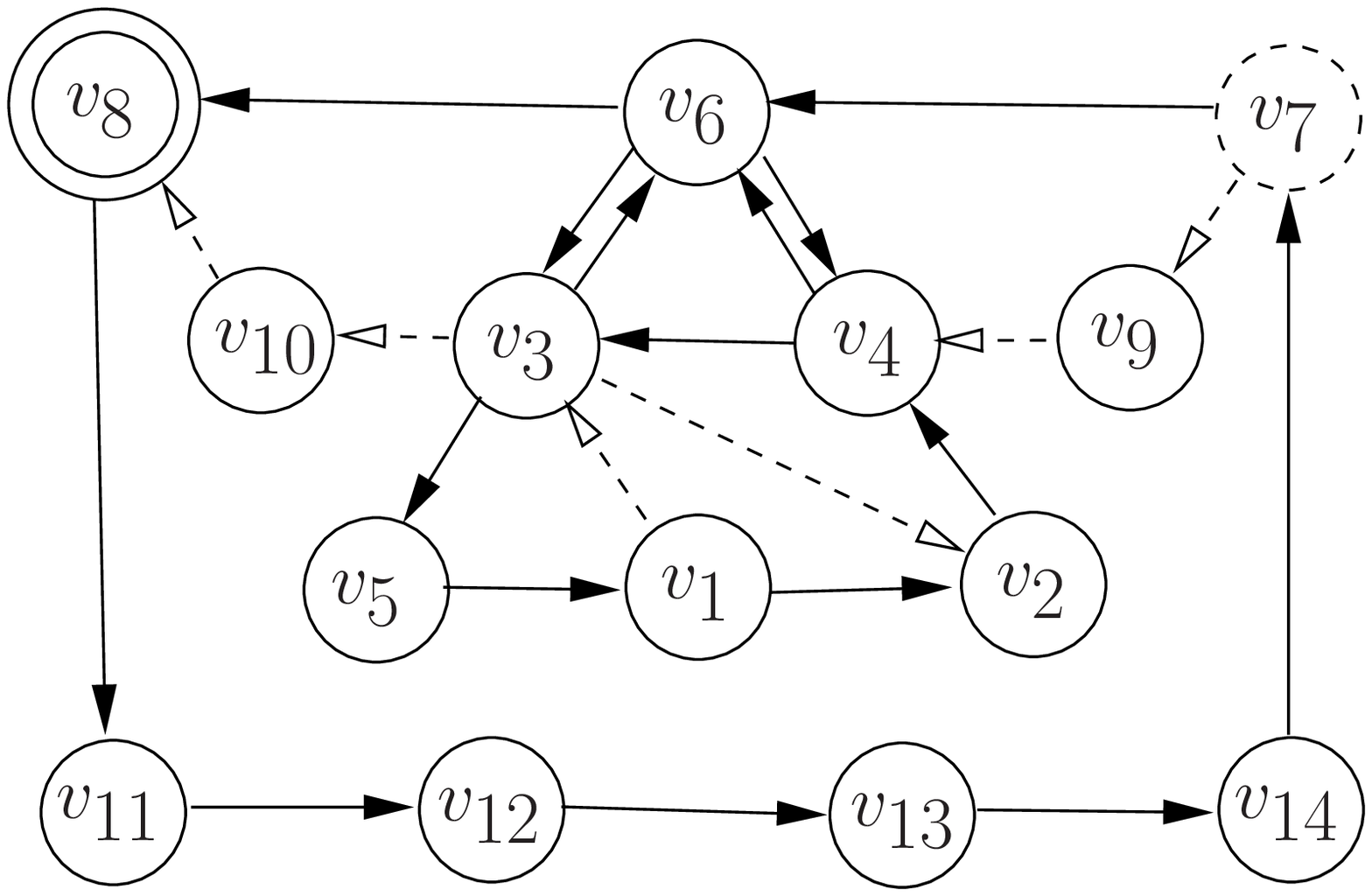}
    \label{fig:greedyR}
    }
\end{tabular}
\end{center}
\vspace*{-0.4cm}
\caption{$\appeulerian(G)$ returned by \greedyD and \greedyR}
\vspace*{-0.4cm}
\label{fig:greedy}
\end{figure}

It is worth noticing that $\appeulerian(G)$ obtained by \greedyR is at
least as good as that obtained by \greedyD.  If each edge in
$G-\appeulerian(G)$ is reversed once, then the $\appeulerian(G)$
obtained by \greedyR is equivalent to that obtained by \greedyD, as
each edge appears in at most one \pnpath. On the other hand, if there
are some edges being reversed more than once, \greedyR performs
better.  Fig.~\ref{fig:greedy} shows the difference between \greedyD
and \greedyR. Since \pnpaths of length 1 and 2 are the same, we only
show the last deleted/reversed \pnpath. In Fig.~\ref{fig:greedyD}, we
delete \pnpath$(v_8,v_7)=(v_8,v_{11},v_{12},$ $v_{13},$ $v_{14},v_7)$. On
the other hand, in Fig.~\ref{fig:greedyR}, we reverse
\pnpath$(v_8,v_7)$ $=(v_8,v_{10},v_3,v_4,v_9,v_7)$. Here edge
$(v_4,v_3)$ is reversed twice. $\appeulerian(G)$ returned by \greedyR
consists of solid lines, which is better than that returned by
\greedyD.

\subsection{The Refine Algorithm}
\label{sec:refine}

With the greedy Eulerian subgraph $\appeulerian(G)$ found, we have
insight on $G$ because we know $G = \appeulerian(G) \cup
\widetilde{G}_D$ where $\widetilde{G}_D$ is a DAG (acyclic), and can
design a \refine algorithm based on such insight, to reduce the number
of times to update $dst(u)$, which reduces the cost of
relaxing.  The \refine algorithm (Algorithm~\ref{alg:OR}) is designed
based on the similar idea given in \DFSEVEN using \DFSSPFA with two
following enhancements.

\comment{
we can apply
\DFSEVEN to obtain the maximum Eulerian subgraph $\eulerian(G)$ by
initializing edge weight $w(v_i, v_j)=-1$ for each edge $(v_i, v_j)
\in G-\appeulerian(G)$ and reversing edges $(v_i, v_j) \in
\appeulerian(G)$ to be $(v_j, v_i)$ with edge weight $w(v_j,
v_i)=1$. However, benefit from the fact that $G-\appeulerian(G)$ is
acyclic and $\appeulerian(G)$ approximates $\eulerian(G)$ well, we can
design subtle algorithms to accelerate refinement. In this section,

we
propose a refinement algorithm called \refine which refines the
Eulerian subgraph obtained by \greedy, i.e. $\appeulerian(G)$, to the
maximum Eulerian subgraph $\eulerian(G)$.
}

\begin{algorithm}[t]
\caption{\refine($\appeulerian(G)$, $G$)}
\label{alg:OR}
{\small
{\bf Input}: A graph $G$, and the Eulerian subgraph obtained by \greedy,
$\appeulerian(G)$ \\
{\bf Output}: Two subgraphs of $G$, $\eulerian(G)$ and $G_D$ ($G = \eulerian(G) \cup
G_D$)
\label{alg:refine}
\begin{algorithmic}[1]
\FOR {each edge $(v_i, v_j)$ in $E(G)$}
     \IF {$(v_i, v_j) \in \appeulerian(G)$}
        \STATE reverse the edge to be $(v_j, v_i)$ in $G$;
               $w(v_j, v_i) \leftarrow +1$;
     \ELSE
        \STATE $w(v_i, v_j) \leftarrow -1$;
     \ENDIF
\ENDFOR
\STATE Assign $dst(u)$ for every $u \in V(G)$ based on
Eq.~(\ref{eq:dstinit});
\STATE {\bf for} each vertex $u$ in $V(G)$ {\bf do}
        $relax(u) \leftarrow true$, $pos(u)
       \leftarrow 0$;
%
%
\STATE Enqueue every vertex $u$ in $V(G)$ into a queue ${\cal Q}$;
\STATE $u \leftarrow {\cal Q}$.front();
\WHILE {${\cal Q} \neq \emptyset$}
       \STATE $S_V \leftarrow \emptyset$, $S_E \leftarrow \emptyset$,
              $NV \leftarrow \emptyset$;
       \IF {$relax(u)=true$ and \DFSSPFA($G$, $u$)}
           \STATE Reverse negative cycle and change the edge weights
           using $S_V$ and $S_E$ (refer to Algorithm~\ref{alg:FMES});
           \STATE ${\cal Q} \leftarrow {\cal Q} \cup S_V$;
       \ELSE
           \STATE ${\cal Q}$.pop(); $u \leftarrow {\cal Q}$.front();
\ENDIF
\ENDWHILE
\STATE $G_D$ is a subgraph that contains all edges with a weight of -1;
\STATE $\eulerian(G)$ is a subgraph that contains the edges reversed for all
edges with a weight of +1;
\end{algorithmic}
}
\end{algorithm}


First, we utilize $G = \appeulerian(G) \cup \widetilde{G}_D$ to
initialize the edge weight $w(v_i, v_j)$ for every edge $(v_i, v_j)$
and $dst(u)$ for every vertex $u$ in $G$.
The edge weights are initialized in Line~1-7 in Algorithm~\ref{alg:OR}
based on $\appeulerian(G)$ which is a greedy Eulerian subgraph.
We also make use of $\widetilde{G}_D$ to initialize $dst(u)$ based on
Eq.~(\ref{eq:dstinit}) in Line~8.

\begin{small}
\begin{equation}
\label{eq:dstinit}
dst(u) = \begin{cases}
            0& \text{if $d_I(u)$ in $\widetilde{G}_D$ is 0}, \\
            \min\{dst(v)-1|(v,u) \in \widetilde{G}_D\} & \text{$u \in
              \widetilde{G}_D$}, \\
            0& \text{otherwise}
         \end{cases}
\end{equation}
\end{small}

\noindent
Some comments on the initialization are made below. Following
Algorithm~\ref{alg:FMES}, $dst(u)$ can be initialized as
$dst(u)=0$. In fact, consider Lemma~\ref{lem:-4}. No matter what
$dst(v_i)$ is for a vertex $v_i$ ($1 \leq i \leq k-1$) in a negative
cycle $C=(v_1, v_2, \dots, v_k=v_1)$, the negative cycle can
be identified because there is at least one edge $(v_i,v_{i+1})$ that
can be relaxed. Based on it,
%
%
if we initialize $dst(u)$ in a way such that $dst(u) \leq
dst(v)+w(v,u)$, then $u$ cannot be relaxed through $(v,u)$ before
updating $dst(v)$. It reduces the number of times to update $dst(u)$, and
improves the efficiency. We explain it further.  Because for any edge, $(v,
u) \in \widetilde{G}_D$, $u$ can never be relaxed through edge $(v,
u)$ before $dst(v)$ being updated, \DFSSPFA($G,u$) will relax edges
along a path with a few branches to identify a \ncycle.
The variables such as $relax(u)$ and $pos(u)$ are initialized in
Line~9 as done in Algorithm~\ref{alg:FMES}.

\comment{
First, since $\widetilde{G}_D$ is acyclic, every vertex $u \in
\widetilde{G}_D$ can be initialized effectively with a different
$dst(u)$. We initialize $dst(u)=0$ for vertices $u \in \widetilde{G}_D$
with $d_I(u)=0$ and remove all their out-going edges. Repeat this
process, other vertices $u$ are initialized with
$dst(u)=min\{dst(v)-1|(v,u) \in G-\appeulerian(G) \}$. Consequently,
for each edge $(u,v) \in G-\appeulerian(G)$, $v$ can never be relaxed
through edge $(u,v)$ before $dst(u)$ being updated, making
\DFSSPFA($G,u$) relax edges along a path with a few branches,
efficiently identifies a \ncycle or terminates by returning
false.
}

Second, we use a queue ${\cal Q}$ to maintain candidate vertices, $u$,
from which there may exist \ncycles, if $relax(u)=true$.
%
%
Initially, all vertices are enqueued into ${\cal Q}$.
%
%
In each iteration, when invoking \DFSSPFA($G,v$), let $V'$ be the set
of vertices relaxed. Among $V'$, for any vertex $w \in S_V \setminus \{v\}$,
$dst(w)$ has been updated and it has only relaxed partial
out-neighbors when finding the negative cycle. On the other hand, for any
vertex $w \in V' \setminus S_V$, all of the out-neighbors of $w$ have
been relaxed and cannot be relaxed before updating $dst(w)$. We
exclude $w \in V' \setminus S_V$ from ${\cal Q}$ implicitly by setting
$relax(w)=false$ in \DFSSPFA($G,w$).

\begin{example}
Suppose we have a greedy Eulerian subgraph $\appeulerian(G)$
(Fig.~\ref{fig:fig2}) of $G$ (Fig.~\ref{fig:fig1}) by \greedyD, and
will refine it to the optimal $\eulerian(G)$ using \refine.
Initially, all edges (solid lines) in $\appeulerian(G)$ are reversed
with initial +1 edge weight, and all remaining edges in
$\widetilde{G}_D$ are initialized with -1 edge weight.
$dst(v_1)=-2, dst(v_3)=-1, dst(v_7)=-5, dst(v_{11})=-1,
dst(v_{12})=-2, dst(v_{13})=-3, dst(v_{14})=-4$, and other vertices
$u$ have $dst(u)=0$. In the while loop, \DFSSPFA($G,v_1$) relaxes
$dst(v_5)=-1$ and returns false. This makes
$relax(v_1)=relax(v_5)=false$ by which $v_1$ and $v_5$ are dequeued
from ${\cal Q}$. Afterwards, none of $v_2, v_3, v_4, v_6$ can
relax any out-neighbors, and all are dequeued from ${\cal
  Q}$. \DFSSPFA($G,v_7$) relaxes all vertices, finds a negative cycle
$(v_7, v_9, v_4, v_2, v_3,$ $v_{10}, v_8, v_{11}, v_{12}, v_{13},
v_{14}, v_7)$, and adds $v_2, v_3, v_4$ into ${\cal Q}$ as new
candidates. Then, no vertices from $v_8$ to $v_{14}$ can relax any
out-neighbors until \DFSSPFA($G,v_2$) finds the last negative cycle
$(v_2, v_4,$ $v_3, v_2)$.
For most cases, \DFSSPFA($G,u$) relaxes a few of $u$'s out-neighbors.
%
\end{example}

We discuss the time complexity of \refine. The initialization
(Lines~1--9) is $O(n+m)$. Since $\appeulerian(G)$ approximates
$\eulerian(G)$, the number of \ncycles found by \refine will
be no more than $|E(\eulerian(G))|-|E(\appeulerian(G))|$,
%
%
and vertices $u$ will have $dst(u)$ updated less than
$|E(\eulerian(G))|-|E(\appeulerian(G))|$ times. This implies the while
loop costs $O(|E(\eulerian(G))|-|E(\appeulerian(G))|\cdot m)$.  Time
complexity of \refine is $O(cm^2)$, where $c \ll 1$, as confirmed in
our testing.

\subsection{The Bound between Greedy and Optimal}
\label{sec:bound}

We discuss the bound between $\appeulerian(G)$ obtained by \greedy
and the maximum Eulerian subgraph $\eulerian(G)$. To simplify our
discussion, below, a graph $G$ is a graph with multiple edges
between two vertices but without self loops, and every edge $(v_i,
v_j)$ is associated with a weight $w(v_i, v_j)$, which is
initialized to be -1.  Given a graph $G$, we use $\overline{G}$ to
represent the reversed graph of $G$ such that $V(\overline{G}) =
V(G)$ and $E(\overline{G})$ contains every edge $(v_j, v_i)$ if
$(v_i, v_j) \in E(G)$, and $w(v_j, v_i)=-w(v_i, v_j)$. In addition,
we use two operations, $\oplus$ and $\ominus$, for two graphs $G_i$
and $G_j$. Here, $G_{ij} = G_i \oplus G_j$ is an operation that
constructs a new graph $G_{ij}$ by union of two graphs, $G_i$ and
$G_j$, such that $V(G_{ij}) = V(G_i) \cup V(G_j)$, and $E(G_{ij}) =
E(G_i) \cup E(G_j)$.  And $G' = G_i \ominus G_j$ is an operation
that constructs a new graph $G'$ by removing a subgraph $G_j$ from
$G_i$ ($G_j \subseteq G_i$) such that $V(G') = V(G_i)$ and $E(G') =
E(G_i) \setminus E(G_j)$. Given two Eulerian subgraphs, $G_i$ and
$G_j$, it is easy to show that $G_i \oplus G_j$ and $G_i \ominus
G_j$ are still Eulerian graphs. Given any graph $G$, $G \oplus
\overline{G}$ is an Eulerian graph. Note that assume that there is a
cycle with two edges, $(v_i, v_j)$ and $(v_j, v_i)$, between two
vertices, $v_i$ and $v_j$, in $G$. there will be four edges in $G
\oplus \overline{G}$, i.e., two edges are from $G$ and two
corresponding reversed edges from $\overline{G}$.

\comment{

\begin{lemma}
An Eulerian graph $G$ can be divided into several edge disjoint simple
cycles.
\end{lemma}

\proofsketch It can be proved if there is a process that we can
repeatedly remove edges from a cycle found in an Eulerian graph $G$, and $G$ does not have
any edges after the last cycle being removed. Note that in $G$, $d_I(u) =
d_O(u)$ for every $u$ in $G$. Let a subgraph of $G$, denoted as $G_c$,
be such a cycle found in $G$. $G_c$ is an Eulerian subgraph, and $G
\ominus G_c$ is also an Eulerian subgraph. The lemma is established. \eop

}

\comment{
In this part, we will introduce a structure named \kcycle that can
regard refinement as finding and reversing several edge-disjoint
\kcycles. Thus, each edge can be reversed at most once during
refinement. In addition, with some extra constraints, \kcycles can be
utilized to get a theoretical upper bound of $|Opt(E_1)|-|Init(E_1)|$,
indicating the high quality of $Init(E_1)$ providing some empirical
results.
}

\begin{figure}[t]
\begin{center}
\begin{tabular}[t]{c}
    \hspace*{-0.5cm}
    \subfigure[$\appeulerian(G)$]{
    \includegraphics[width=0.5\columnwidth,height=2.5cm]{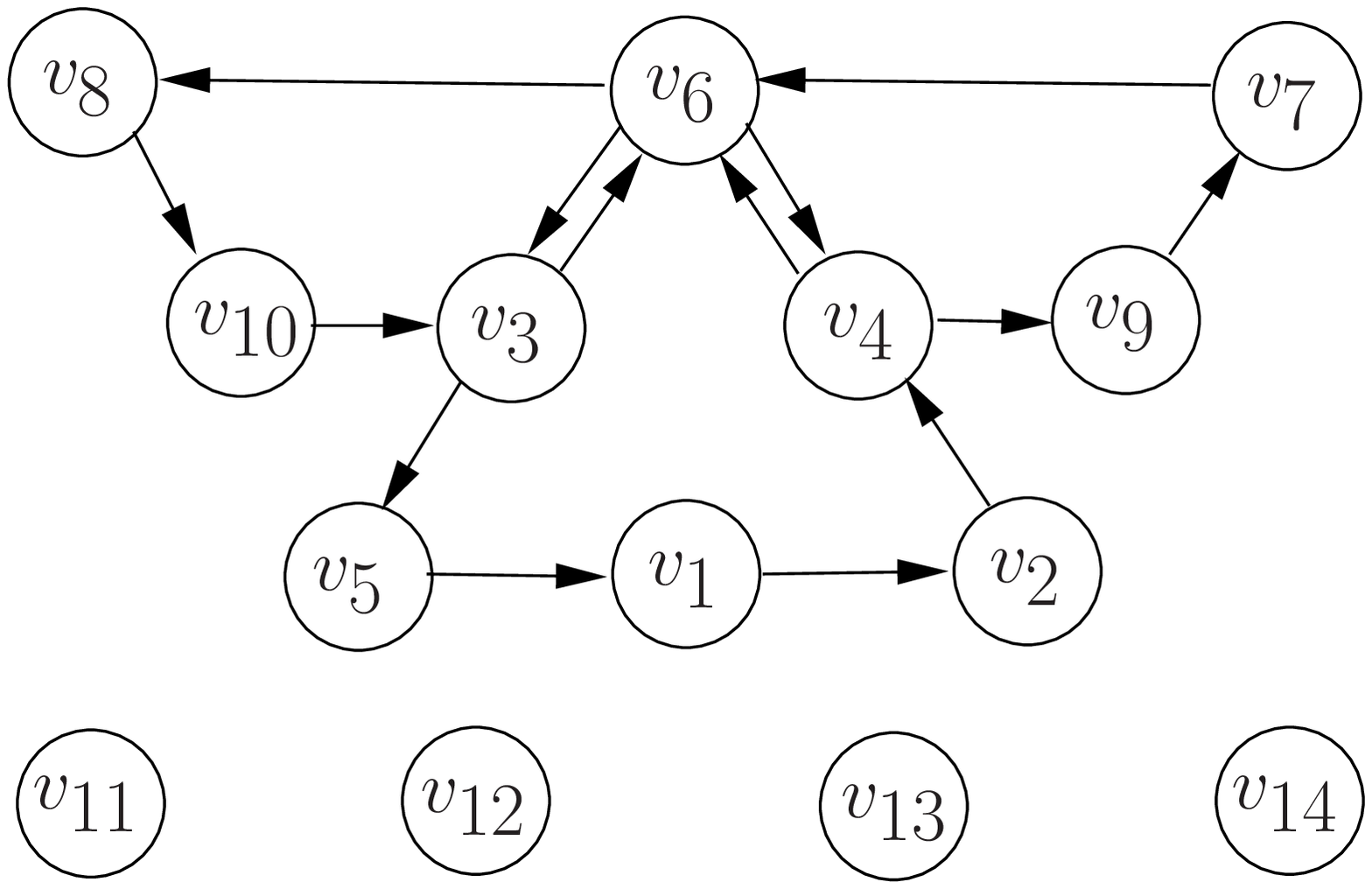}
    \label{fig:fig9_1}
    }
    \hspace*{-0.5cm}
    \subfigure[$G_P = G \ominus \appeulerian(G)$]{
    \includegraphics[width=0.5\columnwidth,height=2.5cm]{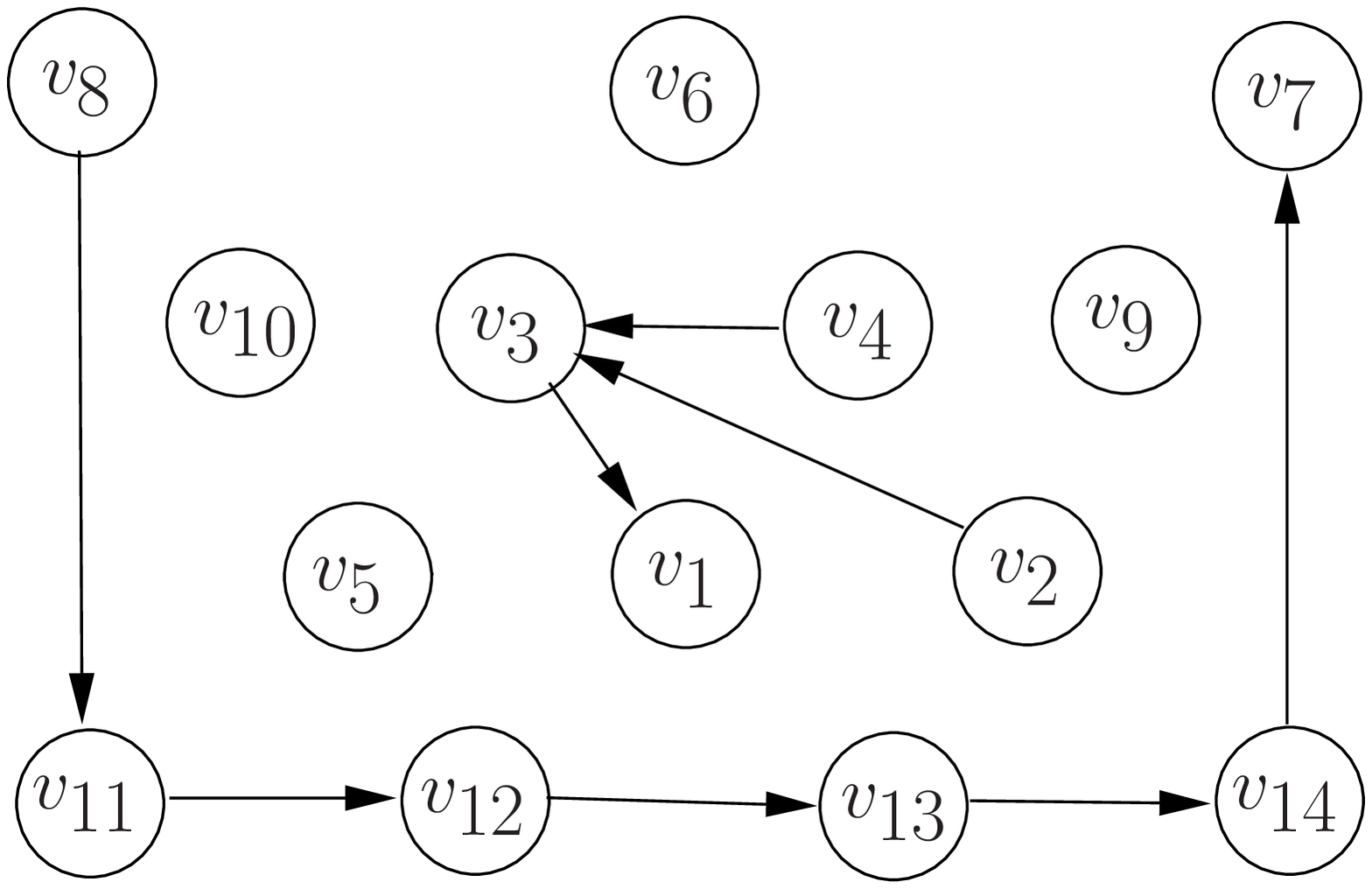}
    \label{fig:fig9_2}
    }
    \\ \vspace*{-0.1cm}
    \hspace*{-0.5cm}
    \subfigure[$\eulerian(G)$]{
    \includegraphics[width=0.5\columnwidth,height=2.5cm]{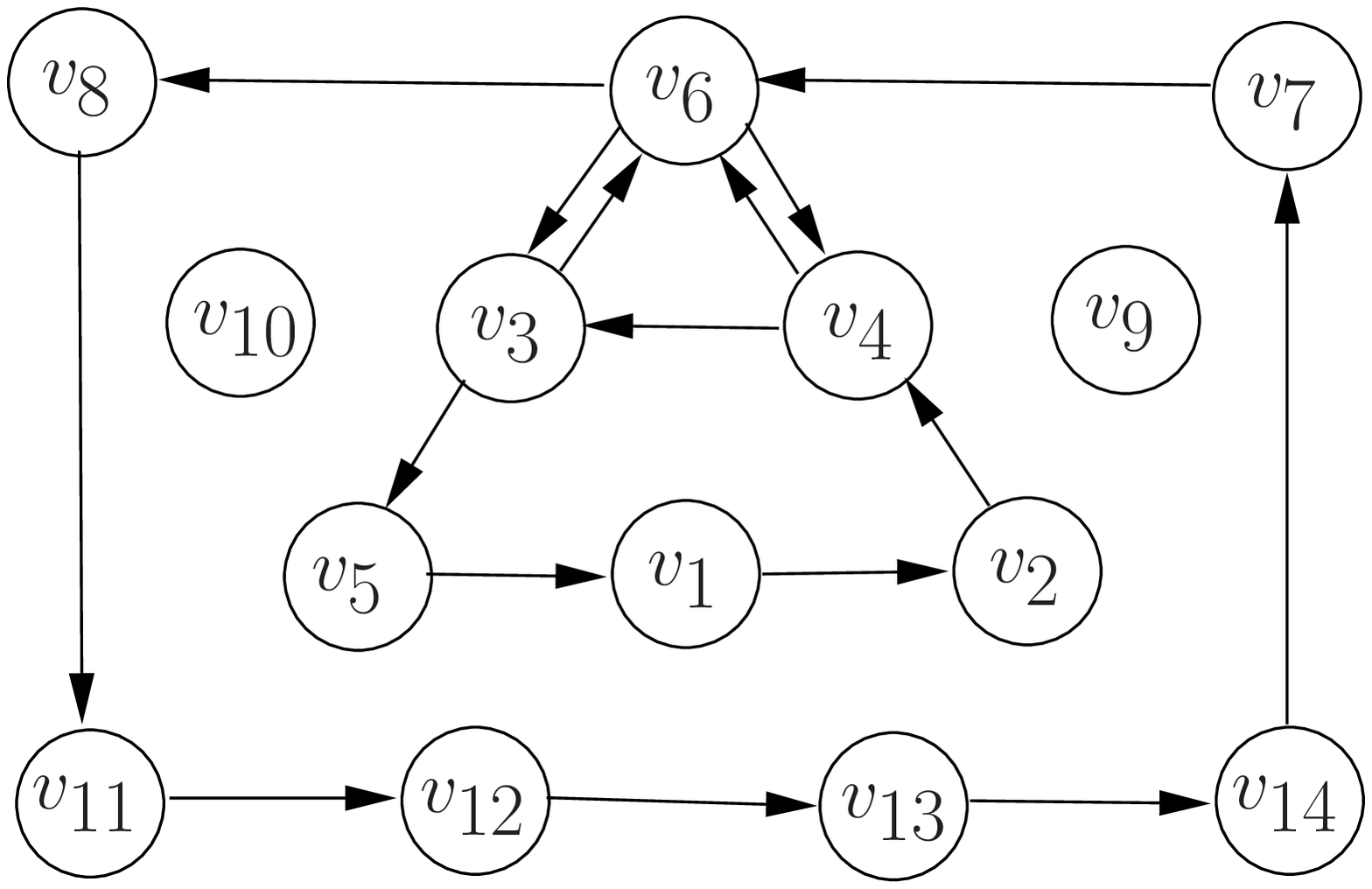}
    \label{fig:fig9_3}
    }
    \hspace*{-0.5cm}
    \subfigure[$G_N = G \ominus \eulerian(G)$]{
    \includegraphics[width=0.5\columnwidth,height=2.5cm]{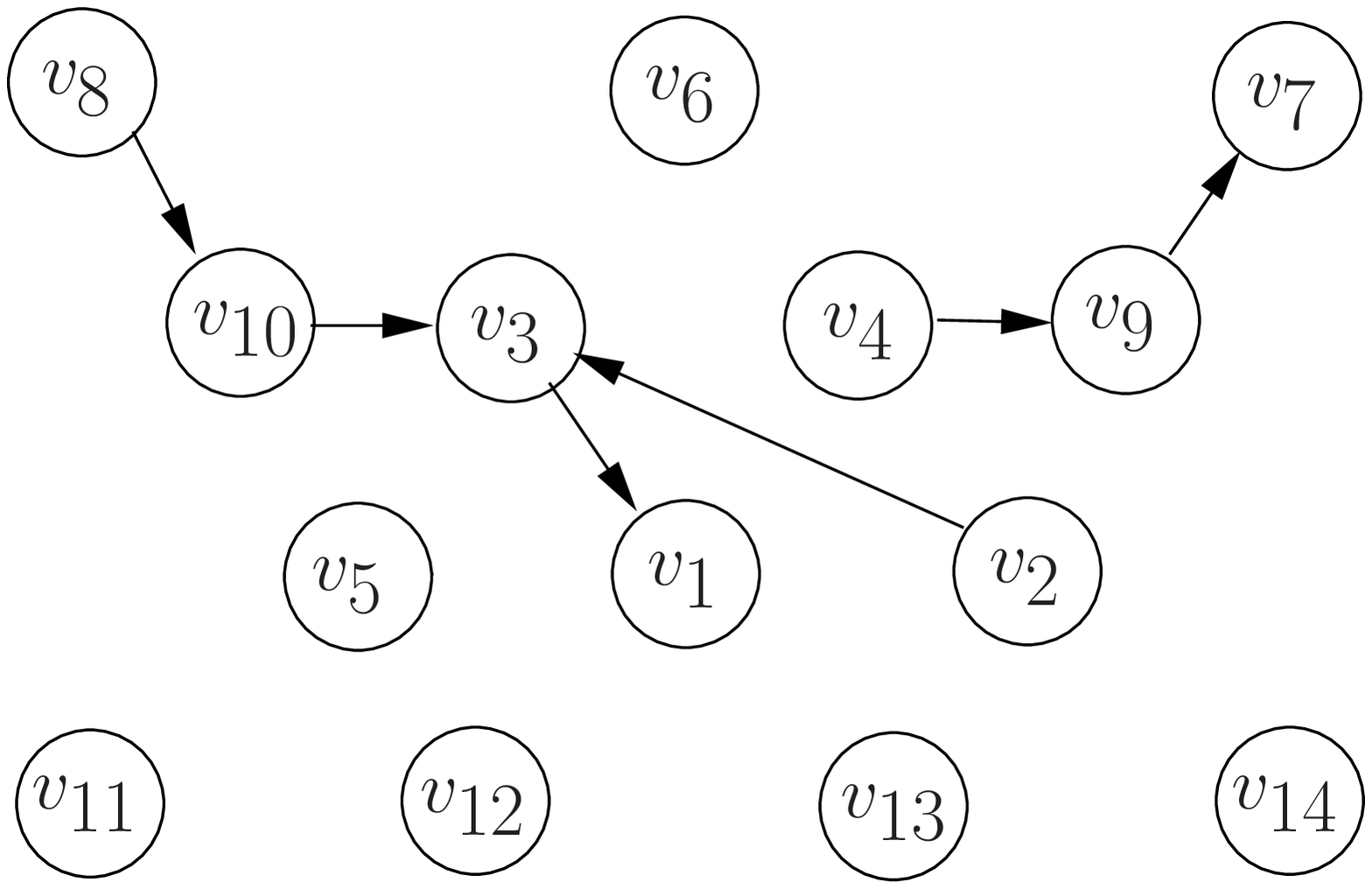}
    \label{fig:fig9_4}
    }
\end{tabular}
\vspace*{-0.2cm}
\caption{$\appeulerian(G)$ and $\eulerian(G)$ of $G$ in Fig.~\ref{fig:fig1}}
\vspace*{-0.2cm}
\end{center}
\label{fig:fig9}
\end{figure}

\begin{figure}[t]
\centering
\includegraphics[width=2in,height=2.5cm]{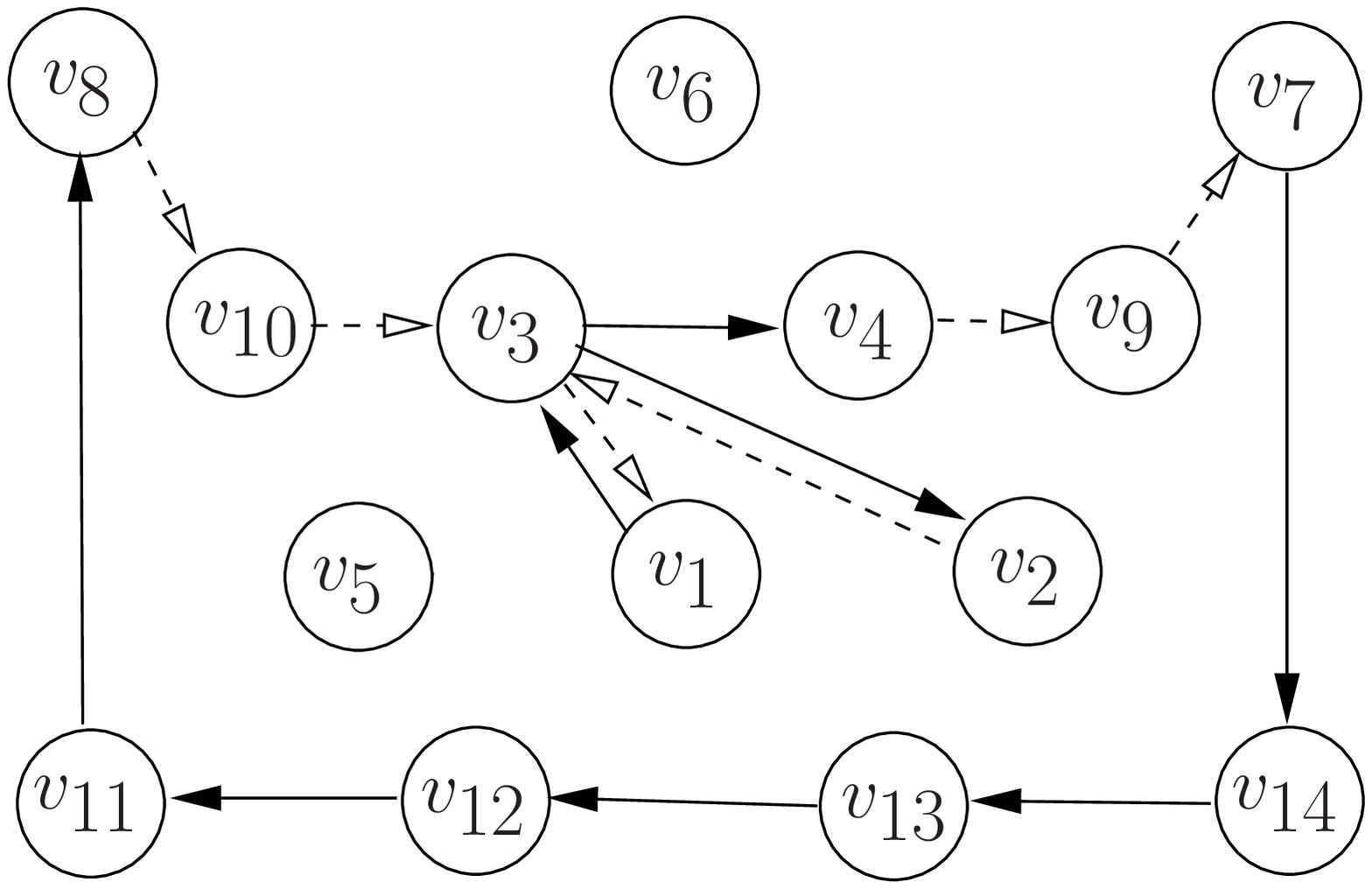}
\vspace*{-0.2cm}
\caption{${\cal G} =\overline{G_P} \oplus G_N$ where
$G_P = G \ominus \appeulerian(G)$ and $G_N = G
  \ominus \eulerian(G)$.}
\vspace*{-0.4cm}
\label{fig:fig10}
\end{figure}


We discuss the bound using an Eulerian graph ${\cal G} =\overline{
G_P }\oplus G_N$, where $G_P = G \ominus \appeulerian(G)$ and $G_N =
G \ominus \eulerian(G)$.  We call every edge in $G_N$ a negative
edge (\nedge), and a path in $G_N$ a negative path (\npath). We also
call every edge in $\overline{G_P}$ a positive edge (\pedge), and a
path in $\overline{G_P}$ a positive path (\ppath).  It is important
to note that \pedges are given for $\overline{G_P}$ but not for
$G_P$, all \nedges are with a weight of -1, while all \pedges are
with a weight of +1, because they are the reversed edges in $G_P$.
Here, ${\cal G}$ is a graph with multiple edges between a pair of
vertices.

\begin{example}
Consider the example graph $G$ in Fig.~\ref{fig:fig1}. The Eulerian subgraph
obtained by \greedy, i.e. $\appeulerian(G)$ is shown in
Fig.~\ref{fig:fig9_1}.  It is worth noting that we make use of the
resulting graph of \greedyD, since that obtained by \greedyR is actually
the maximum Eulerian subgraph in this case. Fig.~\ref{fig:fig9_3} shows
the maximum Eulerian subgraph $\eulerian(G)$.  As observed, some
edges in $\appeulerian(G)$ do not appear in
$\eulerian(G)$, while some edges that do not appear in $\appeulerian(G)$
appear in $\eulerian(G)$.
Fig.~\ref{fig:fig9_2} and Fig.~\ref{fig:fig9_4} show $G_P = G
\ominus \appeulerian(G)$ and $G_N = G \ominus \eulerian(G)$,
respectively.  Fig.~\ref{fig:fig10} shows ${\cal G}
=\overline{G_P}\oplus G_N$. In Fig.~\ref{fig:fig10}, the solid edges
represent the \pedges from $\overline{G_P}$, and the dashed edges
represent the \nedges from $G_N$.
\end{example}

Since ${\cal G}$ is Eulerian, it can be divided into several edge disjoint simple cycles as given by Lemma~\ref{lem:1}. Among these cycles,
there are no cycles in ${\cal G}$ with only
\nedges, because they must be in $\eulerian(G)$ if exist. And there
are no cycles in ${\cal G}$ with only \pedges, because all such
cycles have been moved into $\appeulerian(G_i)$ in \refineAlg
(Algorithm~\ref{alg:gr}, Line~4).

Next, let a cycle be a \pcycle if the total weight of the edges in this
cycle $>0$, and let it be a \ncycle if its total weight of edges $<0$.
We show there are no \ncycles in ${\cal G}$.

\begin{lemma}
There does not exist a \ncycle in ${\cal G}$.
\label{lemma:pcycle}
\end{lemma}

\proofsketch Assume there is a \ncycle in ${\cal G}$, denoted
as $G_{cyc}$. Since there are no cycle with only \pedges or \nedges,
 there are \pedges and \nedges in $G_{cyc}$. We divide
$G_{cyc}$ into two subgraphs, $G_p$ and $G_n$. Here $G_p$ consists
of all \pedges, where each \pedge is with a +1 weight, and $G_n$
consists of all \nedges, where each \nedge is with a -1 weight.
Clearly, $|E(G_p)| < |E(G_n)|$, since it assumes that $G_{cyc}$ is a
\ncycle.  Note that $\eulerian(G) \ominus G_p \oplus G_n$,
which is equivalent to $\eulerian(G) \oplus G_{cyc}
\ominus (G_p \oplus G_p)$, is Eulerian, and it contains more edges than
$\eulerian(G)$, resulting in a contradiction. Therefore, there does
not exist a \ncycle in ${\cal G}$. \eop

Lemma~\ref{lemma:pcycle} shows all cycles in ${\cal G}$ are
non-negative.  Since there are no cycles with only \pedges or
\nedges, each cycle in ${\cal G}$ can be partitioned into an
alternating sequence of $k$ \ppaths and $k$ \npaths, and represented
as $(v_1^+,v_1^-,v_2^+,\ldots,v_k^+,v_k^-,v_1^+)$, where
$(v_i^+,v_i^-)$, for $i=1,2,\dots,k$, are \npaths, and
$(v_i^-,v_{i+1}^+)$, for $i=2,$ $\dots,$ $k-1$, $k$, plus
$(v_k^-,v_1^+)$ are \ppaths.  We call such cycle a \kcycle.
Fig.~\ref{fig:cycle} shows an example of \kcycle, and an arrow
presents a path. \ppaths are in solid lines while \npaths are in
dashed lines.

\comment{
\begin{figure}[t]
\begin{center}
\includegraphics[scale=0.3]{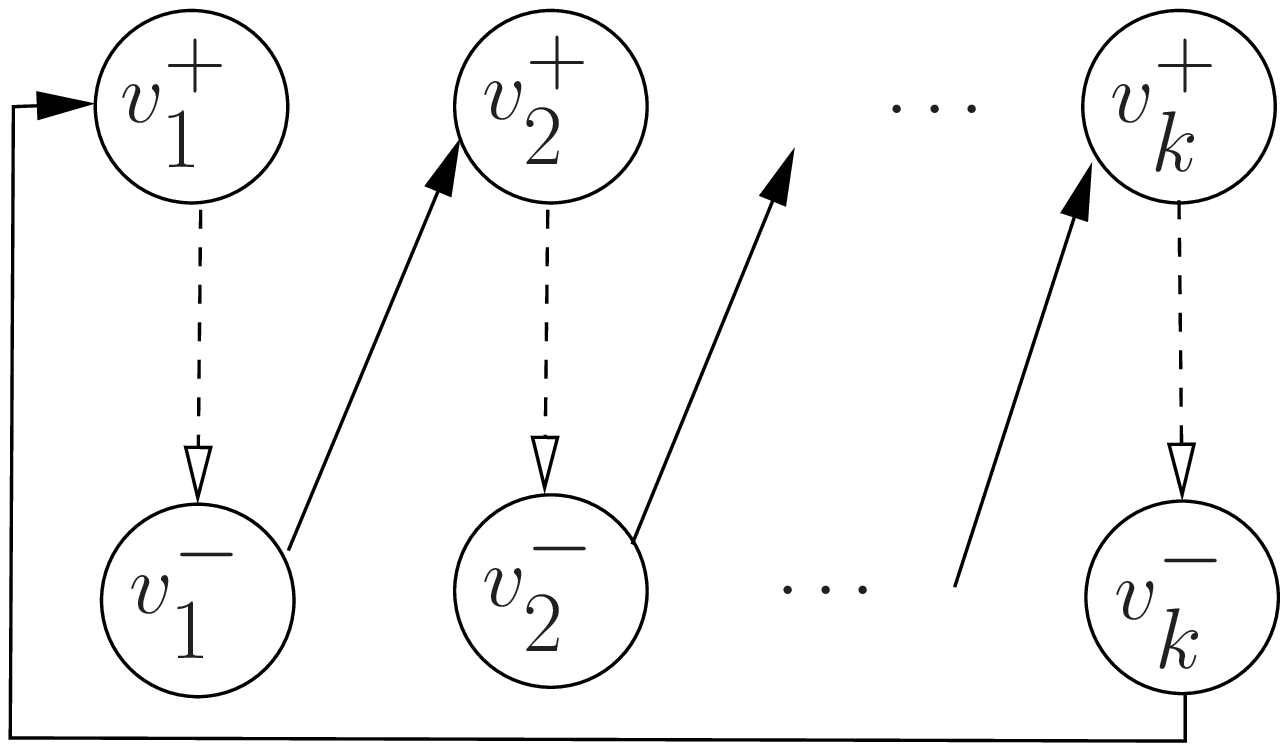}
\end{center}
\vspace*{-0.4cm}
\caption{\kcycle}
\vspace*{-0.2cm}
\label{fig:cycle}
\end{figure}
}

\begin{figure}[t]
\begin{center}
\begin{tabular}[t]{c}
    \subfigure[\kcycle]{
         \includegraphics[width=0.36\columnwidth,height=2.2cm]{fig/cycle}
         \label{fig:cycle}
    }
    \subfigure[2-cycle]{
         \includegraphics[width=0.3\columnwidth,height=2.2cm]{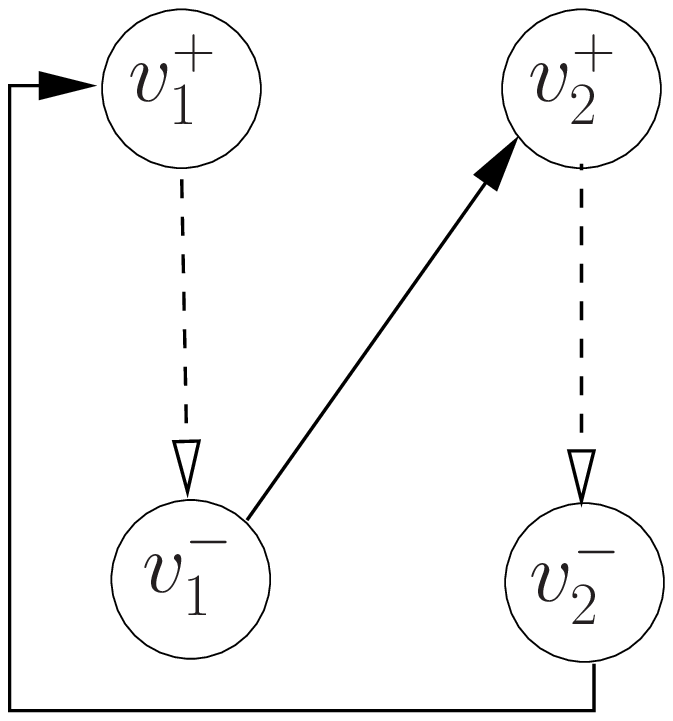}
         \label{fig:2-cycle}
    }
    \subfigure[3-cycle]{
    \includegraphics[width=0.33\columnwidth,height=2.2cm]{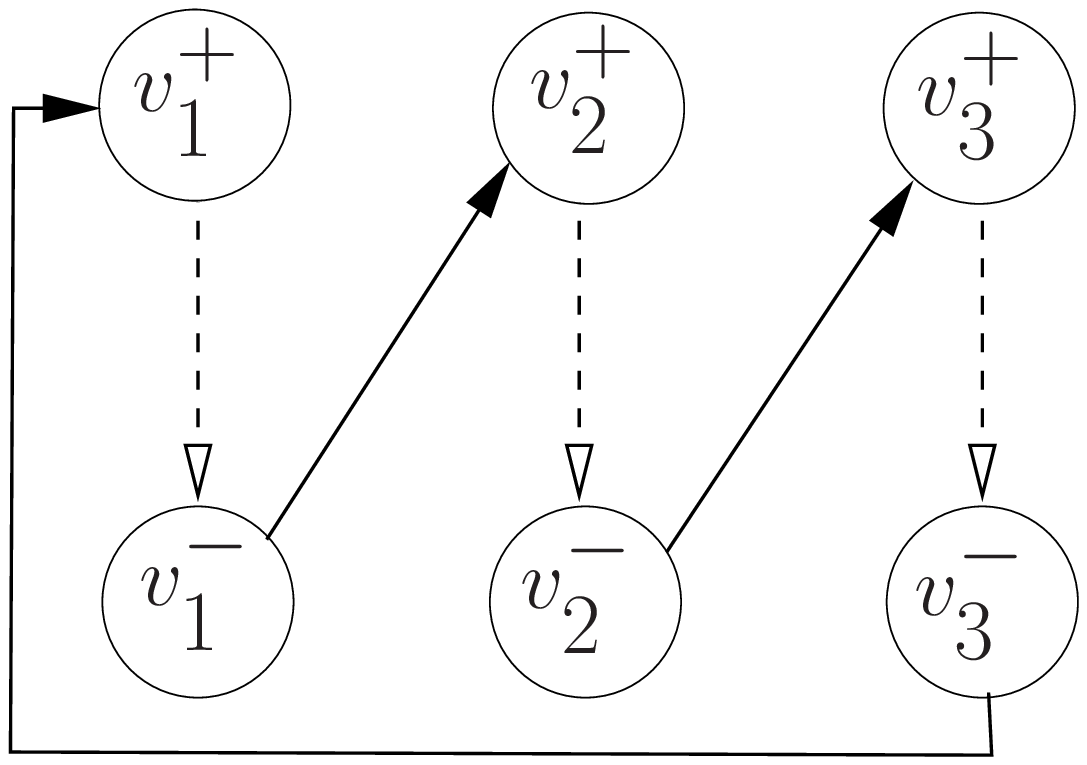}
    \label{fig:3-cylce}
    }
\end{tabular}
\end{center}
\vspace*{-0.4cm}
\caption{\kcycle}
\vspace*{-0.4cm}
\label{fig:kcyle}
\end{figure}

The difference $|E(\eulerian(G))| - |E(\appeulerian(G))|$ is equal
to $|E(G \ominus \appeulerian(G))| - |E(G \ominus \eulerian(G))| =
|E(G_P)| - |E(G_N)| = |E(\overline{G_P})| - |E(G_N)|$,
becomes the total number of edges in $G_P$ minus the total
number of edges in $G_N$. On the other hand, the difference
$|E(\eulerian(G))| - |E(\appeulerian(G))|$ can be considered as the
total weight of all \kcycles in ${\cal G}$. Recall that all edges in
$G$ are with weight -1 and the edges in $\overline{G}$ are with
weight +1 by our definition. Assume that ${\cal
G}=\{C_1,C_2,\cdots\}$, where $C_i$ is a \kcycle. The total weight
of ${\cal G}$ regarding all \kcycles is $w({\cal G})=\sum_i w(C_i)$.
Below, we bound $|E(\eulerian(G))| - |E(\appeulerian(G))|$ using
\kcycles.


%
Consider ${\cal G}$ in Fig.~\ref{fig:fig10}, there are 3 \kcycles.
$C_1 = (v_3,v_1,v_3)$ and $C_2 = (v_3,v_2,v_3)$ with weight 0, and
$C_3 = (v_8,v_{11},v_{12},v_{13},$ $v_{14},$ $v_7,$ $v_9,$ $v_4,$
$v_3,v_{10},v_8)$ with weight 2.  This means that it needs at most 2 more
iterations to get the maximum Eulerian subgraph from the greedy solution.

For a \kcycle $(v_1^+,v_1^-,v_2^+,\ldots,v_k^+,v_k^-,v_1^+)$, we use
$\triangle_k$ and $\triangle_k'$ to represent the total weight of
\nedges \footnote{For \nedges, we take the absolute value of total weight.} and \pedges, i.e. $\triangle_k=\sum_{i=1,\ldots,k}w(v_i^+,v_i^-)$ and
$\triangle_k'=\sum_{i=1,\ldots,k-1}$ $w(v_i^-,v_{i+1}^+)$ $+$
$w(v_k^-,v_1^+)$. Because $\triangle_k$ is determined by the optimal
in $|E(G_N)|$ $= |E(G \ominus \eulerian(G))|$, the bound is obtained when
getting the maximum of $\triangle_k'$.

\begin{figure}[t]
\begin{center}
\begin{tabular}[t]{c}
    \subfigure[Step-1]{
         \includegraphics[width=0.33\columnwidth,height=2cm]{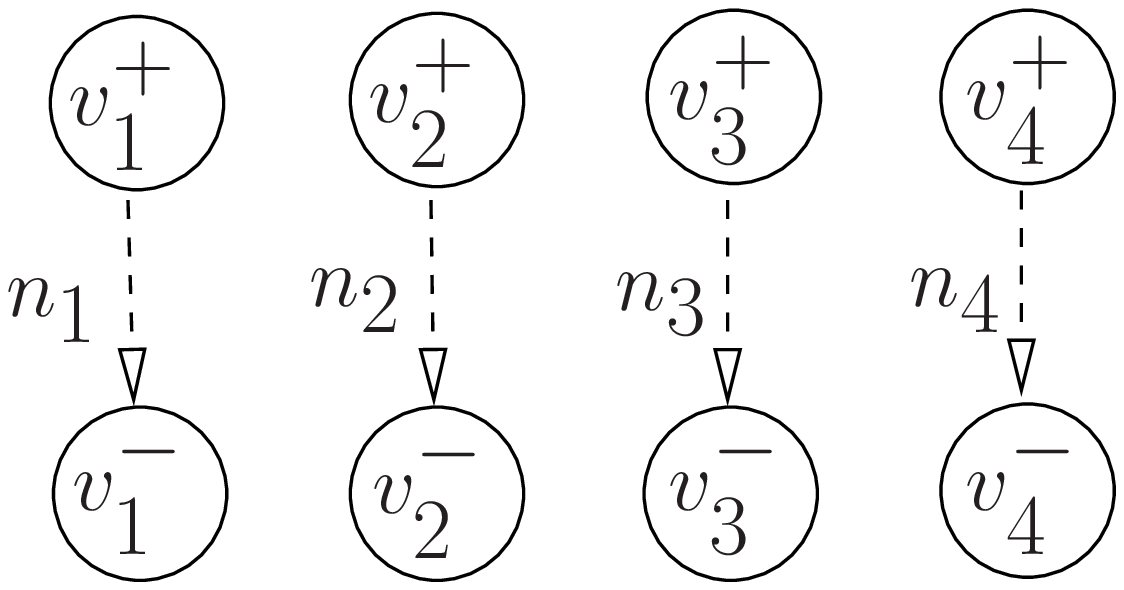}
         \label{fig:gene1}
    }
    \subfigure[Step-2]{
    \includegraphics[width=0.33\columnwidth,height=2cm]{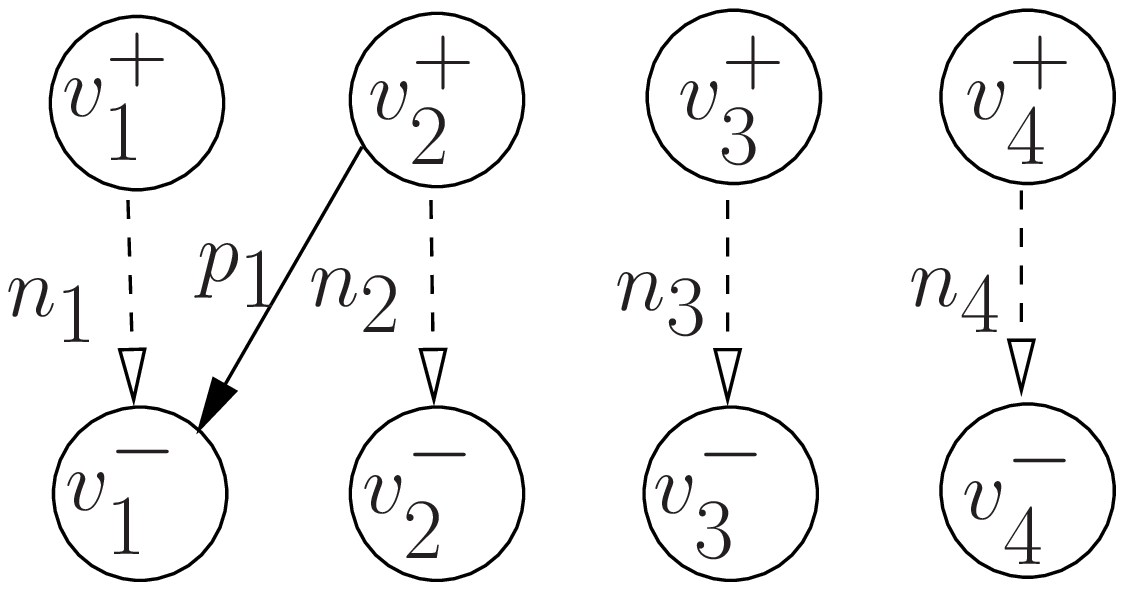}
    \label{fig:gene2}
    }
    \subfigure[Step-3]{
    \includegraphics[width=0.33\columnwidth,height=2.5cm]{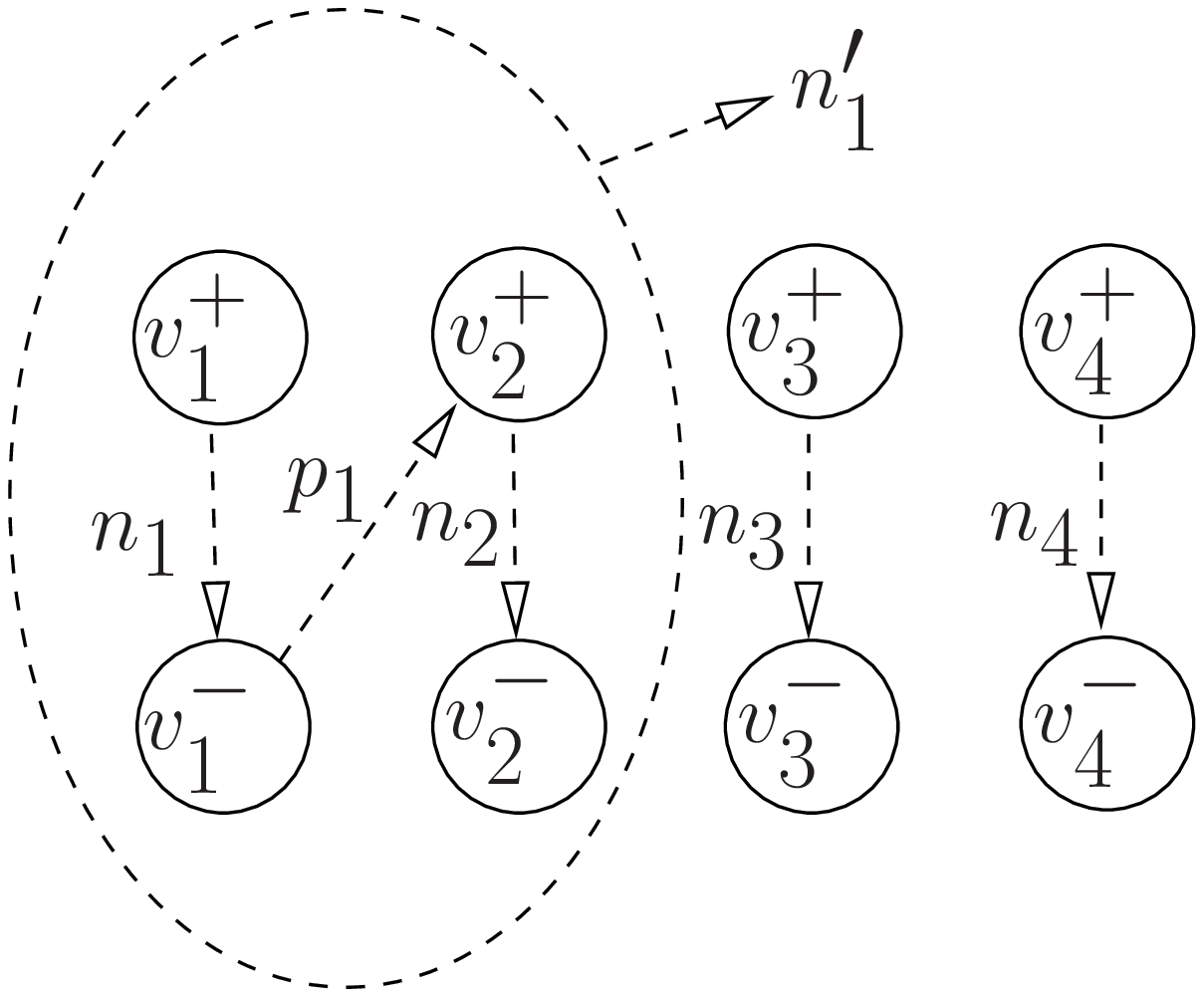}
    \label{fig:gene3}
    }
\end{tabular}
\end{center}
\vspace*{-0.4cm}
\caption{\kcycle generated by \greedyR}
\vspace*{-0.4cm}
\label{fig:generation}
\end{figure}

\begin{theorem}
The upper bound of the total weight of \pedges in a \kcycle with
specific $k$ is $k$ times that of \nedges, i.e.,
$\triangle_k' \leq k \cdot \triangle_k$
\label{them:upper}
\end{theorem}

\proofsketch The proof is based on the way \kcycles constructed by \greedyR.  For
simplicity, we first assume that each \npath\xspace and \ppath\xspace
is a \pnpath\xspace itself, and we will deal with general cases
later. Based on \greedyR, a \kcycle is constructed as shown in
Fig.~\ref{fig:generation}, which is a 4-cycle. Initially, there are
$4$ \npaths, $n_i=(v_i^+,v_i^-)$, $i=1,2,\dots,4$, as
Fig.~\ref{fig:gene1} shows.  \greedyR deals with \pnpaths of length
$l$ from a small $l$ to a large $l$. First, \greedyR finds a path
$p_1$ = \pnpath$(v_2^+,v_1^-)$, and combines $p_1$ with two separated
\npaths, $n_1$ and $n_2$ into a new \npath $n_1'$. Here, $\len(p_1)$
is no larger than any $\len(n_i)$.  \greedyR will repeat this process
to add all \ppaths, $p_i$ into \kcycle in an ascending order of their
lengths. The last \ppath $(v_k^-,v_1^+)$ to be added to \kcycle should
be the longest one among all \ppaths. Then its upper bound is
$\sum_{i=1,\ldots,k}w(v_i^+,v_i^-)+
\sum_{i=1,\ldots,k-1}w(v_i^-,v_{i+1}^+)$. Otherwise, its upper bound
should be $max\{w(v_i^-,v_{i+1}^+)\}$. Below, we prove
Theorem.~\ref{them:upper}.

\comment{
Each time we find \pnpath$(v_i^+,v_{i-1}^-)$, and
combine two separate \ppaths $(v_{i'-},v_{i+})$ and
$(v_{(i+1)-},v_{(i+1)''-})$ into a new \ppath\xspace
$(v_{i'-},v_{(i+1)''-})$. Since each \npath\xspace is a \pnpath\xspace
itself, its length is no larger than any \ppath\xspace before
combining. Furthermore, it is not difficult to see that \npaths are
added to the \ncycle in an ascending order of its maximum possible
length in order to get a upper bound of
$\triangle_k'-\triangle_k$. Consider the last
\npath\xspace$(v_{k+},v_{1-})$ added to \ncycle, if it is the longest
one among all \npaths, then its upper bound is
$\sum_{i=1,\ldots,k}w(v_{i-},v_{i+})+\sum_{i=1,\ldots,k-1}w(v_{i+},v_{(i+1)-})$,
otherwise its upper bound is
$max\{w(v_{i+},v_{(i+1)-}),w(v_{k+},v_{1-})\}$.  Below, we give a
theorem indicating the relationship between $\triangle_k'$ and
$\triangle_k$.
}
\begin{itemize}
\itemsep-2mm\parsep-2mm
\item For 2-cycle (Fig.~\ref{fig:2-cycle}): Since
%
%
$w(v_1^-,v_2^+) \leq w(v_1^+,v_1^-)$, $w(v_1^-,v_2^+) \leq
  w(v_2^+,v_2^-)$, and $w(v_2^-,v_1^+) \leq
  w(v_1^+,v_1^-)+w(v_1^-,v_2^+)+ w(v_2^+,v_2^-)$ we have,
\begin{eqnarray*}
\triangle_2'&=& w(v_1^-,v_2^+)+w(v_2^-,v_1^+)\\
&\leq& w(v_1^+,v_1^-)+2\cdot w(v_1^-,v_2^+)+ w(v_2^+,v_2^-) \\
&\leq& 2 \cdot \triangle_2
\end{eqnarray*}
\item For 3-cycle (Fig.~\ref{fig:3-cylce}): Since
\begin{eqnarray*}
w(v_1^-,v_2^+) &\leq& w(v_1^+,v_1^-), w(v_2^+,v_2^-), w(v_3^+,v_3^-) \\
w(v_2^-,v_3^+) &\leq& w(v_1^+,v_1^-)+w(v_1^-,v_2^+)+w(v_2^+,v_2^-)\\
w(v_2^-,v_3^+) &\leq&   w(v_3^+,v_3^-) \\
w(v_3^-,v_1^+) &\leq&
  w(v_1^+,v_1^-)+w(v_1^-,v_2^+)+w(v_2^+,v_2^-) + \\
     && w(v_2^-,v_3^+)+w(v_3^+,v_3^-)
\end{eqnarray*}
we have,
\begin{eqnarray*}
\triangle_3' &=& w(v_1^-,v_2^+)+w(v_2^-,v_3^+)+w(v_3^-,v_1^+)\\
&\leq& \triangle_3+ 2\cdot (w(v_1^-,v_2^+)+w(v_2^-,v_3^+))\\
&\leq& 2 \cdot \triangle_3+ 3 \cdot w(v_1^-,v_2^+)
\leq 3 \cdot \triangle_3
\end{eqnarray*}

\item Assume that it holds for \kcycles when $k<l$, we prove that it also holds
  when $k=l$. Suppose that the shortest \ppath is $(v_1^-,v_2^+)$,
  combine $(v_1^+,v_1^-)$, $(v_1^-,v_2^+)$ and $(v_2^+,v_2^-)$ into a
  single \ppath $(v_1^+,v_2^-)$, then we get a \kcycle as
  $k=l-1$. As a result, $\triangle_k' \leq (k-1) \cdot
  (\triangle_k+w(v_1^-,v_2^+))+w(v_1^-,v_2^+) \leq k \cdot
  \triangle_k$. \eop
\end{itemize}


Let $\triangle_{C_i}'$ and $\triangle_{C_i}$ denote the total weight of
\pedges and \nedges in a \kcycle $C_i$. Bounding $|E(\eulerian(G))|$-
$|E(\appeulerian(G))|$ can be formulated as an LP (linear programming)
problem.
\begin{displaymath}
\begin{aligned}
\max \quad &\sum_{C_i}(\triangle_{C_i}'-\triangle_{C_i}) \\
s.t.\quad  & \mbox{(Cond-1) } \triangle_{C_i}' > \triangle_{C_i},
     \quad \forall i, \\
&  \mbox{(Cond-2) } \triangle_{C_i}' \leq k_i \cdot \triangle_{C_i},
\mbox{for a \kcycle $C_i$ with $k$-value $k_i$,} \\
& \mbox{(Cond-3) } \sum_{C_i}(\triangle_{C_i}'+\triangle_{C_i}) \leq
|E({\cal G})| \leq |E|
\end{aligned}
\end{displaymath}

\begin{figure}[t]
\begin{center}
\begin{tabular}[t]{c}
    \subfigure[Theoretical Upper Bound]{
         \includegraphics[width=0.45\columnwidth,height=3cm]{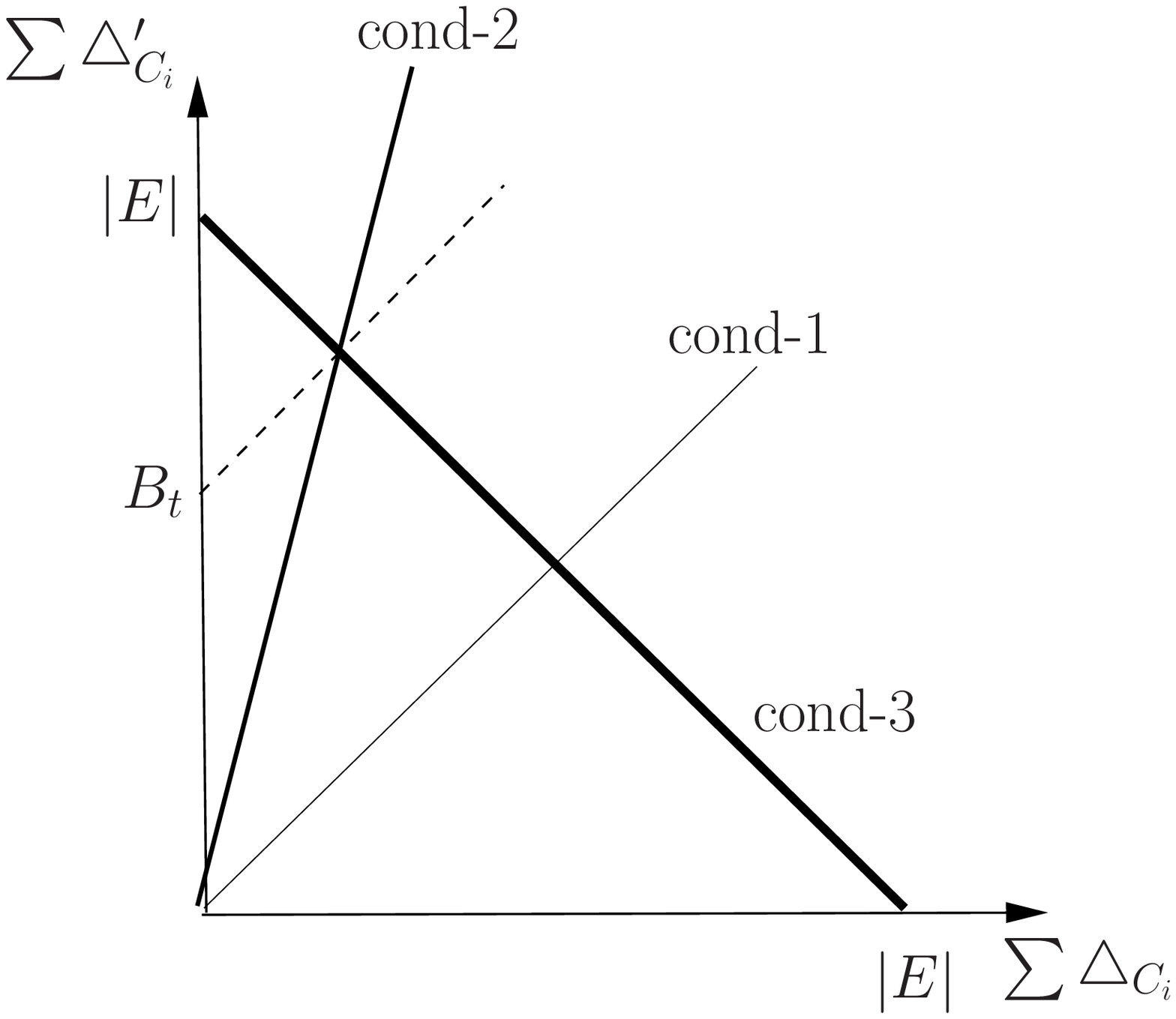}
         \label{fig:appro1}
    }
    \subfigure[Empirical Upper Bound]{
        \includegraphics[width=0.45\columnwidth,height=3cm]{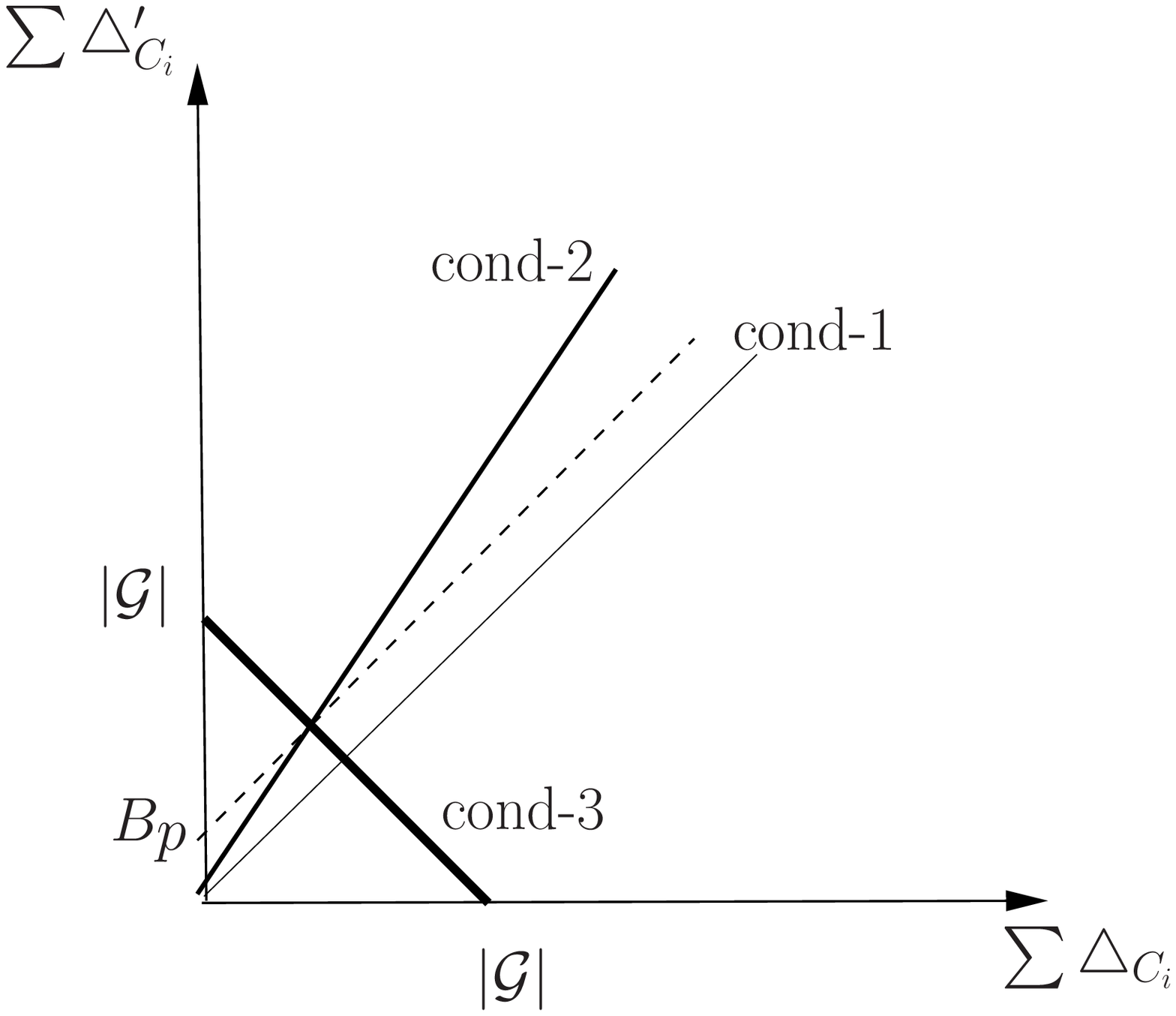}
        \label{fig:appro2}
    }
\end{tabular}
\end{center}
\vspace*{-0.4cm}
\caption{Upper Bounds}
\vspace*{-0.4cm}
\label{fig:upper_bounds}
\end{figure}

\comment{
\begin{figure}[t]
\centering
\includegraphics[width=2.5in,height=2in]{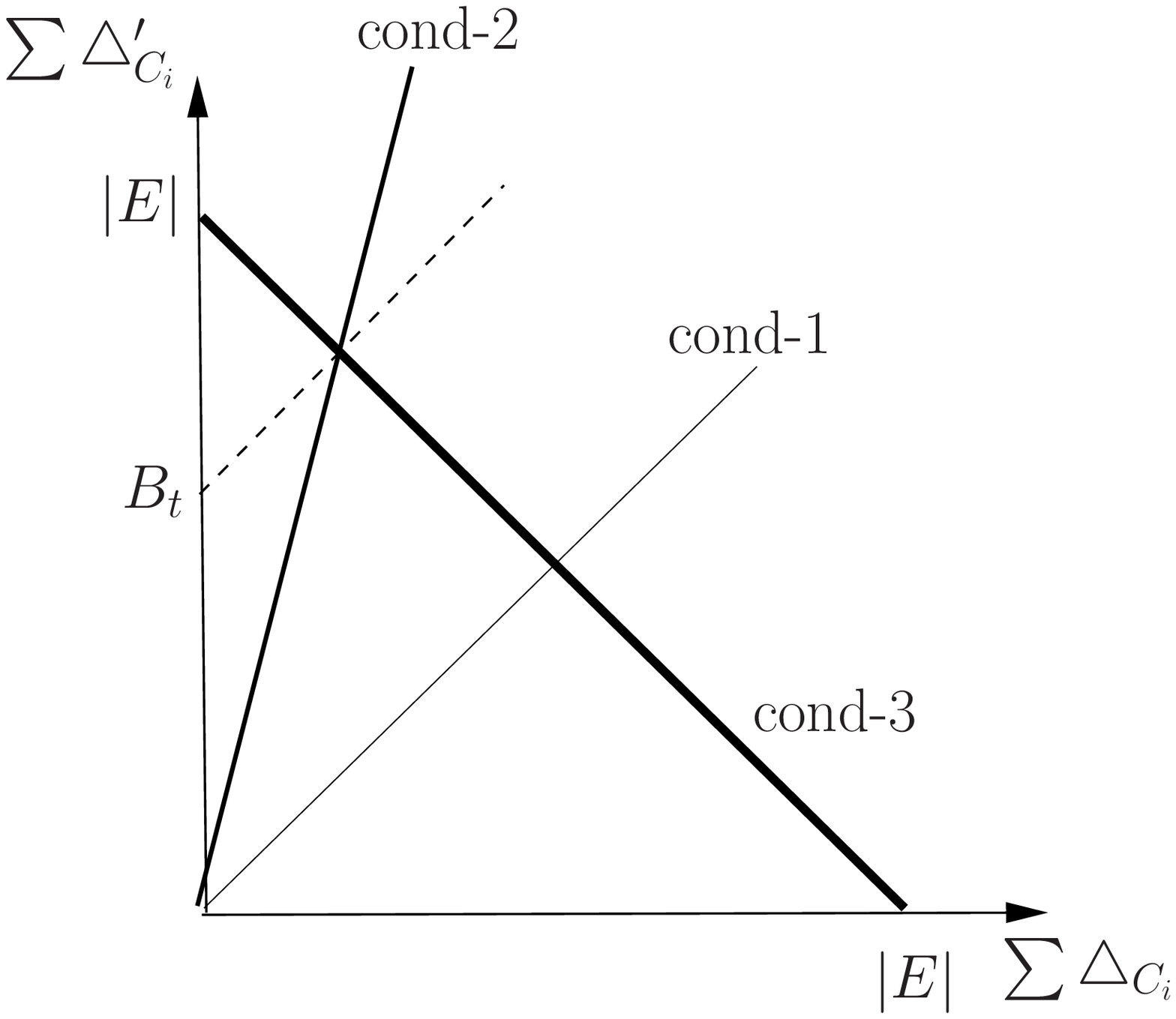}
\caption{Theoretical Upper Bound} \label{fig:appro1}
\end{figure}

\begin{figure}[t]
\centering
\includegraphics[width=2.5in,height=2in]{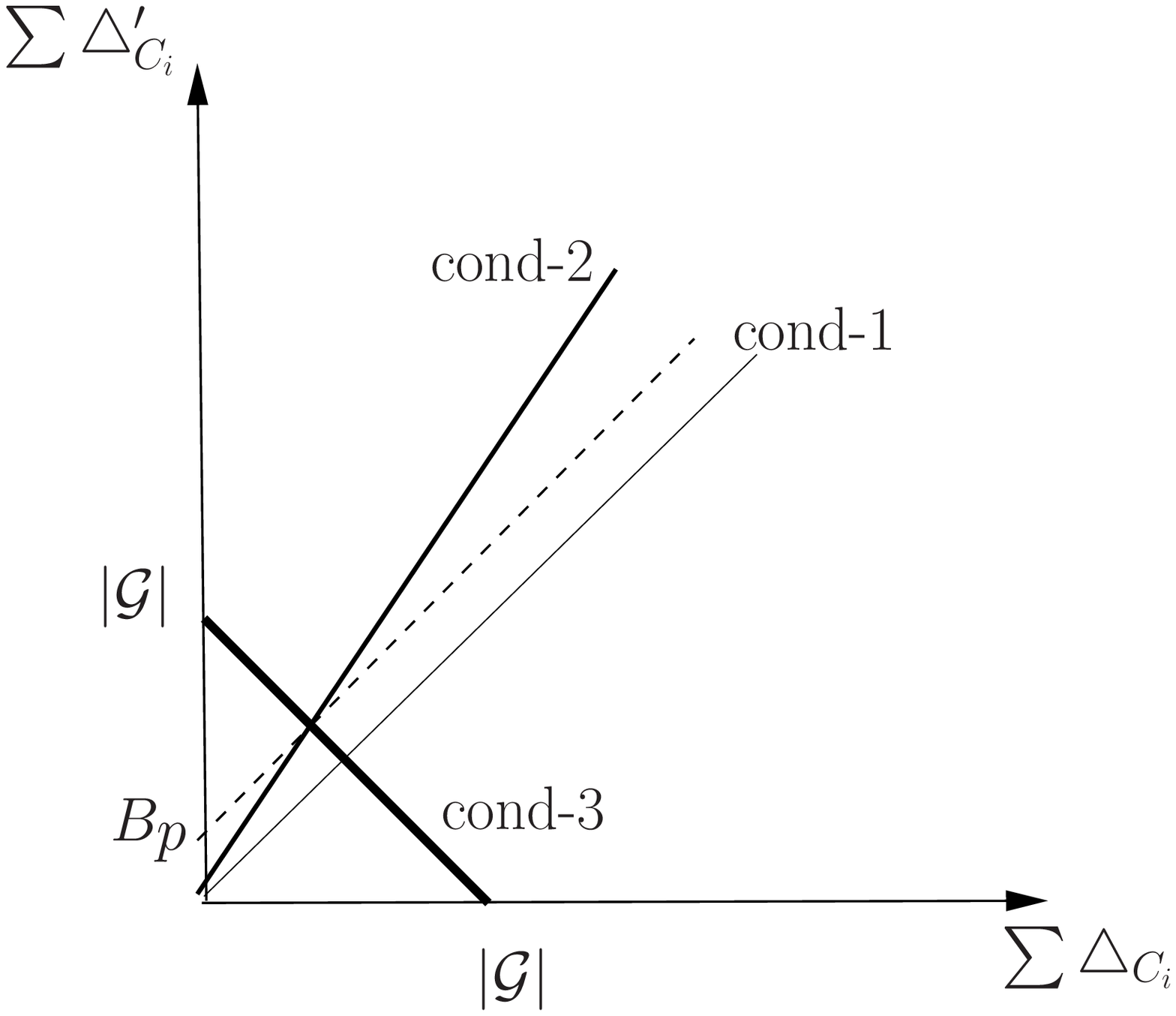}
\caption{Empirical Upper Bound} \label{fig:appro2}
\end{figure}
}

\comment{
In Fig.~\ref{fig:appro1}, $B_t$ at y-axis illustrates the theoretical
upper bound of $|E(\eulerian(G))|$-
$|E(\appeulerian(G))|=\frac{K-1}{K+1}|E|$ by solving the LP problem.
%
%
Here, $K$ is the maximum among all $k$ values.
This theoretical upper bound is far from tight. First, $|{\cal G}| \ll
|E|$. Second, for most \kcycles, $\triangle_k'=(1+\epsilon)\cdot
\triangle_k$, $0<\epsilon<1$, since most \ppaths in a \kcycle are far
from the upper bound it can get. A more tight empirical upper bound is
$B_p$ at y-axis in Fig.~\ref{fig:appro2}. We will discuss it the
experiments.
}

In Fig.~\ref{fig:appro1}, $B_t$ at y-axis illustrates the
theoretical upper bound of
$|E(\eulerian(G))|-|E(\appeulerian(G))|=\frac{K-1}{K+1}|E|$ by
solving the LP problem, where the three solid lines represent the
three conditions in the above LP problem, respectively. Here, $K$ is
the maximum among all $k$ values. The theoretical upper bound is far
from tight. First, $|E({\cal G})| \ll |E|$, which is a tighter upper
bound of $\sum_{C_i}(\triangle_{C_i}'+\triangle_{C_i})$, moving
Cond-3 towards the origin. Second, for most \kcycles,
$\triangle_k'=(1+\epsilon)\cdot \triangle_k$, $0<\epsilon<1$, since
most \ppaths in a \kcycle are far from the upper bound it can
get. This leads Cond-2 moving towards x-axis. Therefore, a tighter
empirical upper bound is $B_p$ at y-axis in Fig.~\ref{fig:appro2}.
We will show it in the experiments.

\begin{figure}[t]
\begin{center}
\includegraphics[scale=0.3]{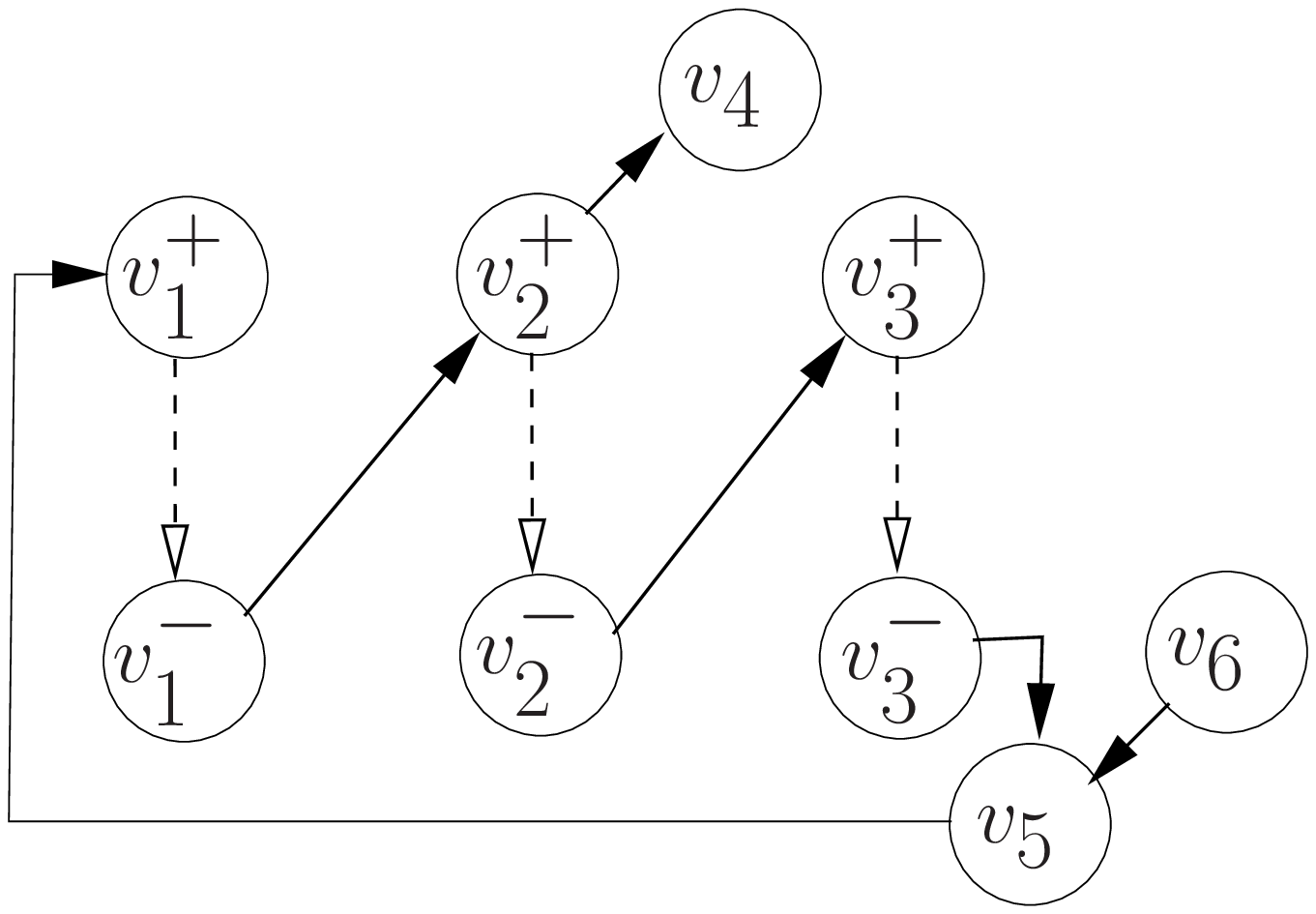}
\end{center}
\vspace*{-0.4cm}
\caption{General \ppaths}
\vspace*{-0.4cm}
\label{fig:fig8}
\end{figure}

We have proved Theorem~\ref{them:upper} for the case \ppaths and
\npaths are \pnpaths,
which shows that each \ppath in a \kcycle has an implicit upper
bound. In general, there are cases where \ppaths are not \pnpaths.
In fact, each \ppath in a \kcycle can be classified into two
classes.
(a) A \ppath is a part of a \pnpath, including the case
  that the \ppath is a \pnpath.
(b)
A \ppath can be divided into several sub-paths, each
is a part of a \pnpath.
%
In Fig.~\ref{fig:fig8}, there are three \ppaths in the \kcycle,
$(v^-_{1},v^+_{2})$ is a part of \pnpath$(v_1^-,v_4)$, $(v^+_{2},v^-_{3})$
itself is \pnpath$(v^-_{2},v^+_{3})$ and $(v^-_{3},v^+_{1})$ consists of
two sub-paths: $(v^-_{3},v_5)$ and $(v_5,v^+_{1})$, and each of them is
a part of a \pnpath or itself is a \pnpath.

For the cases when a \ppath in a \kcycle is not a \pnpath, we use
$w_p$ and $w_u$ to denote its practical weight and the theoretical
upper bound it can reach when itself is a \pnpath, respectively.
Since we concentrate on weight of \ppaths, we treat such a \ppath as
a \pnpath with weight $w_p$ if $w_p < w_u$, and treat it as a
\pnpath with weight $w_u$ if $w_p>w_u$ and add the difference
$w_p-w_u$ to a global variable $W$. We will show in
Section~\ref{sec:performance} that $W$ is very small compared with
$|E(\eulerian(G))|$.

\stitle{Time complexity}: Revisit \refineAlg (Algorithm~\ref{alg:gr}),
it includes four parts: \scc decomposition (Line~1), \greedy (Line~3),
cycle moving (Line~4) and \refine (Line~5). \scc decomposition can be
accomplished in 2 \DFS, in time $O(n+m)$. As analyzed in
Section~\ref{sec:gr}, \greedy invokes $l_{max}$ times \lengthL, and
each \lengthL needs 2 \BFS (\lsubgraph) and 1 \DFS (remove/reverse
\pnpaths). Since $l_{max}$ is small ($< 100$ in our extensive
experiments), the time complexity of \greedy is $O(n+m)$. Regarding
moving cycles from $G_i-\appeulerian(G_i)$ to $\appeulerian(G_i)$, it
is equivalent to moving cycles from non-trivial \sccs of
$G_i-\appeulerian(G_i)$ to $\appeulerian(G_i)$. Based on the fact that
$G_i-\appeulerian(G_i)$ is near acyclic, there are a few cycles in
$G_i-\appeulerian(G_i)$, cycle moving is in $O(n+m)$. The time
complexity of \refine, as given in
Section~\ref{sec:refine} is $O(cm^2)$,
because most \DFSSPFA($G,u$) relax edges along a path with a few branches
and vertices $u$ will have $dst(u)$ updated less than
$|E(\eulerian(G))|-|E(\appeulerian(G))|$ times.
%

\comment{
\stitle{Time complexity}: Revisit Algorithm~\ref{alg:gr}, it includes
three parts: \scc computing, \greedy, and \refine. Both \scc computing
and \greedy can be accomplished in $O(n+m)$ time as analyzed in
previous sections. For the \refine part, we have given both
theoretical and empirical upper bounds of the gap between the greedy
solution $\appeulerian(G)$ and the optimal
$\eulerian(G)$. Approximately, we can regard $|E(\eulerian(G))|$-
$|E(\appeulerian(G))|$ as $c \cdot m$ where $c$ is a small constant,
ranging from 1/100 to 1/10 as illustrated in the
experiments. Therefore, the \refine part dominates
Algorithm~\ref{alg:gr} and the time complexity of
Algorithm~\ref{alg:gr} can be approximated as $O(c \cdot m^2)$, and
our work is to make the constant $c$ small and make
Algorithm~\ref{alg:gr} practical.
}

\section{Performance Studies}
\label{sec:performance}

We conduct extensive experiments to evaluate two proposed \refineAlg
algorithms. One is \refineAlgD using \greedyD
(Algorithm~\ref{alg:greedyD}) and \refine (Algorithm~\ref{alg:OR}),
and the other is \refineAlgR using \greedyR
(Algorithm~\ref{alg:greedyR}) and \refine (Algorithm~\ref{alg:OR}). We
do not compare our algorithms with \simpleAlg in
\cite{gupte2011finding}, because \simpleAlg is in $O(nm^2)$ and is too
slow.  We use our \DFSEVEN as the baseline algorithm, which is
$O(m^2)$.  We show that \greedy produces an answer which is very close
the the exact answer.  In order to confirm \greedy is of time
complexity $O(n+m)$, we show the largest iteration $l_{max}$ used in
\greedy is a small constant by showing that the longest \pnpath (the
same as $l_{max}$) deleted/reversed by \greedy is small. In addition,
we confirm the constant $c$ of $O(c\cdot m^2)$ for \refine
is very small by showing statistics of ${\cal G}$, $W$, and \kcycles.
We also confirm the scalability of \refineAlg as well as \greedy and
\refine.
%
%

All these algorithms are implemented in C++ and complied by gcc 4.8.2,
and tested on machine with 3.40GHz Intel Core i7-4770 CPU, 32GB RAM
and running Linux. The time unit used is second.

\begin{table}[t]
{\scriptsize 
\begin{center}
    \begin{tabular}{|l||r|r|r|r|} \hline
    {\bf Graph}  & $|V|$ & $|E|$ & $|V(\eulerian(G))|$ & $|E(\eulerian(G))|$ \\ \hline\hline
    wiki-Vote    &  7,115    &    103,689 &     1,286 &     17,676\\\hline
    Gnutella     & 62,586    &    147,892 &    11,952 &     18,964\\\hline
    Epinions     & 75,879    &    508,837 &    33,673 &    264,995\\\hline
    Slashdot0811 & 77,360    &    828,159 &    70,849 &    734,021\\\hline
    Slashdot0902 & 82,168    &    870,159 &    71,833 &    748,580\\\hline
    web-NotreDame& 325,729   &  1,469,679 &    99,120 &    783,788\\\hline
    web-Stanford & 281,903   &  2,312,497 &   211,883 &    691,521\\\hline
    amazon       &   403,394 &  3,387,388 &   399,702 &  1,973,965\\\hline
    Wiki-Talk    & 2,394,385 &  5,021,410 &   112,030 &  1,083,509\\\hline
    web-Google   &   875,713 &  5,105,039 &   461,381 &  1,841,215\\\hline
    web-BerkStan &   685,230 &  7,600,595 &   478,774 &  2,068,081\\\hline
    Pokec        & 1,632,803 & 30,622,560 & 1,297,362 & 20,911,934\\\hline

    \end{tabular}
\end{center}
}
\vspace*{-0.6cm}
\caption{Summary of real Datasets}
\vspace*{-0.2cm}
\label{tbl:realdataset}
\end{table}

\stitle{Datasets:} We use 14 real datasets. Among the datasets,
Epinions, wiki-Vote, Slashdot0811, Slashdot0902, Pokec, Google+, and
Weibo are social networks; web-NotreDame, web-Stanford, web-Google,
and web-BerkStan are web graphs; Gnutella is a peer-to-peer network;
amazon is a product co-purchasing network; and Wiki-Talk is a
communication network. All the datasets are downloaded from Stanford
large network dataset collection (\url{http://snap.stanford.edu/data})
except for Google+ and Weibo. The detailed information of the datasets
are summarized in Table~\ref{tbl:evolving} and
Table~\ref{tbl:realdataset}.  In the tables, for each graph, the 2nd
and 3rd columns show the numbers of vertices and edges\footnote{for
  each dataset, we delete all self-loops if exist.}, respectively, and
the 4th and 5th columns show the numbers of vertices and edges of its
maximum Eulerian subgraph, respectively.

\begin{table}[t]
{\scriptsize 
\begin{center}
    \begin{tabular}{|@{}l@{}||@{}r|@{}r||@{}r|@{}r||@{}r| |r|} \hline
    {\bf Graph}   & {\bf \refine} & {\bf \refineAlgD} & {\bf \refine} & {\bf \refineAlgR} & {\bf \DFSEVEN} & {$c$} \\ \hline\hline
    wiki-Vote     &   0.1 &    0.1 &   0.1 &    0.1 &      1.0  &0.100 \\\hline
    Gnutella      &   0.5 &    0.5 &   0.4 &    0.4 &      1.6 &0.250 \\\hline
    Epinions      &  15.9 &   16.1 &  15.2 &   15.4 &    414.4 &0.037\\\hline
    Slashdot0811  & 80.6 &  80.8 &  70.9 &   71.0 & 12,748.6 &0.006 \\\hline
    Slashdot0902  & 87.3 &  87.5 & 76.6 &  76.8 & 14,324.5  &0.005 \\\hline
    web-NotreDame &   2.6 &    3.0 &   2.4 &    2.7 &    370.4 &0.007 \\\hline
    web-Stanford  &  21.5 &   25.7 &  16.7 &   24.9 &  2,780.0 &0.009 \\\hline
    amazon        & 126.5 &  133.5 & 124.8 &  130.5 & 44,865.0 &0.003 \\\hline
    Wiki-Talk     & 504.3 &  504.9 & 487.3 &  487.9 & 9,120.1 &0.053 \\\hline
    web-Google    & 100.2 &  110.3 & 78.6 &  84.6 & 35,271.7 &0.002 \\\hline
    web-BerkStan  & 129.7 &  137.7 &  67.8 &   75.9 & 7,853.9 &0.010 \\\hline
    Pokec      & 30,954.5 & 30,983.7 & 30,120.4 & 30,140.5 & - &- \\\hline
    Gplus2        & 363.5  &364.2  &360.5    &361.2  &39,083.8  &0.009   \\\hline
    Weibo0        & 206.5 &  207.3 &202.4  &203.3   &8,004.6  &0.025  \\\hline
    \end{tabular}
\end{center}
}
\vspace*{-0.6cm}
\caption{Efficiency of \refineAlgD, \refineAlgR and \DFSEVEN}
\vspace*{-0.4cm}
\label{tbl:efficiency}
\end{table}

\stitle{Efficiency}: Table~\ref{tbl:efficiency} shows the efficiency
of these three algorithms, i.e., \refineAlgD, \refineAlgR, and
\DFSEVEN, over 14 real datasets.  For \refineAlgD, the 2nd column
shows the running time of \refine and the 3rd column shows the total
running time of \refineAlgD. As can be seen, for \refineAlgD, the
running time of \refine dominates that of \greedyD. The 4th and 5th
columns show the running time of \refine and the total running time of
\refineAlgR, respectively. Likewise, the \refine algorithm is the most
time-consuming procedure in \refineAlgR. It is important to note that
both \refineAlgD and \refineAlgR significantly outperform \DFSEVEN. In
most large datasets, \refineAlgD and \refineAlgR are two orders of
magnitude faster than \DFSEVEN. For instance, in web-Stanford dataset,
\refineAlgR takes 25 seconds to find the maximum Eulerian subgraph,
while \DFSEVEN takes 2,780 seconds, which is more than 100 times
slower. In addition, it is worth mentioning that in Pokec dataset,
\DFSEVEN cannot get a solution in two weeks. In the 6th column, $c$ is
the $c$ value in \refine's time complexity $O(cm^2)$, by comparing
running time of \refineAlgR and \DFSEVEN. In all graphs, $c \ll 1$.
Note \simpleAlg is very slow, for example, \simpleAlg takes more than
30,000 seconds to handle the smallest dataset wiki-Vote, while our
\refineAlg takes only 0.1 second.

\begin{figure}[t]
\begin{center}
    \includegraphics[width=0.9\columnwidth,height=2.5cm]{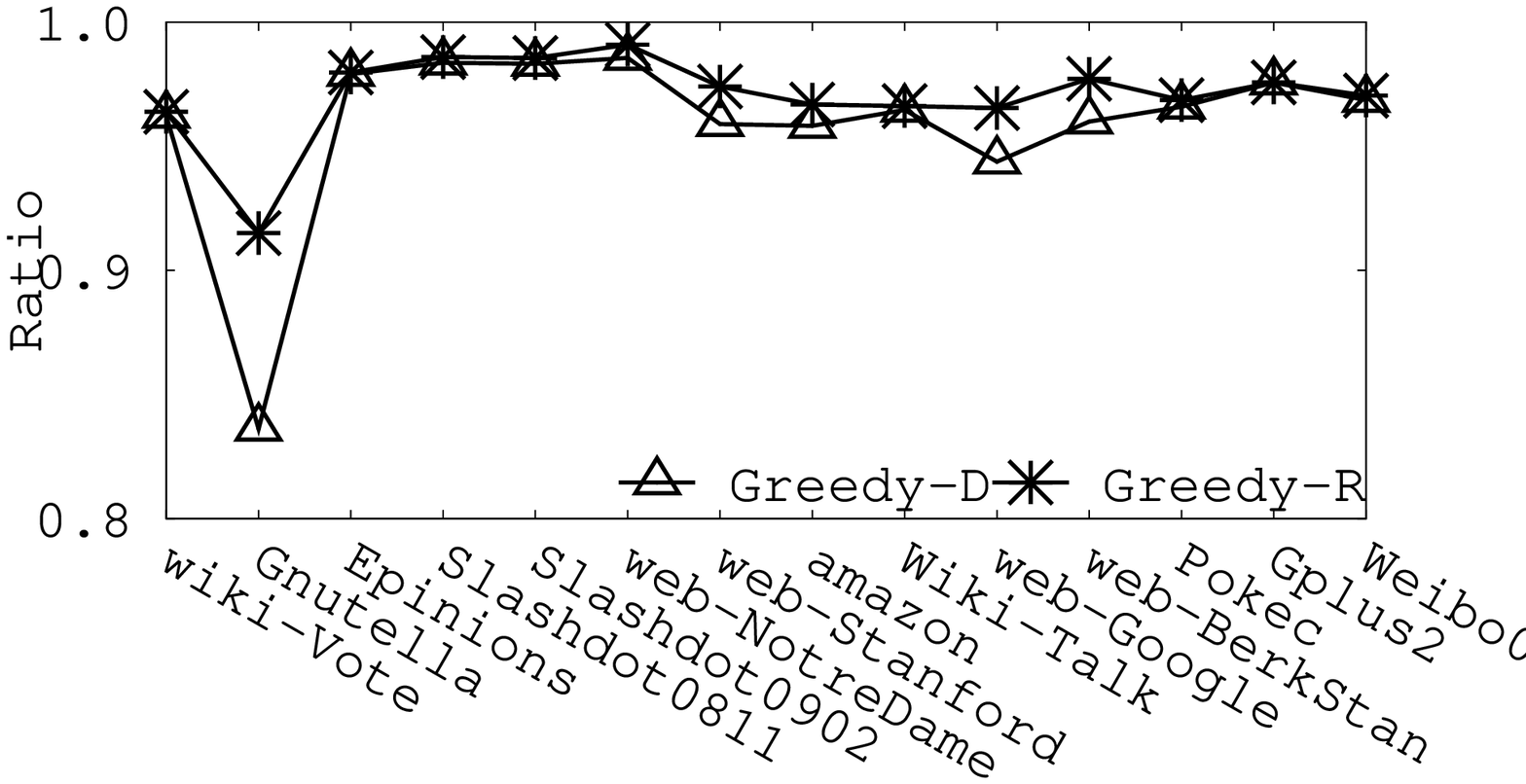}
\end{center}
\vspace*{-0.4cm}
\caption{$|E(\appeulerian(G))|/|E(\eulerian(G))|$}
\vspace*{-0.4cm}
\label{fig:near_optimal}
\end{figure}

\begin{table}[t]
{\scriptsize 
\begin{center}
    \begin{tabular}{|l@{}||r|@{}r||r|@{}r||@{}r|} \hline
    {\bf Graph}  & {\bf IRD} & {\bf ISD \%} & {\bf IRR} & {\bf ISR \%}  & {\bf IR\_DSU} \\ \hline\hline
    wiki-Vote     &    659 & 95.4 & 629 & 95.6 & 14,361 \\    \hline
    Gnutella      &  2,504 & 69.5 &1,410 & 82.8 & 8,202 \\    \hline
    Epinions      &  5,466 & 97.4 & 5,334 & 97.4 & 207,124\\    \hline
    Slashdot0811 & 11,464 & 97.9 &9,990& 98.2 & 541,970  \\    \hline
    Slashdot0902 & 12,036 & 97.8 &10,426 & 98.1 & 554,163\\    \hline
    web-NotreDame &  9,030 & 98.1 & 6,119 & 98.7 & 486,240 \\    \hline
    web-Stanford  & 23,427 & 94.8 & 15,721 & 96.5 & 448,960 \\    \hline
    amazon        & 75,104 & 94.1 & 61,818 & 95.2 & 1,282,326\\    \hline
    Wiki-Talk     & 37,662 & 95.7 & 36,139 & 95.9 & 871,020 \\    \hline
    web-Google        & 90,375 & 92.4 & 59,387 & 95.0 & 1,196,616 \\    \hline
    web-BerkStan  & 69,078 & 95.2 & 41,703 & 97.1 & 1,437,188 \\ \hline
    Pokec         & 686,765& -    & 635,286 & -   & - \\ \hline
    Gplus2        &18,766    &96.9      &18,721  &96.9    &613,008 \\ \hline
    Weibo0        &25,991 &96.2      &24,550    &96.4    &686,765	\\ \hline
    \end{tabular}
\end{center}
}
\vspace*{-0.4cm}
\caption{The numbers of Iterations}
\vspace*{-0.2cm}
\label{tbl:iteration}
\end{table}

\stitle{Effectiveness of \greedy}: To evaluate the effectiveness of
the greedy algorithms, we first study the size of Eulerian subgraph
obtained by \greedyD and \greedyR. Fig.~\ref{fig:near_optimal} depicts
the results. In Fig.~\ref{fig:near_optimal}, $|E(\appeulerian(G))|$
denotes the size of Eulerian subgraph obtained by the greedy
algorithms, $| E(\eulerian(G))|$ denotes the size of the maximum
Eulerian subgraph, and $|E(\appeulerian(G))|/| E(\eulerian(G))|$
denotes the ratio between them. The ratios obtained by both \greedyD
and \greedyR are very close to 1 in most datasets. That is to say,
both \greedyD and \greedyR can get a near-maximum Eulerian subgraph,
indicating that both \greedyD and \greedyR are very effective.  The
performance of \greedyR is slightly better than that of \greedyD,
which supports our analysis.
%
%
The ratio of Gnutella dataset using \greedyD is slightly lower than
others. One possible reason is that Gnutella is much sparser than
other datasets, thus some inappropriate \pnpath deletions may result
in enlarging other \pnpaths, and this situation can be largely
relieved in \greedyR.

Second, we investigate the numbers of iterations used in \refineAlgD,
\refineAlgR, and \DFSEVEN. Table~\ref{tbl:iteration} reports the
results. In Table~\ref{tbl:iteration}, the 2nd and 4th columns `IRD'
and `IRR' denote the numbers of iterations used in the refinement
procedure (i.e., \refine, Algorithm~\ref{alg:OR}) of \refineAlgD and
\refineAlgR, respectively. The last column `IR\_DSU' reports the total
number of iterations used in \DFSEVEN. From these columns, we can see
that in large graphs (e.g., web-NotreDame dataset), the numbers of
iterations used in \refine of \refineAlgD and \refineAlgR are at least
two orders of magnitude smaller than those used in \DFSEVEN. In
addition, it is worth mentioning that in Pokec dataset, \DFSEVEN
cannot get a solution in two weeks. The 3rd and 5th columns report the
percentages of iterations saved by \refineAlgD and \refineAlgR,
respectively. Both \greedyD and \greedyR can reduce at
least 95\% iterations in most datasets.  Similarly, the results
obtained by \refineAlgR are slightly better than those obtained by
\refineAlgD. 
%

\comment{
\begin{figure}[t]
\begin{center}
\begin{tabular}[t]{c}
    \subfigure[\greedyD]{
         \includegraphics[width=0.45\columnwidth,height=2.5cm]{chart/pnpath-GreedyD}
         \label{fig:disD}
    }

    \subfigure[\greedyR]{
        \includegraphics[width=0.45\columnwidth,height=2.5cm]{chart/pnpath-GreedyR}
        \label{fig:disR}
    }
\end{tabular}
\vspace*{-0.2cm}
\caption{Distribution of paths deleted/reversed by \greedy}
\label{fig:disGreedy}
\vspace*{-0.4cm}
\end{center}
\end{figure}
}

\stitle{The largest iteration $l_{max}$:} We show the largest
iteration $l_{max}$ in \greedy by showing the longest \pnpaths
deleted/reversed, which is the numbers of \lengthLD/\lengthLR invoked
by \greedyD/\greedyR using the real datasets. Below, the first/second
number is the longest \pnpaths deleted/reversed.
wiki-Vote (9/9),
Gnutella (29/22),
Epinions  (12/10),
Slashdot0811 (6/6),
Slashdot0902 (8/8),
web-NotreDame (96/41),
web-Stanford (275/221),
amazon (57/37),
Wiki-Talk (9/7),
web-Google (93/37),
web-BerkStan (123/85),
Pokec  (14/13),
Gplus2 (9/8), and
Weibo0 (12/10).
%
%
The longest \pnpaths deleted or reversed are always of small size,
especially compared with $|E|$. Therefore, the time complexity of
\greedy can be regarded as $O(n + m)$.

\begin{table}[t]
{\scriptsize
\begin{center}
    \begin{tabular}{|l@{}|r|@{}r|@{}r|} \hline
    {\bf Graph} &  $|E(\eulerian(G))|$ & $|E({\cal G})|$ & $W$ \\ \hline\hline
    wiki-Vote   &17,676  &3,214 &20 \\ \hline
    Gnutella   &18,964  &6,906  &10 \\ \hline
    Epinions   &264,995   &30,997 &129 \\ \hline
    Slashdot0811  &734,021  &45,315 &118 \\ \hline
    Slashdot0902  &748,580  &46,830 &145 \\ \hline
    web-NotreDame  &783,788 &10,439  &3,963 \\ \hline
    web-Stanford  &691,521  & 35,402 &6,168 \\ \hline
    amazon    &1,973,965  &202,513  &12,994 \\ \hline
    Wiki-Talk  &1,083,509 &158,848  &331 \\ \hline
    web-Google  &1,841,215  & 149,425  &22,361 \\ \hline
    web-BerkStan  &2,068,081 &105,569  &16,991 \\ \hline
    Pokec   & 20,911,934   & 3,003,797    &8,964  \\ \hline
    Gplus2  & 770,854      & 117,641     & 80   \\ \hline
    weibo0  & 850,136      & 124,395     & 384   \\ \hline
    \end{tabular} 
\end{center}
}
\vspace*{-0.4cm}
\caption{Statistics of $|{\cal G}|$ and $W$}
\label{tbl:statg}
\vspace*{-0.4cm}
\end{table}

\comment{
\begin{figure}[t]
\begin{center}
\hspace*{-1cm}
\includegraphics[width=1.2\columnwidth,height=4cm]{chart/kc-resnew.eps}
\vspace*{-2em}
\caption{Distributions of \kcycles for each $k$}\label{fig:kcylenew}
\end{center}
\vspace*{-0.5cm}
\end{figure}
}

\comment{
\begin{figure}[t]
\begin{center}
\begin{tabular}[t]{c}
    \subfigure[Epinions]{
         \includegraphics[width=0.5\columnwidth,height=3cm]{chart/kc-Epinions}
    }
    \subfigure[web-Stanford]{
        \includegraphics[width=0.5\columnwidth,height=3cm]{chart/kc-web-Stanford}
    }
\end{tabular}
\end{center}
\vspace*{-0.4cm}
\caption{Distributions of \kcycles for each $k$}
\vspace*{-0.4cm}
\label{fig:kcylenew}
\end{figure}
}

\begin{figure}[t]
\begin{center}
\includegraphics[width=\columnwidth]{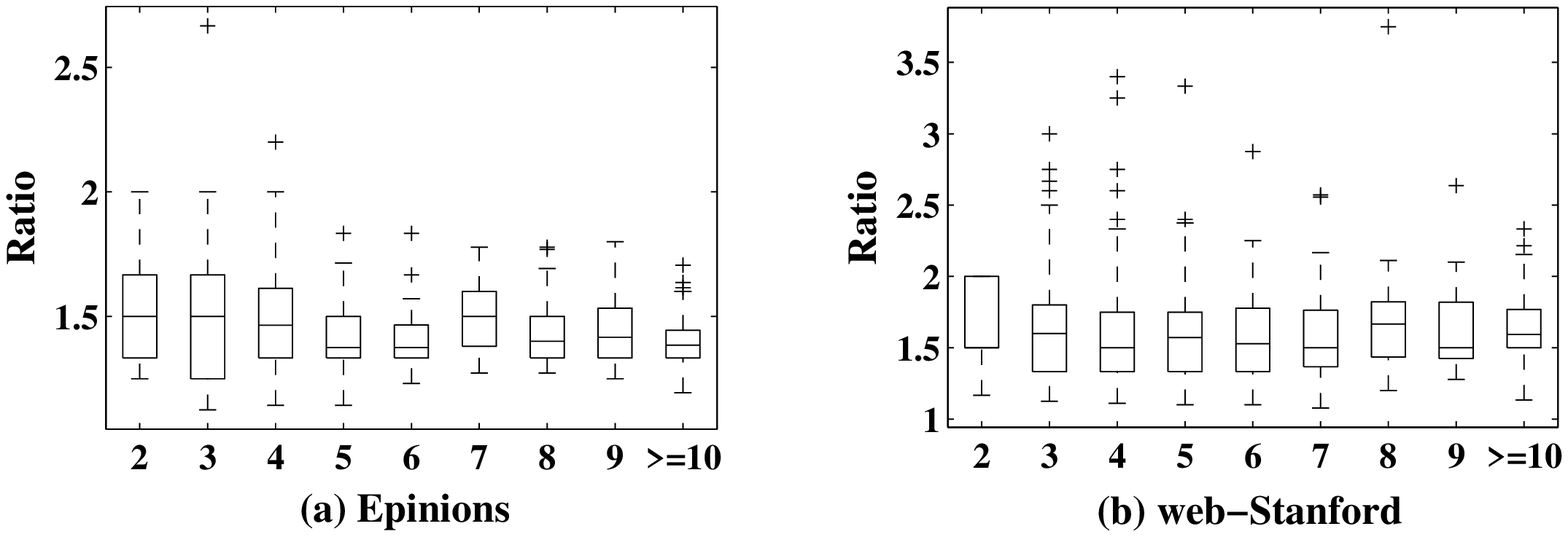}
\end{center}
\vspace*{-0.8cm}
\caption{Distributions of \kcycles for each $k$}
\vspace*{-0.4cm}
\label{fig:kcylenew}
\end{figure}

\stitle{The support to a small $c$:} We show the support that $c$
given in $O(cm^2)$ for \refine is small by giving statistics of ${\cal
  G}$, $W$, and \kcycles. We first show the statistics of $|{\cal
  G}|~{}(= \overline{G_P} \oplus G_N)$ and $W$ discussed in
Section~\ref{sec:bound}. Table~\ref{tbl:statg} reports the
results. From Table~\ref{tbl:statg}, we can find that for each graph,
$|E(\cal {G})|$ and $W$ are small compared with
$|E(\eulerian(G))|$. These results confirm our theoretical analysis in
Section~\ref{sec:bound}. Second, we study the statistics of
\kcycles. The results of Epinions and web-Stanford datasets are
depicted in Fig.~\ref{fig:kcylenew}, and similar results can be
observed from other datasets. In Fig.~\ref{fig:kcylenew}, y-axis
denotes the ratio between the total weights of \pedges and the total
weights of \nedges (i.e., $\Delta_k^\prime / \Delta_k$ defined in
Section~\ref{sec:bound}), and the x-axis denotes $k$ for \kcycles,
where $k=2, 3, \cdots, >=10$. As can be seen, for all \kcycles, the
ratios are always smaller than 2 in both Epinions and web-Stanford
datasets. These results confirm our analysis in
Section~\ref{sec:bound}.

\comment{
\begin{figure}[t]
\begin{center}
\begin{tabular}[t]{c}
    \subfigure[web-NotreDame]{
\includegraphics[width=0.45\columnwidth,height=2.5cm]{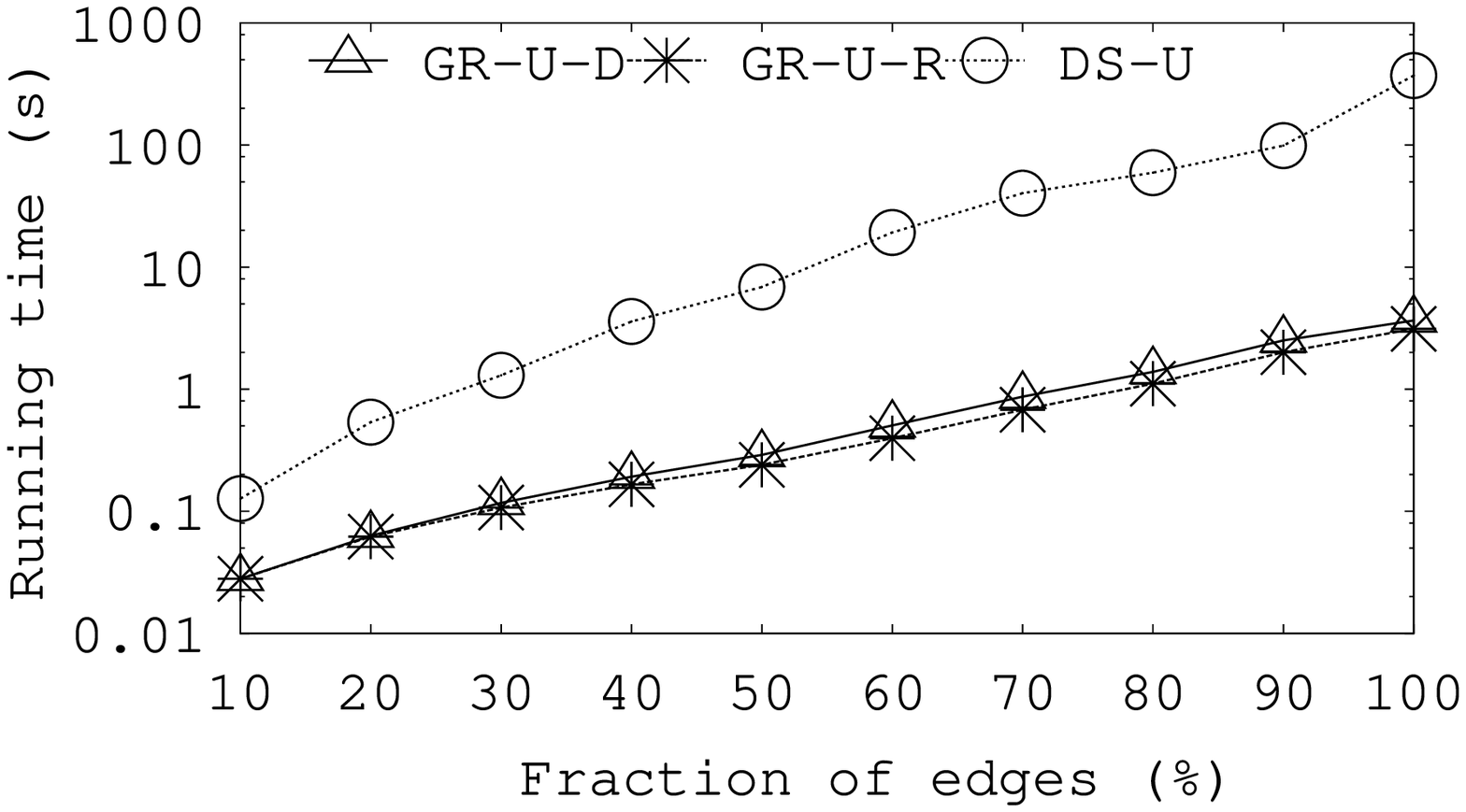}
         \label{fig:ScalaNotreDame}
    }
    \subfigure[web-Stanford]{
\includegraphics[width=0.45\columnwidth,height=2.5cm]{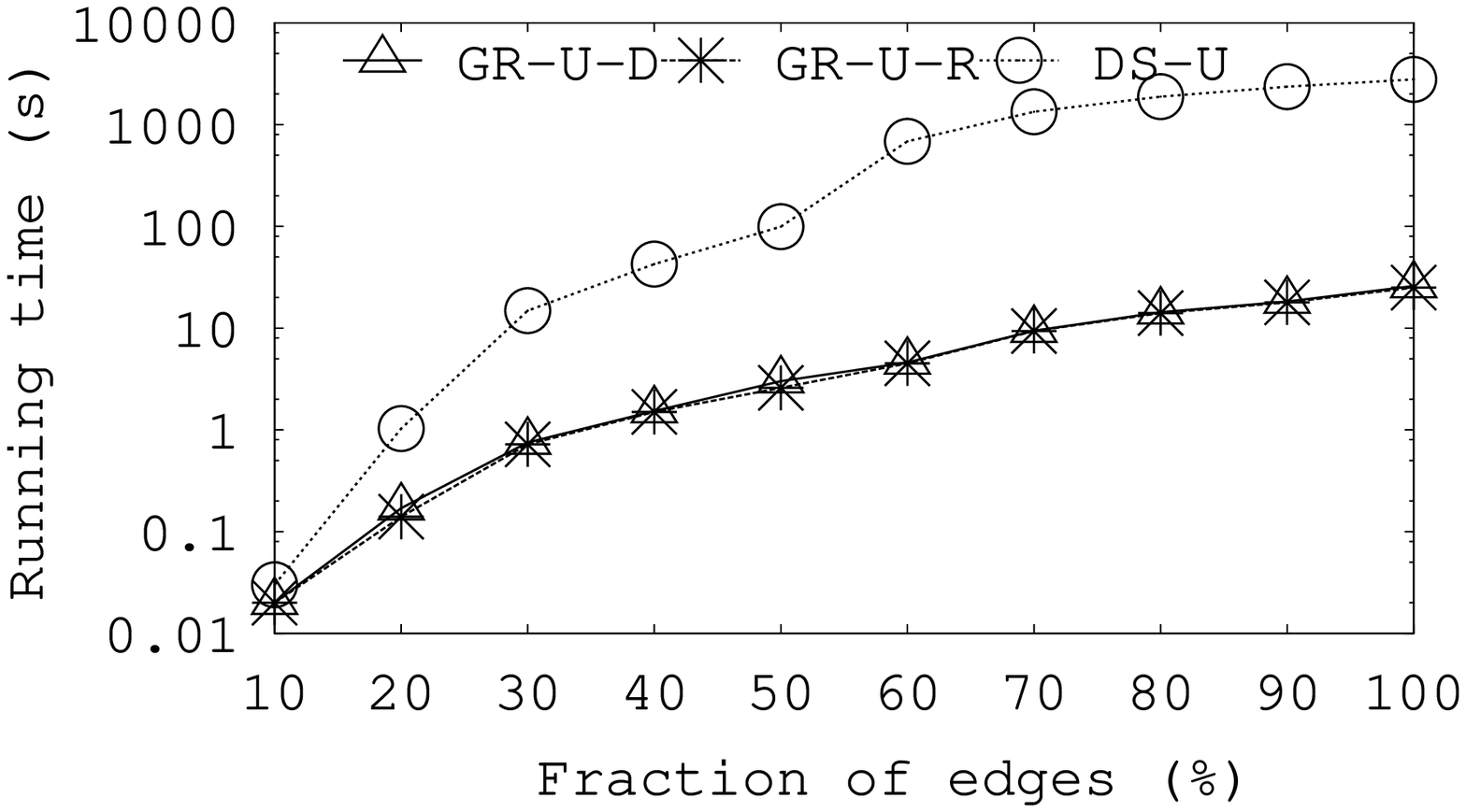}
       \label{fig:ScalaStanford}
    }
\end{tabular}
\end{center}
\vspace*{-0.6cm}
\caption{Scalability testing on web-NotreDame and web-Stanford}
\vspace*{-0.4cm}
\label{fig:scale}
\end{figure}

\begin{figure}[t]
\begin{center}
\begin{tabular}[t]{c}
    \subfigure[\greedy]{
\includegraphics[width=0.45\columnwidth,height=2.5cm]{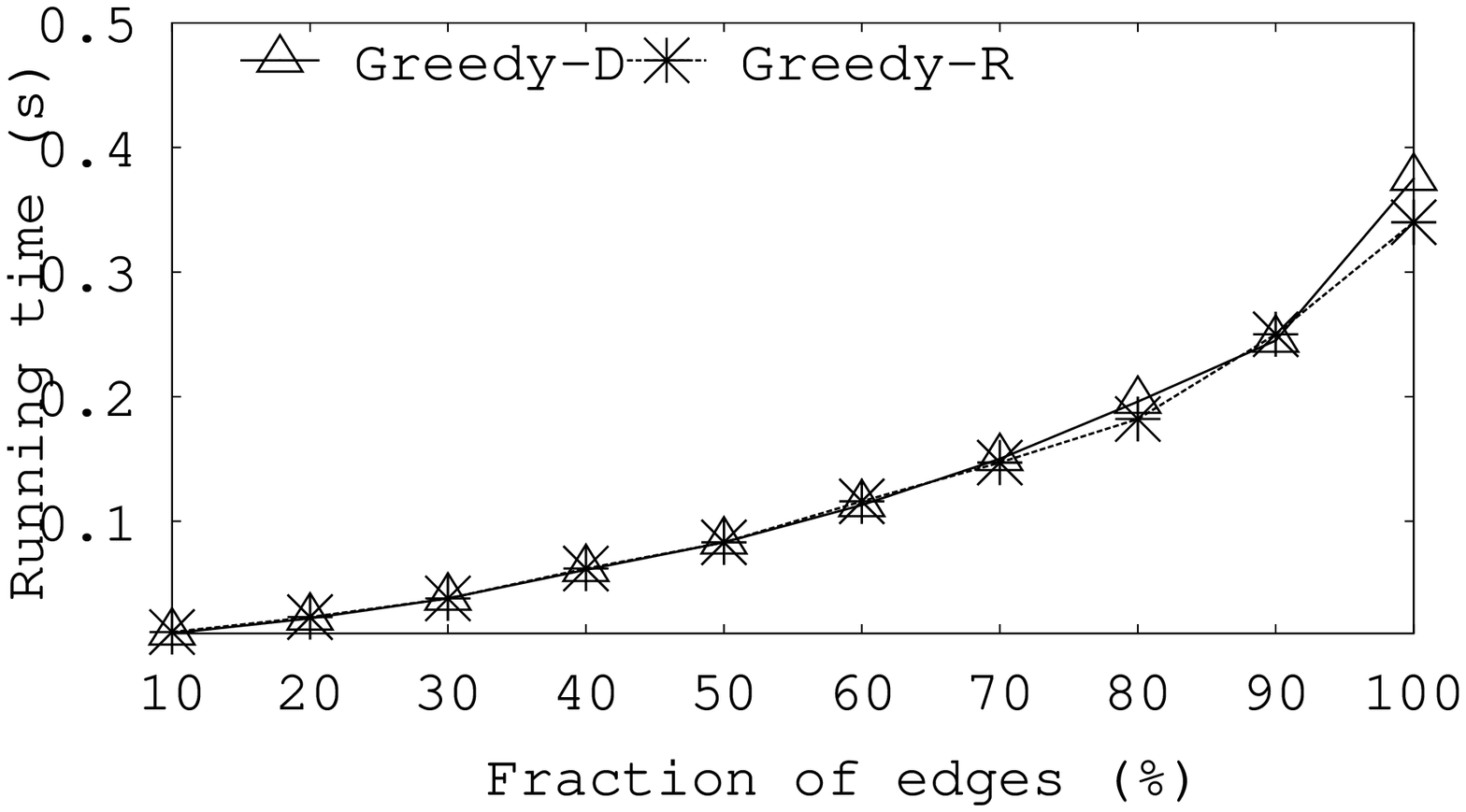}
         \label{fig:ScalaGreedy}
    }
    \subfigure[\refine]{
\includegraphics[width=0.45\columnwidth,height=2.5cm]{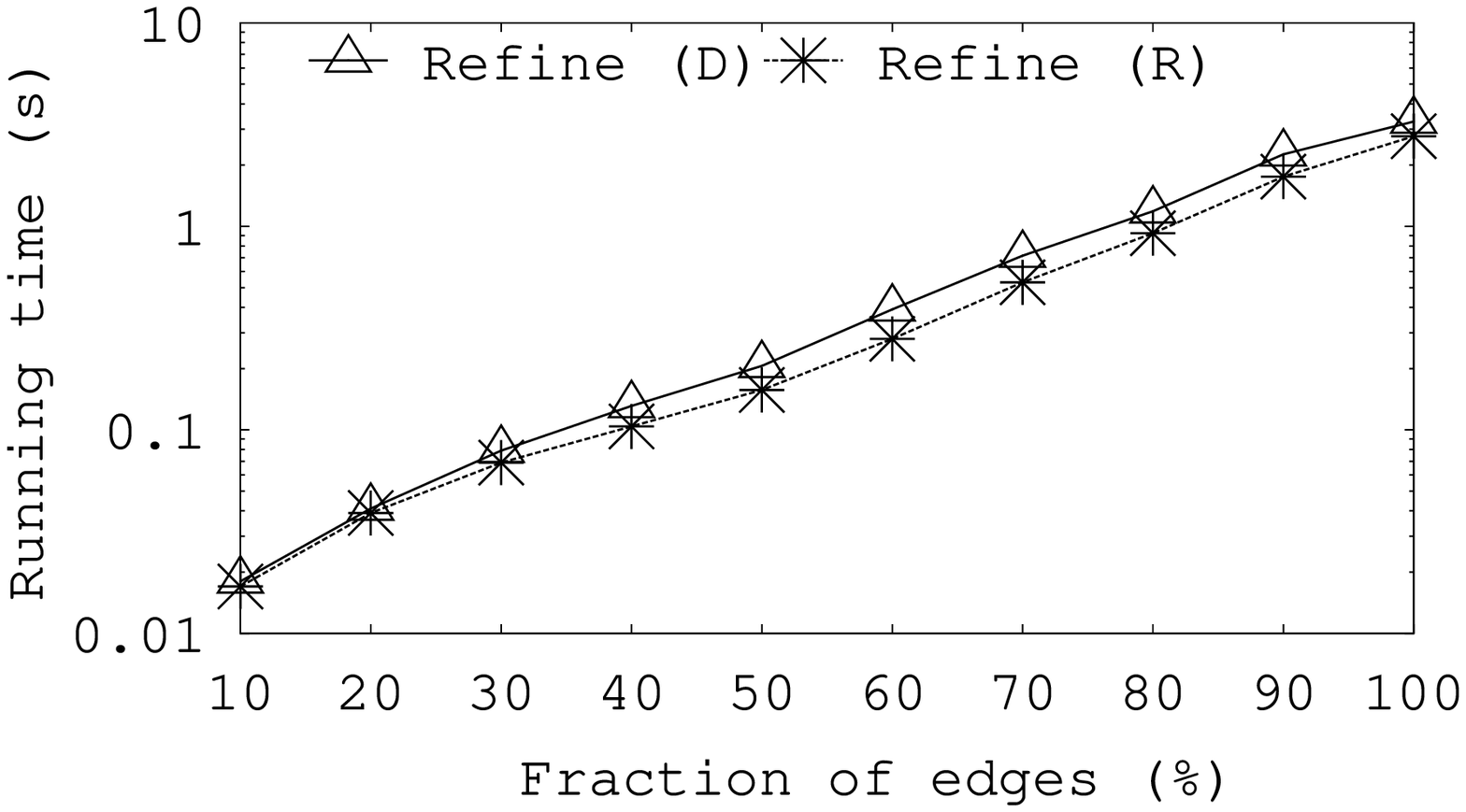}
       \label{fig:ScalaRefine}
    }
\end{tabular}
\end{center}
\vspace*{-0.6cm}
\caption{Scalability of \greedy and \refine on web-NotreDame}
\vspace*{-0.4cm}
\label{fig:scalability}
\end{figure}
}

\begin{figure}[t]
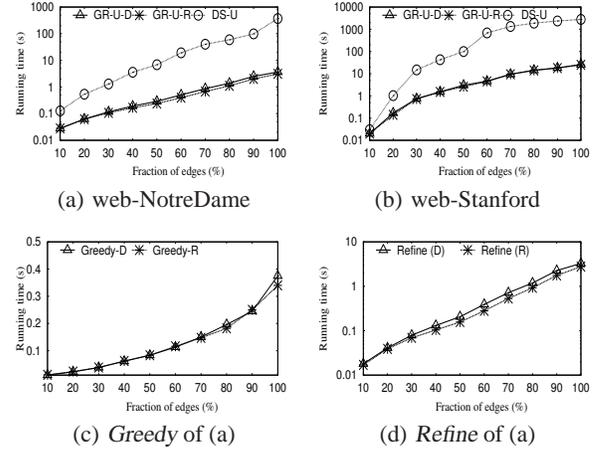

\begin{center}
\begin{tabular}[t]{c}
    \subfigure[web-NotreDame]{
\includegraphics[width=0.45\columnwidth,height=2.5cm]{chart/scalability-NotreDame}
         \label{fig:ScalaNotreDame}
    }
    \subfigure[web-Stanford]{
\includegraphics[width=0.45\columnwidth,height=2.5cm]{chart/scalability}
       \label{fig:ScalaStanford}
    }
\\
    \subfigure[\greedy of (a)]{
\includegraphics[width=0.45\columnwidth,height=2.5cm]{chart/TC-Greedy-NotreDame}
         \label{fig:ScalaGreedy}
    }
    \subfigure[\refine of (a)]{
\includegraphics[width=0.45\columnwidth,height=2.5cm]{chart/TC-Refine-NotreDame}
       \label{fig:ScalaRefine}
    }
\end{tabular}
\end{center}
\vspace*{-0.6cm}
\caption{Scalability: web-NotreDame and web-Stanford}
\vspace*{-0.4cm}
\label{fig:scalability}
\end{figure}

\stitle{Scalability}: We test the scalability for \refineAlgR,
\refineAlgD, and \DFSEVEN.  We report the results for web-NotreDame
and web-Stanford in Fig.~\ref{fig:scalability}. Similar results are
observed for other real datasets. To test the scalability, we sample
10 subgraphs starting from 10\% of edges, up to 100\% by 10\%
increments.
Fig.~\ref{fig:ScalaNotreDame} and Fig.~\ref{fig:ScalaStanford} show
both \refineAlgR and \refineAlgD scale well.
%
%
For web-NotreDame, we further show the performance of \greedy and
\refine in Fig.~\ref{fig:ScalaGreedy} and Fig.~\ref{fig:ScalaRefine}.
In Fig.~\ref{fig:ScalaGreedy}, \greedy seems to be not really linear.
We explain the reason below. Revisit Algorithm~\ref{alg:gr}, the
efficiency of \greedy is mainly determined by two factors, the graph
size (or more precisely the size of the largest \scc) and the number
of times invoking \lengthL (i.e. $l_{max}$). When a subgraph is
sparse, both \scc size and $l_{max}$ tend to be small (the smallest
sample graph with 10\% edges contains a largest \scc with 1,155
vertices and 4,317 edges, and $l_{max}=30/16$ for \greedyD/\greedyR),
whereas, both the size of the largest \scc and $l_{max}$ tend to be
large in dense subgraphs (the entire graph contains a largest \scc
with 53,968 vertices and 296,228 edges, and $l_{max}=96/41$ for
\greedyD/\greedyR).

\section{Conclusion}
\label{sec:conclusion}

In this paper, we study social hierarchy computing to find a social
hierarchy $G_D$ as DAG from a social network represented as a directed
graph $G$. To find $G_D$, we study how to find a maximum Eulerian
subgraph $\eulerian(G)$ of $G$ such that $G = \eulerian(G) \cup
G_D$. We justify our approach, and give the properties of $G_D$ and
the applications.  The key is how to compute $\eulerian(G)$.
%
We propose a \DFSEVEN algorithm to compute $\eulerian(G)$,
and develop a novel two-phase Greedy-\&-Refine
algorithm, which greedily computes an Eulerian subgraph and then
refines this greedy solution to find the maximum Eulerian subgraph.
The quality of our greedy approach is high which can be used to
support social mobility and recover the hidden directions.
We conduct extensive experiments to confirm the efficiency of our
Greedy-\&-Refine approach.

{\small
\bibliographystyle{abbrv}
\bibliography{ref}
}

\end{document}